\newif\ifXHTML\XHTMLfalse % for normal latex
\ifXHTML
       \documentclass[12pt]{article}
       \usepackage[authoryear]{natbib}
\else
       \documentclass[rmp,twocolumn,nofootinbib]{revtex4}
\fi
\usepackage{graphics}
\usepackage{epsfig}
\usepackage{graphicx}% Include figure files
\usepackage{graphics}% Include figure files
\usepackage{dcolumn}% Align table columns on decimal point
\usepackage{bm}% bold math
\usepackage{makeidx}
\usepackage[usenames]{color}

\newcommand{\sigmav}{\mbox{\boldmath$\sigma$}}
                                           
\begin{document}
\title{Theory of ferromagnetic (III,Mn)V semiconductors}

\ifXHTML
\author{%
T. Jungwirth\\
{\small Institute of Physics  ASCR, Cukrovarnick\'a~10,}\\
{\small 162~53 Praha~6, Czech Republic }\\
{\small School of Physics and Astronomy, University of Nottingham,}\\
{\small Nottingham NG7~2RD, UK}\\[2em]
Jairo Sinova\\
{\small Department of Physics, Texas A\&M University,}\\
{\small  College Station, TX 77843-4242}\\[2em]
J. Ma\v{s}ek\\
{\small Institute of Physics  ASCR, Na Slovance~2,}\\
{\small  182~21 Praha~8, Czech Republic}\\[2em]
J. Ku\v{c}era\\
{\small Institute of Physics  ASCR, Cukrovarnick\'a~10,}\\
{\small 162~53 Praha~6, Czech Republic}\\[2em]
A.~H. MacDonald\\
{\small Department of Physics, University of Texas at Austin,}\\
{\small  Austin TX 78712-1081}
}

\maketitle

\else

\author{T. Jungwirth}
\affiliation{ Institute of Physics  ASCR, Cukrovarnick\'a 10, 162 53
Praha 6, Czech Republic }
\affiliation{School of Physics and Astronomy, University of Nottingham,
Nottingham NG7 2RD, UK}
\author{Jairo Sinova}
\affiliation{Department of Physics, Texas A\&M University, College Station, TX 77843-4242}
\author{J. Ma\v{s}ek}
\affiliation{Institute of Physics  ASCR, Na Slovance 2, 182 21
Praha 8, Czech Republic }
\author{J. Ku\v{c}era}
\affiliation{Institute of Physics  ASCR, Cukrovarnick\'a 10, 162 53
Praha 6, Czech Republic}
\author{A.H. MacDonald}
\affiliation{Department of Physics, University of Texas at Austin, Austin TX 78712-1081}
\fi

\begin{abstract}  
The body of research on (III,Mn)V diluted magnetic semiconductors initiated during the 1990's has concentrated on three major fronts: i) the microscopic origins and fundamental physics of the  ferromagnetism that occurs in these systems, 
ii) the materials science of growth and defects and iii) the development of spintronic devices with new functionalities.
This article reviews the current status of the field, concentrating on the first two, more mature research 
directions.  From
the fundamental point of view,  
(Ga,Mn)As and several other (III,Mn)V DMSs are now regarded as textbook examples of a rare
class of robust ferromagnets with dilute magnetic moments coupled by delocalized charge carriers.
Both local moments and itinerant holes are provided by Mn, which makes the systems particularly
favorable for realizing this unusual ordered state. Advances in growth and post-growth treatment
techniques have played a central role in the field, often pushing the limits of  dilute Mn moment densities and
the uniformity and purity of materials far  beyond those allowed by equilibrium thermodynamics. In (III,Mn)V compounds,
material quality and magnetic properties are intimately connected.
In the review we focus on the theoretical
understanding of  the origins of ferromagnetism and basic structural, magnetic, magneto-transport, and magneto-optical
characteristics  of simple (III,Mn)V epilayers, with the main emphasis on (Ga,Mn)As. The
conclusions we arrive at are based on an extensive literature covering results of complementary
{\em ab initio} and effective Hamiltonian 
computational techniques, and on comparisons 
between theory and experiment. 
The  
applicability of ferromagnetic semiconductors in microelectronic technologies requires
 increasing Curie temperatures from the current record of 173~K in (Ga,Mn)As
epilayers to beyond  room temperature. 
The issue of whether or not this is a 
realistic expectation for (III,Mn)V DMSs is a central question in the field 
and motivates many of the analyses presented in this review. 
\end{abstract}                                                                 

\ifXHTML\cleardoublepage\else\maketitle\fi \tableofcontents

\section{Introduction}
\label{intro}

Semiconductor physics and magnetism are established subfields
of condensed matter physics that continue to reveal a rich variety 
of new phenomena, often in new types of solid state materials.
The properties of semiconductors are 
extraordinarily sensitive to impurity atoms, defects, and charges
on external gates.  Magnetism is a collective electronic phenomenon with an ordered state
that is often stable to exceptionally high temperatures.  Magnetic order, when
it is present, has a large impact on other material properties 
including transport and optical properties.  In both semiconductor
and magnetic cases, sophisticated and economically important  
technologies have been developed to exploit the unique electronic 
properties, mainly for information processing in the case of semiconductors and 
 for information storage and retrieval in the case of magnetism.  
 
The realization of materials that combine semiconducting behavior with 
robust magnetism has long been a dream of material physics.  One strategy
for creating systems that are simultaneously semiconducting and magnetic,
initiated in the late 1970's \cite{Jaczynski:1978_a,Gaj:1978_a},
is to introduce local moments into well-understood semiconductors. The result is a new class of materials  
now known as diluted magnetic semiconductors (DMSs).  Over the past 
fifteen years, building on a series of pioneering publications in the 1990's
\cite{Munekata:1989_a,Ohno:1992_a,Munekata:1993_a,Ohno:1996_a,Hayashi:1997_a,VanEsch:1997_a,Ohno:1998_a}, 
it has been established that several (III,V) compound semiconductors 
become ferromagnetic when heavily doped with Mn, and that 
the ferromagnetic transition temperatures can be well above 100~K. 
In semiconductors like GaAs and InAs, Mn has been shown to act both as an acceptor and as a source of 
local moments. These (III,Mn)V materials 
are examples of ferromagnetic semiconductors, a phrase we reserve for magnetic 
systems in which ferromagnetism is due primarily to coupling between magnetic
element moments that is mediated by conduction-band electrons or valence-band holes. 
This definition implies that in ferromagnetic semiconductors, magnetic properties can 
be influenced by the same assortment of engineering variables that are available
for other more conventional semiconductor electronic properties.  
In the best understood arsenide DMSs,  semiconductor valence-band carriers participate in the magnetic order. The
materials require participation of valence band holes for the formation of a ferromagnetic state.  Efforts to increase their 
critical temperatures further run into incompletely understood fundamental 
limits on the ratio of the magnetic transition temperature to the Fermi temperature 
of the free-carrier systems and the role of disorder in these heavily doped materials.  
The tension between achieving high Curie
temperatures and the desire for low, and therefore gateable, carrier densities,
is among the major issues in the study of these materials.

In this article we review the considerable theoretical  progress that has been
made in understanding the very broad range of properties that occur in 
(III,Mn)V ferromagnetic semiconductor epilayers in different 
regimes of Mn content and defect density.  The main focus of this article is on the extensively
studied (Ga,Mn)As ferromagnetic semiconductor, but we also make frequent comments
on other (III,Mn)V DMSs. Comparisons to experimental data are made throughout the article.
In Section~I we 
review progress that has been achieved in the effort to realize 
useful DMS materials for spintronics (or magnetoelectronics).  In Section~II we discuss the 
properties of dilute Mn atoms in a (III,V) crystal, and the various mechanisms that 
can couple the orientations of distinct moments and lead to ferromagnetism.  
In Section~III we discuss several different strategies that can be used to elevate material 
modeling from a qualitative to a more quantitative level.  Sections~IV-VII address a variety 
of different characteristics of (III,Mn)V layers, including their structural, 
magnetic, magneto-transport, and magneto-optical properties.   
Finally in Section~VIII we discuss the ferromagnetic ordering physics in (III,Mn)V DMSs in the broad
context of magnetic interactions in systems with coupled local and itinerant moments, and then
extrapolate from (III,Mn)V 
materials to comment on the effort to find high temperature ferromagnetism in other 
DMS materials.  We conclude in Section~IX with a brief summary. 

To  partially remedy  omissions in the bibliography that originate from our incomplete coverage of this topic,
we refer to  an extended  
database   of   published   work and preprints maintained
at  http://unix12.fzu.cz/ms. 
The structure of the database is similar
to the structure of this review and we encourage the
reader in need of a more detailed bibliography to use this resource.

A number of review articles  on various  aspects of  the
physics of DMSs
have been published previously and may  help the  reader who seeks a broader scope than we are able to supply 
in this review. The extensive body of research on DMSs in the 1980's, focused mostly on
(II,Mn)VI alloys, is reviewed in \cite{Furdyna:1988_a,Furdyna:1988_b,Dietl:1994_a}.
Several  extended papers  cover the experimental properties of (III,Mn)V DMSs, particularly 
(Ga,Mn)As and (In,Mn)As, 
interpreted within the carrier-mediated ferromagnetism model 
 \cite{Ohno:1999_a,Matsukura:2002_a,MacDonald:2005_a}.
Theoretical predictions based on this model for a number of properties of bulk 
DMSs and heterostructures are reviewed in  
 \cite{Dietl:2002_b,Lee:2002_a,Konig:2003_a,Dietl:2003_a}. 
 A detailed description of wide band gap and oxide DMSs can be found in 
 \cite{Pearton:2003_a,Graf:2003_a,Fukumura:2004_a,Fukumura:2005_a,Liu:2005_e}. 
We also mention here several specialized theoretical reviews focusing on the
predictions of density functional  first  principles
calculations    for   (III,Mn)V     DMSs  \cite{Sanvito:2002_b,Sato:2002_a},  on Mn-doped II-VI 
and III-V DMSs in the low
carrier density regime \cite{Bhatt:2002_a}, and on 
effects of  disorder in (Ga,Mn)As \cite{Timm:2003_a}.

\subsection{Functional (III,Mn)V material requirements}
\label{intro-why}
III-V materials are among the most widely used semiconductors.
There is little doubt that ferromagnetism in these materials would enable
a host of new microelectronics device applications if the 
following criteria were met: i) the ferromagnetic transition
temperature should safely exceed room temperature, ii) the mobile charge carriers
should respond strongly to changes in the ordered magnetic state, 
and iii) the material should retain fundamental semiconductor 
characteristics, including sensitivity to doping, electric fields produced by gate charges, and sensitivity to light.
For more than a decade these three key issues have been the focus of
intense experimental, and theoretical research into the material 
properties of Mn-doped III-V compounds.  At first sight, fundamental obstacles
appear to make the simultaneous achievement of these objectives
unlikely. Nevertheless, interest in this quest remains high because of the  
surprising progress that has been achieved. Highlights of this 
scientific endeavor are briefly reviewed in this introductory chapter. 

%%%%%%%%%%%%%%%%%%%%%%%%%%%%%%%%%%%%%%%%%%%%%%%%%%%%%%%%%%%%%%%%%%%%%%%%%%
\subsection{Search for high transition temperatures}
\label{transtemp}
Under equilibrium growth 
conditions the incorporation of magnetic Mn ions
into III-As semiconductor crystals is limited to approximately 
0.1\%.  Beyond this doping level, surface segregation and phase
separation occur. To circumvent 
the solubility problem 
a non-equilibrium, low-temperature molecular-beam-epitaxy
(LT-MBE) technique was applied and led to first successful
growth of (In,Mn)As
and (Ga,Mn)As DMS ternary
alloys with more than 1\% Mn. Since the first report in 1992
of a ferromagnetic transition in $p$-type
(In,Mn)As at a critical temperature $T_c=7.5$~K \cite{Ohno:1992_a}, the
unfolding story of critical temperature limits in (III,Mn)V DMSs
has gone through different stages. Initial experiments
in (In,Mn)As suggested an intimate relation between the ferromagnetic 
transition and carrier localization, reminiscent of the behavior of manganites 
(perovskite (La,A)MnO$_3$ with A=Ca, Sr, or Ba) in which ferromagnetism
arises from a Zener double exchange process associated with d-electron hopping
between Mn ions \cite{Coey:1999_a}.  (We comment at greater depth on qualitative pictures 
of the ferromagnetic coupling in Sections~\ref{micro-general} and \ref{discussion-general}.)  
This scenario was corroborated by a pioneering theoretical
ab-initio study of the (In,Mn)As ferromagnet \cite{Akai:1998_a} and the mechanism was also  
held responsible for mediating ferromagnetic Mn-Mn coupling in some of the first ferromagnetic
(Ga,Mn)As samples with $T_c$'s close to 50~K \cite{VanEsch:1997_a}.  

In  1998 the  Tohoku University  group announced  a jump  of  $T_c$ in
$p$-type  (Ga,Mn)As  to  110~K  \cite{Ohno:1998_a} and  pointed  out  that  the  critical
temperature  value was  consistent  with the kinetic-exchange mechanism for 
ferromagnetic coupling, also first proposed by Zener (see Section~\ref{micro-general}).  In its simplest form,
ferromagnetism in this picture follows \cite{Dietl:1997_a} from Ruderman-Kittel-Kasuya-Yosida 
(RKKY) indirect coupling between Mn $d$-shell moments mediated by 
induced spin-polarization in a free-hole itinerant carrier system (see 
Sections~\ref{micro-general}, \ref{bulk-mag-tc-sw} and \ref{discussion-general}).
Zener proposed this mechanism  originally for
transition metal ferromagnets for which the applicability of this 
picture is now known to be doubtful because of the itinerant character
of transition metal $d$-electrons.  The model of Mn($d^5$) local
moments  that  are  exchange-coupled  to  itinerant  $sp$-band  carriers
does however provide a good  description  of  
Mn-doped  IV-VI  and II-VI  DMSs \cite{Dietl:1994_a}.  The key difference between (III,Mn)V 
materials like (Ga,Mn)As and IV-VI and II-VI DMSs is that Mn substituting for the trivalent
cation (Ga) is simultaneously 
an acceptor and a source of magnetic moments.  Theoretical
critical temperature  calculations based  on the kinetic-exchange  model predict
room  temperature ferromagnetism  in (Ga,Mn)As with  10\%  Mn content.   In spite of these
optimistic predictions, the goal of breaking the 110~K record in (Ga,Mn)As remained 
elusive for nearly four years. Only recently has progress in  MBE
growth and in the development of post-growth annealing techniques 
\cite{Hayashi:2001_a,Yu:2002_a,Edmonds:2002_b,Chiba:2003_b,Ku:2003_a,Eid:2005_a}
made 
it possible to suppress extrinsic effects, pushing $T_c$ in (Ga,Mn)As up to 173~K \cite{Wang:2004_c,Jungwirth:2005_b}.
$T_c$  trends in  current high quality (Ga,Mn)As  epilayers are consistent
with the Zener kinetic-exchange model \cite{Jungwirth:2005_b}. 
The current $T_c$ record should be broken if DMS material with a higher concentration of 
substitutional Mn ions can be grown.  

Based on the few experimental and theoretical studies
reported to date, (III,Mn)Sb DMSs are expected to fall into the same 
category as (Ga,Mn)As and (In,Mn)As DMS. The kinetic-exchange model calculations predict
$T_c$'s that are small compared to their arsenide counterparts 
\cite{Dietl:2000_a,Jungwirth:2002_b}.
This difference, confirmed by experiment
\cite{Abe:2000_a,Wojtowicz:2003_a,Panguluri:2004_a,Csontos:2005_b}, 
is caused by the weaker $p-d$ exchange and smaller magnetic susceptibility (smaller effective mass)
of itinerant holes in the larger unit-cell antimonides. Also consistent with the kinetic-exchange model is
the remarkable observation of an increase of $T_c$ in (In,Mn)Sb by 25\% induced by the applied hydrostatic
pressure \cite{Csontos:2005_b}. 

Moving in the opposite direction in the periodic table towards (III,Mn)P and (III,Mn)N appears to be the natural route 
to high $T_c$ ferromagnetic semiconductors. The kinetic-exchange model predicts
$T_c$'s far above room temperature in these smaller lattice constant materials, 
in particular in (Ga,Mn)N \cite{Dietl:2000_a}. Also the 
solubility limit of Mn is much larger than in arsenides, making it possible in principle 
to grow highly Mn-doped DMSs under or close to equilibrium conditions.
However, the nature of the magnetic interactions in Mn-doped phosphides and nitrides is not completely
understood either theoretically or experimentally \cite{Liu:2005_e}. 
As the valence band edge moves closer to the Mn $d$-level 
and the $p-d$ hybridization increases with
increasing semiconductor gap width and decreasing lattice constant, charge fluctuations of the $d$-states may become large
\cite{Sanyal:2003_a,Wierzbowska:2004_a,Sandratskii:2004_a}. With increasing ionicity
of the host crystal, the Mn
impurity may also undergo a transition  
from a $d^5$ divalent acceptor  to a $d^4$ trivalent neutral 
impurity \cite{Luo:2004_a,Kreissl:1996_a,Schulthess:2005_a}.  
In either case, the picture of ferromagnetism 
based on the Zener kinetic-exchange model needs to be revisited in these materials. 

Experimental critical temperatures close to 1000~K have been reported
in some (Ga,Mn)N samples \cite{Sasaki:2002_b}.
It is still unclear, however, whether the high-temperature ferromagnetic phase should be attributed
to a (Ga,Mn)N ternary alloy or to the presence of ferromagnetic metal
precipitates embedded in the host GaN lattice. Reports of (Ga,Mn)N epilayers synthesized in cubic and hexagonal crystal structures, of p-type and n-type ferromagnetic 
(Ga,Mn)N, and of multiple ferromagnetic phases in one material all add to the complex phenomenology
of these wide-gap DMSs
\cite{Korotkov:2001_a,Graf:2002_a,Graf:2003_b,Arkun:2004_a,Edmonds:2004_d,Hwang:2005_b,Edmonds:2005_c,Sawicki:2004_c}. 

Uncertainties apply also to the interpretation of ferromagnetism seen in the
(Ga,Mn)P samples studied to date, which have been prepared by
post MBE ion-implantation of Mn followed by rapid thermal
\cite{Theodoropoulou:2002_a,Poddar:2005_a} or pulse laser melting annealing
\cite{Scarpulla:2005_a}.
Experiments in these materials 
have not yet established unambiguously the nature of magnetic interactions
in the (III,Mn)P compounds. However, a comparative study of (Ga,Mn)P and
(Ga,Mn)As prepared by the post MBE ion-implantation and pulse laser melting annealing 
suggests carrier mediated origin of ferromagnetism in the (Ga,Mn)P material \cite{Scarpulla:2005_a}.

Our current understanding of the material physics of (III,Mn)V DMS epilayers suggests that synthesis of  room temperature
ferromagnetic semiconductor will require a level of doping and defect control comparable to what has now
been achieved in 
high quality (Ga,Mn)As samples, Mn densities of order 10\%, and may require the use of wider gap III-V alloys. 

Finally, we note that efforts to enhance Curie temperature
in Mn-doped (III,V) semiconductors have also led to material research in more complex semiconductor
heterostructures with highly Mn-doped monolayers ($\delta$-doped layers), 
showing promising results \cite{Nazmul:2005_a,Kawakami:2000_a,Chen:2002_a,Fernandez-Rossier:2002_a,Nazmul:2003_a,Sanvito:2003_a,Myers:2003_a,Vurgaftman:2001_b}.  
%%%%%%%%%%%%%%%%%%%%%%%%%%%%%%%%%%%%%%%%%%%%%%%%%%%%%%%%%%%%%%%%%%%%%%%%%%%
\subsection{Conventional spintronics}
\label{mag-transp}
Spintronic devices exploit the electron spin 
to manipulate the flow of electrons and therefore require materials in which the
charge and spin degrees of freedom of carriers are strongly coupled \cite{Wolf:2001_a,DeBoeck:2002_a,Zutic:2004_a}.
The most robust, and currently the most useful, spintronic devices rely on
the collective behavior of many spins in ferromagnetic 
materials to amplify the coupling of external magnetic fields to electronic spins, a coupling that is very weak for individual electrons. The intrinsically large spin-orbit  interaction in III-Sb and III-As
valence band states makes these hosts ideal candidates for exploring
various spintronic functionalities.
In (Ga,Mn)As DMS epilayers, for example, the measured anisotropic magneto-resistance
(AMR) effect (the relative difference between longitudinal
resistivities for different magnetization orientations) 
can reach $\sim 10$\%  
\cite{Wang:2002_a,Baxter:2002_a,Jungwirth:2002_c,Tang:2003_a,Jungwirth:2003_b,Matsukura:2004_a,Wang:2005_c,Goennenwein:2004_a}.

\begin{figure}
\ifXHTML\Picture{review/figures/Fig01.png}\else\includegraphics[width=1.8in,angle=-90]{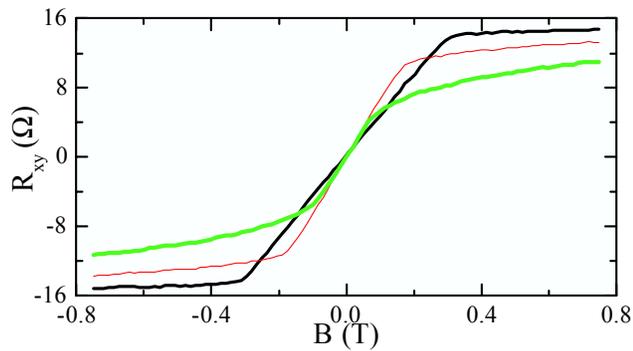}\fi
\caption{Hall resistance versus external
magnetic field for annealed Ga$_{0.94}$Mn$_{0.06}$As at 110 K (black),
130 K (red), and 140K (green). From \protect\cite{Edmonds:2002_b}.
} 
\label{ahe_nott}
\end{figure}

A particularly strong manifestation of valence-band spin-orbit coupling
occurs in the antisymmetric off-diagonal element of the resistivity 
tensor.
The anomalous Hall effect (AHE) shown in Fig.~\ref{ahe_nott}, which completely dominates the low-field
Hall response in (Ga,Mn)As and some other III-V DMSs,  
has become one of the key tools used to detect the
paramagnetic/ferromagnetic transition 
\cite{Ohno:1992_a,Ohno:1998_a}. Its large value is due to the spin-polarization of holes and provides  strong 
evidence for the participation
of mobile charge carriers in the ordered magnetic state of these DMSs.

In metals, the current response to changes in the magnetic state is
strongly enhanced in layered structures consisting of alternating ferromagnetic
and non-magnetic materials. The giant magneto-resistance effect \cite{Baibich:1988_a}
which is widely exploited in current technology, for example in field-sensors 
and magnetic random-access memories, reflects the large difference between resistivities
in configurations with parallel and antiparallel polarizations of ferromagnetic layers in
magnetic superlattices or trilayers like spin-valves and magnetic tunnel junctions \cite{Gregg:2002_a}.
The effect relies on transporting spin information between layers
and therefore is sensitive to spin-coherence times in the system.
Despite strong spin-orbit coupling which reduces spin coherence in 
DMSs, functional spintronic trilayer devices can be built, as
demonstrated by the measured large MR effects in (Ga,Mn)As based tunneling
structures \cite{Tanaka:2001_a,Chiba:2004_a,Saito:2005_a,Mattana:2005_a}. 
The coercivities of individual DMS layers can be tuned via exchange-biasing to an antiferromagnet \cite{Eid:2003_a}
which is a standard technique used in metal giant-magnetoresistance devices \cite{Gregg:2002_a}. 

%%%%%%%%%%%%%%%%%%%%%%%%%%%%%%%%%%%%%%%%%%%%%%%%%%%%%%%%%%%%%%%%%%%%%%%%%%
\subsection{Magneto-semiconducting properties and related new spintronics effects}
\label{semi}
DMS ferromagnets possess all the properties that are exploited in conventional spintronics.  
They qualify as ferromagnetic semiconductors to the extent that their magnetic and other 
properties can be altered by the usual semiconductor electronics engineering variables.
The achievement of ferromagnetism in an ordinary III-V semiconductor that  includes 
several per cent of Mn demonstrates on its own the sensitivity of
magnetic properties to doping. Remarkably, doping profiles and,
correspondingly magnetic properties can be grossly changed,
even after growth, by annealing. Early studies of (Ga,Mn)As indicated
that annealing at temperatures above the growth temperature
leads to a reduction of magnetically and electrically active Mn ions
and, at high enough annealing temperatures, to the formation of MnAs clusters \cite{VanEsch:1997_a}.
On the other hand, annealing at temperatures below the growth temperature 
can substantially improve magnetic and transport properties of the thin
DMS layers due to the out-diffusion of charge and moment compensating
defects, now identified as interstitial Mn \cite{Yu:2002_a,Edmonds:2002_b,Chiba:2003_b,Ku:2003_a,Eid:2005_a}.

(In,Mn)As based field effect transistors were built to study 
electric field control of ferromagnetism in DMSs. It has been demonstrated
that changes in the carrier density and distribution in thin film DMS
systems due to an applied bias voltage can reversibly induce the
ferromagnetic/paramagnetic transition \cite{Ohno:2000_a}. Another remarkable effect
observed in this magnetic transistor is
electric field assisted magnetization reversal \cite{Chiba:2003_a}.
This novel
functionality is based on the dependence of the width of the hysteresis
loop on bias voltage, again through the modified charge density
profile in the ferromagnetic semiconductor thin film.

\begin{figure}
\ifXHTML\Picture{review/figures/Fig02.png}\else\includegraphics[width=2.2in,angle=0]{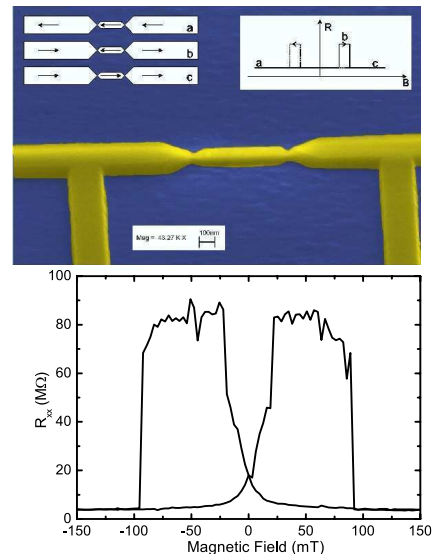}\fi
\caption{Top panel: False-color SEM picture (side-view) of a double constriction
showing part of the outer wires with the voltage
leads. The
insets show the relative magnetization of the different parts (left) and
the resulting schematic magnetoresistance trace for sweep-up (solid line) and
sweep-down (dashed line). Bottom panel: Measured magnetoresistance in a sample with tunnel barriers at the constrictions.
 From \protect\cite{Ruester:2003_a}.
} 
\label{point_cont}
\end{figure} 
Experiments in which ferromagnetism in a (III,Mn)V DMS system
is turned on and off optically add to the list of  functionalities
that result from the realization of carrier-induced ferromagnetism in a semiconductor
host material \cite{Munekata:1997_a,Koshihara:1997_a}.  
The observed emission of circularly polarized light from a 
semiconductor heterostructure, in which electrons (holes) injected from one side
of the structure recombine with spin-polarized holes (electrons) emitted
from a DMS layer \cite{Fiederling:1999_a,Ohno:1999_b}, is an example of phenomena that 
may lead to  novel magneto-optics applications. 

Tunneling  anisotropic  magnetoresistance  (TAMR) is  another novel
spintronic effect  observed in (Ga,Mn)As \cite{Gould:2004_a,Brey:2004_b,Ruester:2004_a,Saito:2005_a}. 
TAMR, like AMR, arises from  spin-orbit  coupling
and  reflects the  dependence  of the  tunneling density  of
states  of   the  ferromagnetic  layer  on  the   orientation  of  the
magnetization  with respect  to the current direction or
the crystallographic  axes. 

The larger characteristic electronic length
scales in DMSs compared to ferromagnetic metals make it possible to lithographically define
lateral structures with independent magnetic areas coupled 
through depleted regions that act as tunnel barriers and magnetic weak links.
The electrical response to magnetization reversals in 
these spintronic nanodevices can lead to MR effects with  magnitudes of order 1000\% \cite{Ruester:2003_a},
as shown in Fig.~\ref{point_cont},
and with a rich
phenomenology \cite{Giddings:2004_a}.
Wider lateral constrictions have been used to demonstrate
controlled domain-wall nucleation and propagation in DMS stripes 
\cite{Ruester:2003_a,Honolka:2005_a},
a prerequisite for developing semiconductor logic gates based on magnetic domain 
manipulation \cite{Allwood:2002_a,Allwood:2005_a}.  
(Ga,Mn)As nanoconstrictions with lateral side gates have revealed a new effect, Coulomb blockade
anisotropic magnetoresistance, which reflects the magnetization orientation dependence of the single-electron
charging energy \cite{Wunderlich:2006_a}. These spintronic single-electron transistors offer a route to non-volatile,
low-field, and highly electro- and magneto-sensitive operation.

%%%%%%%%%%%%%%%%%%%%%%%%%%%%%%%%%%%%%%%%%%%%%%%%%%%%%%%%%%%%%
\section{The origin of ferromagnetism}
\label{micro}
\subsection{General remarks}
\label{micro-general}
The magnetic dipole-dipole interaction strength between two
discrete moments separated by a lattice constant in a typical 
solid is only $\sim$ 1 Kelvin, relegating direct magnetic interactions 
to a minor role in the physics of condensed matter magnetic order.
Relativistic effects that lead to
spin-orbit coupling terms in the Hamiltonian provide more plausible
source of phenomena that are potentially useful for spintronics. 
Although these terms are critical for 
specific properties like  magnetic anisotropy,
they are rarely if ever crucial for the onset of the magnetic order
itself.  Instead the universal ultimate origin of ferromagnetism is almost always the
interplay between the electronic spin degree of freedom, repulsive Coulomb
interactions between electrons, and the fermionic quantum statistics of electrons.
The Pauli exclusion principle
correlates spin and orbital parts of the many-electron
wavefunction by requiring the total wavefunction to be
antisymmetric under particle exchange.  Whenever groups of 
electrons share the same spin state, the orbital part of the
many-body wavefunction is locally antisymmetric, lowering the probability 
of finding electrons close together and  hence the interaction energy 
of the system.  Because magnetic order is associated with the strong repulsive 
Coulomb interactions between electrons, it can persist to very high temperatures, 
often to temperatures comparable to those at which crystalline order occurs.
Ferromagnetism can be as strong as chemical bonds.  
Very often the quantum ground state of a many electron 
system has  non-zero local spin-density, either aligned in the same direction in space at 
every point in the system as in simple ferromagnets, or in non-collinear, ferrimagnetic, or 
antiferromagnetic materials in configurations in which the spin direction varies spatially. 

Although this statement on the origin of magnetic order has very general validity,
its consequences for a system of nuclei with a particular spatial arrangement, are
extraordinarily difficult to judge.  Because ferromagnetism is a strong-coupling 
phenomenon, rigorous theoretical analyses are usually not possible.  There 
is no useful universal theory of magnetism.  Understanding magnetic order in
a particular system or class of systems can be among the most challenging of
solid state physics problems \cite{Ashcroft:1976_a,Marder:1999_a}.  For most 
systems, it is necessary to proceed in a partially phenomenological way,
by identifying the local spins that order, and determining the magnitude 
and sign of the exchange interactions that couple them by comparing the 
properties of simplified (often `spin-only') model Hamiltonians with experimental 
observations.  

One approach that is totally free of phenomenological parameters is density functional theory (DFT), 
including its spin-density functional (SDF) generalizations 
in which energy functionals depend on charge and spin densities.  
Although DFT theory is
exact in principle, its application requires that the formalism's exchange-correlation 
energy functional be approximated.  Approximate forms for this 
functional can be partially phenomenological (making a pragmatic retreat from the 
{\em ab initio} aspiration of this approach) and are normally based in part on 
microscopic calculations of correlation effects in the electron gas model system. This is 
the case for the often-used local (spin) density approximation (L(S)DA) \cite{Barth:1972_a}.
For many magnetic metals, in which correlations are somewhat similar to 
those in the electron gas model system, {\em ab initio} LSDA theory provides a
practical and sufficiently accurate solution of the magnetic many-body
problem \cite{Jones:1989_a}.  This is particularly true for elemental transition metal ferromagnets 
Fe, Co, and Ni and their alloys \cite{Moruzzi:1993_a,Marder:1999_a}.  In practice LSDA 
theory functions as a mean-field theory in which the exchange-energy at 
each point in space increases with the self-consistently determined local 
spin density.  With increasing computer power, LSDA theory has been
applied to more complex materials, including DMSs.

As we discuss below, both phenomenological and DFT 
approaches provide valuable insight into (III,Mn)V ferromagnetism.  
Model Hamiltonian theories are likely to remain
indispensable because, when applicable, they provide more transparent 
physical pictures of ferromagnetism and often
enable predictions of thermodynamic, transport, and other 
properties that are sometimes (depending on material complexity), 
beyond the reach of {\em ab initio} theory
techniques. Of particular importance for DMSs is the
capability of model Hamiltonians to describe localized electronic
levels coincident with an itinerant-electron band which, strictly
speaking, is beyond the reach of  the effectively one-particle
band theories of solids 
that emerge from LSDA theory \cite{Anderson:1961_a,Schrieffer:1996_a}.
Interpreting experiments with model Hamiltonian approaches 
can, on the other hand, be misleading if the model is too simplified and 
important aspects of the physics are absent from the model.
What's more, even simplified models usually leave complex
many-body problems that cannot be completely solved.

Magnetism in (Ga,Mn)As and some other (III,Mn)V ferromagnets originates from the Mn local moments.
(As already pointed out, Mn-doped phosphides and nitrides DMSs are less well understood, however,
local Mn moments are likely to play an important role in these materials as well.)
The dependence of the energy of the system on the relative orientation 
of Mn moments  is generally referred to as an {\em exchange} 
interaction.  This terminology is part of the jargon of magnetism and 
recognizes that Fermi statistics is the ultimate origin.  
Several types of qualitative effects that lead to exchange interactions
can be separately identified when addressing magnetic order
in (III,Mn)V DMSs; the applicability of each and the relative importance
of different effects may depend on
the doping regime and on the host semiconductor material. In this section
we first introduce the jargon that is commonly used
in the magnetism literature, by briefly reviewing some of the
effects that can lead to magnetic coupling, most of 
which have been recognized since near the
dawn of quantum mechanics.

For spins carried by itinerant electrons, exchange interactions are 
often most simply viewed from a momentum space rather than a real
space point of view.  Stoner's {\em itinerant-exchange} \cite{Ashcroft:1976_a}
favors spontaneous spin-polarization of the entire electron gas because electrons are 
less likely to be close together and have strongly repulsive 
interactions when they are more likely to 
have the same spin.  Because the band energy is minimized by double-occupation of each
Bloch state, the Stoner ferromagnetic instability occurs in
systems with a large density of states at the Fermi energy. This helps to explain, for example,  
why ferromagnetism occurs in the late {\em 3d} transition elements.
A large density of states makes it possible to gain exchange energy by moving electrons from one
spin-band to the other while keeping the kinetic energy cost
sufficiently low.  Since the key spins in many (III,Mn)V DMS materials are localized the 
Stoner mechanism does not drive ferromagnetism, although we will 
see in Section~\ref{bulk-mag-tc} that it still plays a minor supporting role.  

In many systems, including (III,Mn)V DMSs, both the local nature
of the moments and strong local Coulomb interactions that 
suppress charge (valence) fluctuations play a key role and have
to be included even in a minimal model.
Many mechanisms have been identified that couple localized spins in a 
solid.  The origin of Heisenberg's {\em direct-exchange} \cite{Ashcroft:1976_a}
between two local spins is the difference between the Coulomb energy of a symmetric orbital
wavefunction (antisymmetric singlet spin wavefunction) state and an 
antisymmetric orbital wavefunction (symmetric triplet spin 
wavefunction) state. 
Kramer's {\em
super-exchange} interaction \cite{Anderson:1950_a}, 
applies to local moments that
are separated by a non-magnetic atom. In a crystal environment,
an electron can be transferred from the non-magnetic atom to an
empty shell of the magnetic atom and interact, via direct
exchange, with electrons forming its local moment. The non-magnetic 
atom is polarized and is coupled via direct-exchange
with all its magnetic neighbors. Whether the resulting
super-exchange interaction between the local moments is
ferromagnetic or antiferromagnetic depends on the relative sign of the two
direct-exchange interactions \cite{Goodenough:1958_a,Kanamori:1959_a}.  
In (III,Mn)V materials, superexchange
gives an antiferromagnetic contribution to the interaction between Mn 
moments located on neighboring cation sites. 

Zener's {\em double-exchange} mechanism \cite{Zener:1951_b} also
assumes  an
intermediate non-magnetic atom. In its usual form, this interaction 
occurs when the two isolated magnetic
atoms have a different number of electrons in the
magnetic shell and  hopping through
the intermediate non-magnetic atom involves
magnetic-shell electrons. Combined with the on-shell Hund's rule,
double-exchange couples the magnetic moments
ferromagnetically.  Parallel spin alignment is favored because it 
increases the hopping probability and therefore decreases the kinetic energy of the 
spin-polarized electrons.    A version of double-exchange, 
in which Mn acceptor states form an impurity
band with a mixed $spd$-character, has 
often been referred to in the (III,Mn)V literature. In this picture electrical conduction and 
Mn-Mn exchange coupling are both realized through hopping within an impurity band. 
The potential importance of double-exchange  is greater
at lower Mn doping and in wider-gap (III,Mn)V materials.

Finally, we identify  Zener's {\em kinetic-exchange} \cite{Zener:1951_a}
or indirect exchange interaction.  It
arises in models with local, usually $d$-shell or $f$-shell, moments whose coupling is mediated by s- or
p-band itinerant carriers.  The local moments can have either 
a ferromagnetic direct exchange 
interaction with band electrons on the same site and/or an antiferromagnetic
interaction
due to hybridization between the local moment and band electrons on 
neighboring sites \cite{Bhattacharjee:1983_a,Dietl:1994_a}.  
Polarization of band electrons due to the interaction at one site
is propagated to neighboring sites.  When the coupling is
weak (the band carrier polarization is weak, {\em e.g.}, at temperatures near  Curie temperature), 
the effect is described by  RKKY theory which was originally applied to carrier mediated
indirect coupling between nuclear moments \cite{Frohlich:1940_a,Ruderman:1954_a,Bloembergen:1955_a,Yosida:1957_a}
and between local $d$-shell moments in metals \cite{Zener:1951_a,Kasuya:1956_a,Yosida:1957_a}.
The range of this interaction can be long and the 
interactions between separate local moments
can be either ferromagnetic or antiferromagnetic and tend to vary in space on the
length scale of the itinerant band's Fermi wavelength.  Unlike the double-exchange case,
magnetic order in this case does not lead to a significant change in the width of the 
conducting band.  This type of mechanism certainly does play a role in 
(III,Mn)V ferromagnetism, likely dominating in the case of strongly metallic (Ga,Mn)As, (In,Mn)As, and
Mn-doped antimonides.  
There is no  sharp distinction between impurity band double-exchange 
and kinetic-exchange interactions; the former is simply a strong coupling, narrow band 
limit of the latter.   

The starting point for developing a useful predictive model of 
(III,Mn)V ferromagnetism is achieving a full understanding of the 
electronic state of a single Mn impurity in the
host lattice.  We need to fully understand the character 
of the isolated local moments before we can critically discuss how they
are coupled.  
The character of the local moment need not 
be the same in all (III,Mn)V materials.
The remaining subsections will focus on
properties of a Mn impurity in GaAs and on the nature of ferromagnetic
coupling in (Ga,Mn)As and related arsenide and
antimonide DMSs. At the end of this section, we comment on 
how things might change in wider-gap hosts like
GaP and GaN.

\subsection{Substitutional Mn impurity in GaAs}
\label{micro-subMn} 
\begin{figure}
\ifXHTML\Picture{review/figures/Fig03.png}\else\includegraphics[width=3.0in,angle=-0]{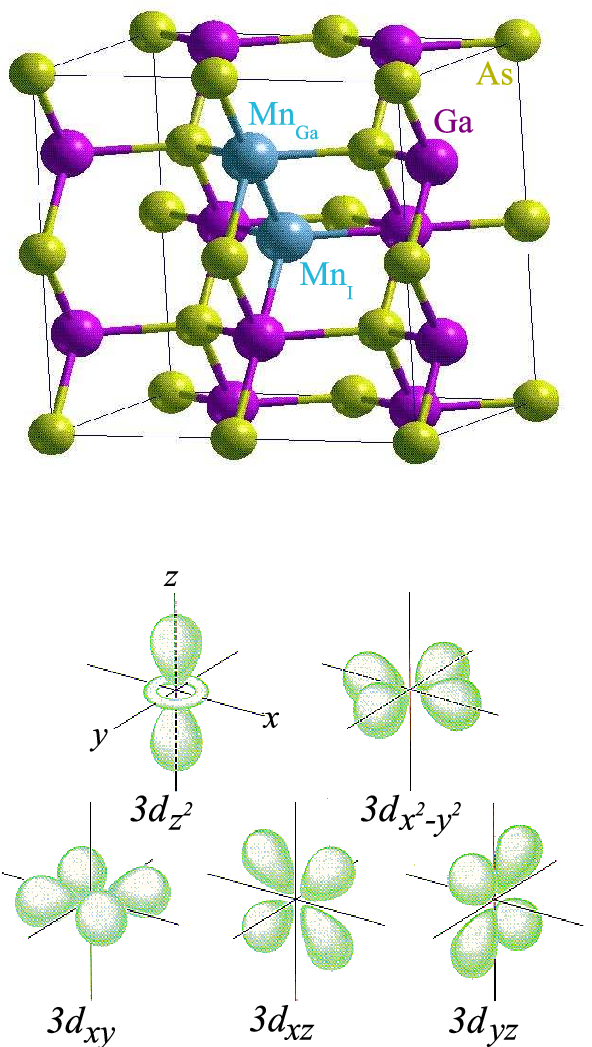}\fi
\caption{Top panel: Substitutional Mn$_{\rm Ga}$ and interstitial
Mn$_{\rm I}$ in GaAs. 
Bottom panel: two $e_g$ $3d$-orbitals and three $t_{2g}$ $3d$-orbitals of Mn.
} 
\label{Mn_Ga}
\end{figure}

Among all (III,V) hosts,
Mn impurity has been studied most extensively in GaAs. 
The elements in the (Ga,Mn)As compound have 
nominal atomic structures 
[Ar]$3d^{10}4s^2p^1$ for Ga, [Ar]$3d^{5}4s^2$ for Mn, and
[Ar]$3d^{10}4s^2p^3$ for As.  This circumstance correctly suggests
that the most stable and,
therefore, most common position of Mn in the GaAs host lattice is
on the Ga site  where its two $4s$-electrons can participate in
crystal bonding in much the same way as the two Ga $4s$-electrons. The substitutional
Mn$_{\rm Ga}$, and the less common interstitial Mn$_{\rm I}$, positions are illustrated
in Fig.~\ref{Mn_Ga}.
Because of the missing valence $4p$-electron,
the Mn$_{\rm Ga}$ impurity acts as an acceptor.
In the electrically neutral state, labeled as $A^0(d^5+hole)$, Mn$_{\rm Ga}$ has the character of
a local moment with zero angular momentum and spin $S=5/2$ (L\'ande g-factor $g=2$) and a 
moderately 
bound hole. The local moment is formed by three occupied
$sp-d$ bonding states with dominant $t_{2g}$ ($3d_{xy}$, $3d_{xz}$, $3d_{yz}$)
character and by two occupied $e_g$ ($3d_{x^2-y^2}$, $3d_{z^2}$) 
orbitals that are split from the $t_{2g}$ states by the tetrahedral
crystal field and do not strongly hybridize with the $sp$-orbitals. 
All occupied $d$-orbitals have the same spin orientation and together 
comprise the $S=5/2$ local moment. The weakly bound hole occupies one of the three
antibonding $sp-d$ levels with dominant As 4$p$ character. 
The charge $-e$ ionized Mn$_{\rm Ga}$
acceptor center, labeled as $A^-(d^5)$, has just the $S=5/2$ local spin character.

\begin{figure}
\ifXHTML\Picture{review/figures/Fig04.png}\else\hspace*{-1cm}\includegraphics[width=3.8in,angle=-0]{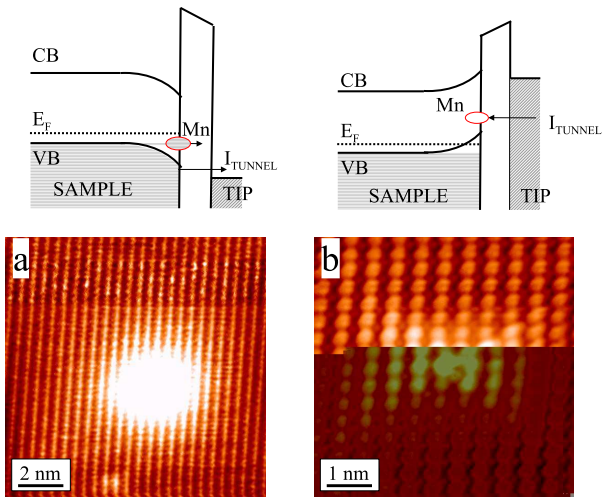}\fi
\caption{STM imaging of a Mn$_{\rm Ga}$ impurity. Top panel:  energy band diagram for the negative (left)
and positive (right) bias.
Bottom panel: Mn$_{\rm Ga}$ impurity in the ionized $A^-(d^5)$ (left) and the neutral $A^0(d^5+hole)$ (right)
state. 
Adapted from \protect\cite{Yakunin:2004_b}.
} 
\label{Mn_STM}
\end{figure}

Electron paramagnetic resonance (EPR) and ferromagnetic resonance (FMR) experiments confirm the
presence of the $A^-(d^5)$ center through the entire range of Mn
concentrations in both bulk and epilayer (Ga,Mn)As
\cite{Almeleh:1962_a,Szczytko:1999_b,Sasaki:2002_a}. The $S=5/2$
local moment on Mn was detected through a resonance line centered
at $g=2$ and, in low Mn-density samples, through a sextet
splitting of the line due to the hyperfine interaction with the $I=5/2$
$^{55}$Mn nuclear spin.  The neutral Mn$_{\rm Ga}$ centers
are more elusive because of nearly full compensation by unintentional
donor impurities at low Mn concentrations and because of the
metal-insulator transition  at high Mn concentrations.
Nevertheless, a multitude of experimental techniques, including
EPR \cite{Schneider:1987_a}, infrared (IR) spectroscopy \cite{Chapman:1967_a,Linnarsson:1997_a}, 
and magnetization measurements \cite{Frey:1988_a},
have detected the $A^0(d^5+hole)$ center in (Ga,Mn)As. 
Strikingly direct evidence was given by an STM experiment 
\cite{Yakunin:2004_b,Yakunin:2004_c,Kitchen:2005_a}, shown in Fig.~\ref{Mn_STM}, in which
the state of a single impurity atom was switched between the
ionized $A^-(d^5)$ and the neutral $A^0(d^5+hole)$ by applying a
bias voltage that corresponded to a binding energy
$E_b\approx0.1$~eV. 

The binding energy $E_b=112.4$~meV inferred from 
IR spectroscopy \cite{Linnarsson:1997_a,Chapman:1967_a} is
consistent with the above STM measurement and
with inferences based on photoluminescence experiments \cite{Lee:1964_a,Schairer:1974_a}. These observations
identify Mn as a moderately shallow
acceptor in GaAs whose band gap is $E_g=1.52$~eV. 
The binding energy, which governs the electrical behavior of the Mn impurity. 
has contributions from both Coulomb attraction between the hole and the $A^-(d^5)$ core (and
a central cell correction) and spin-dependent $p-d$ hybridization.  
The latter effect is  responsible for the exchange interaction on which this review centers. 
We now discuss this parameter in more detail.

\subsubsection{{\em p-d} exchange coupling}
\label{micro-subMn_pd}
The top of the GaAs valence band is dominated by $4p$-levels which are more 
heavily weighted on As than on Ga sites. Direct 
exchange between the holes near the top of the band and the
localized Mn $d$-electrons
is weak since Mn$_{\rm Ga}$ and As belong
to different sublattices.
This fact allows $p-d$ hybridization  to 
dominate, explaining the antiferromagnetic sign of this interaction \cite{Bhattacharjee:1983_a}
seen in experiment \cite{Okabayashi:1998_a}. 

\begin{figure}[h]
\ifXHTML\Picture{review/figures/Fig05.png}\else\includegraphics[width=2.0in,angle=0]{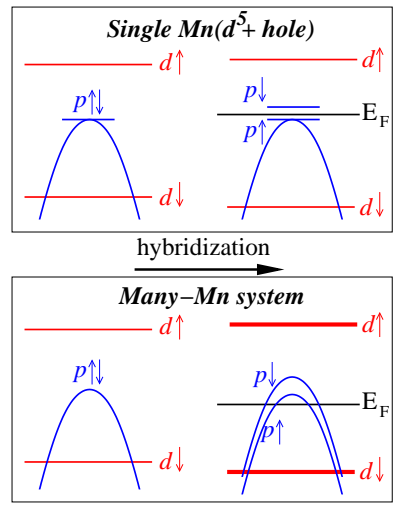}\fi
\caption{Electron-picture cartoon: splitting of the isolated Mn acceptor level (top panel)
and of the top of the valence-band in the many-Mn system (bottom panel) due to p-d 
hybridization.} 
\label{hybrid}
\end{figure}

There is a simple physical picture of the $p-d$ exchange interaction which 
applies when interactions are treated in a mean-field way, and therefore also
applies as an interpretation of LSDA calculations.  Given that the filled, say spin-down, 
Mn $d$-shell level is deep
in the valence band and that the empty spin-up $d$-level is 
above the Fermi level and high in the conduction band, hybridization (level repulsion of like-spin states) 
pushes the energy of spin-down valence band states up relative to the 
energy of spin-up valence band states.  The resulting antiferromagnetic coupling
between valence band states and local Mn spins is illustrated schematically 
in Fig.~\ref{hybrid}.
The same basic picture applies for itinerant valence band states in a 
heavily-doped metallic DMS and for the acceptor state of an isolated Mn$_{\rm Ga}$ impurity. Note that
the cartoon band structure in Fig.~\ref{hybrid}
is plotted in the electron picture while the DMS literature usually refers
to the antiferromagnetic $p-d$ coupling between 
{\em holes} and local Mn
moments. We comment in detail on the equivalent notions of $p-d$ exchange in the physically
direct electron-picture and the computationally more
convenient hole-picture for these $p$-type DMSs in Section \ref{bulk-mag-magnet}. 

We have already mentioned  in Section~\ref{micro-general} the conceptional inadequacy of effective 
single-particle theories, including LSDA, in dealing with local moment levels coincident
with itinerant electron bands.
Anderson suggested a many-body model
Hamiltonian that circumvents this problem by including explicitly
the Coulomb correlation integral of the localized electron states
in the Hamiltonian \cite{Anderson:1961_a,Haldane:1976_a,Fleurov:1976_a}.  The problem is that the 
change in the effective potential when the number of occupied
localized orbitals changes by one, the Hubbard constant $U$, 
can be comparable to or larger than other band parameters,
invalidating any mean-field-like approach.  The consequences 
of this fact can be captured at a qualitative level in models
that include the Hubbard $U$ \cite{Krstajic:2004_a}. In these phenomenological models, the localized orbital part 
of the Hamiltonian generally has an additional   
parameter, the Hund's rule constant $J_H$. This parameter captures the local 
direct exchange physics which favors spin-polarized open shell atomic
states.  For the case of the Mn($d^{5}$) configuration, 
$J_H$ forces all five singly-occupied $d$-orbitals to align their
spins in the ground state. 
Recently considerable effort 
has been devoted to 
developing approaches that combine the local correlation effects 
induced by the $U$ and $J_H$ terms in phenomenological models, with SDF 
theory 
\cite{Anisimov:1991_a,Perdew:1981_a,Park:2000_a,Sandratskii:2004_a,Petit:2004_a,Wierzbowska:2004_a,Filippetti:2003_a,Schulthess:2005_a}.
We comment on these {\em ab initio} techniques in Section~\ref{theory-first}. 

When hybridization \cite{Slater:1954_a,Harrison:1980_a}
between the local moment and band 
electron states is weak it can be treated perturbatively. 
%Using the Schrieffer-Wolff canonical transformation \cite{Schrieffer:1996_a},
%the low-order hybridization is removed.  
%In simple models the hybridization is represented by a single hopping energy $V_{pd}$,
%but the basic ideas can be generalized to more realistic models without 
%conceptual difficulty \cite{Slater:1954_a,Harrison:1980_a,Jungwirth:2003_c,Timm:2004_b}.
%(See Section~\ref{theory-tb} for a discussion of these tight-binding-approximation models.)
The Schrieffer-Wolff transformation  of the Anderson Hamiltonian,
\begin{eqnarray}
H_A&=&\sum_{k,s}\epsilon_kn_{ks}+\sum_s\epsilon_dn_{ds}+Un_{d\uparrow}n_{d\downarrow}
\nonumber \\
&+&\sum_{ks}(V_{kd}c^{\dagger}_{ks}c_{ds}+c.c.)\; ,
\label{Anderson}
\end{eqnarray} 
removes the hybridization (last term in Eq.~(\ref{Anderson}) and 
leads to a model in which the local moment spin 
interacts with the valence band via a spin-spin interaction only,
$\sum_{k^{\prime}k}j_{k^{\prime}k}{\bf s}_{d}\cdot{\bf s}_{k^{\prime}k}$,
with the number of electrons in each band fixed \cite{Schrieffer:1996_a}.  Here 
we assume for simplicity a single localized orbital and a single itinerant band,
we use $k$ to represent band states and $d$
to represent the localized impurity state, ${\bf s}$ labels spin, $\epsilon_{\alpha}$ a single
particle energy, and $n_{\alpha}=c^{\dagger}_{\alpha}c_{\alpha}$ and $c_{\alpha}$ are the
standard second quantization operators.
This procedure is 
normally useful only if the hybridization is relatively weak, in which 
case it is not usually a problem to {\em forget} that the canonical 
transformation should also be applied to operators representing observables. 
%The Schrieffer-Wolff transformation 
%bundles Coulomb correlation and hybridization into a single exchange potential $J_0{\bf
%S}\cdot{\bf s}$, where ${\bf S}$ and ${\bf s}$ are the spin
%operators of the Mn local moment and the itinerant band state, respectively.
Strictly speaking, the Schrieffer-Wolf transformation also leads to a 
spin-independent interaction \cite{Schrieffer:1996_a} which is normally neglected in comparison 
with the stronger spin-independent long-range part of the Coulomb potential.

Since the valence band states 
of interest in GaAs, even in heavily doped samples, are near the 
Brillouin-zone center $\Gamma$, the single phenomenological constant extracted
from experiment for this interaction should be thought of as its value
when both initial and final states are at the $\Gamma$-point.  
The quantity $\sum_{k^{\prime}k}j_{k^{\prime}k}{\bf s}_{d}{\bf s}_{k^{\prime}k}$
is then approximated by $J_0 {\bf s}_{d}\cdot{\bf s}_{k=0}$, where
\begin{equation}
J_0=2|V_{pd}|^2\left(\frac{1}{\epsilon_d+U}-\frac{1}{\epsilon_d}\right)\;, 
\label{anderson}
\end{equation}
$V_{pd}$ represents the As $p$-orbital -- Mn $d$-orbital hybridization potential (neglecting
again the multiple orbital nature of the $p$ and $d$-levels for simplicity),
$\epsilon_d<0$ is the single-particle atomic-level energy of
the occupied Mn $d$-state measured from the top of the valence band, and
$\epsilon_d+U>0$ is the energy cost of adding a second electron to 
this orbital \cite{Schrieffer:1996_a}.
When the hole 
density is large or the holes are more strongly localized near Mn acceptors,
the crystal-momentum dependence of this interaction parameter cannot be
entirely neglected \cite{Timm:2004_b}. 

Since $J_0$ originates from hopping between Mn 
and the neighboring As atoms, the $p-d$ exchange
potential, $J({\bf R}_i -{\bf r})$, produced by Mn impurity at site $R_i$ has a range of order one lattice constant, and 
 \begin{equation}
J_0=\frac{\int_{u.c.}d{\bf r} J({\bf R}_i -{\bf r}) u^{\ast}_n({\bf r})
u_n({\bf r})}{\int_{u.c.}d{\bf r} u^{\ast}_n({\bf r})
u_n({\bf r})}\;.
\label{j0jr}
\end{equation}
In Eq.~(\ref{j0jr})
we assumed that the perfect crystal Bloch function, 
$\psi_{n,{\bf k}}({\bf r})=\exp(i{\bf k} {\bf r})u_{n,{\bf k}}({\bf r})$, is composed of a slowly varying
envelope function and a periodic function
$u_{n,{\bf k}}({\bf r})$ with the normalizations $1/V\int d{\bf r}
\psi^{\ast}_{n,{\bf k}}({\bf r})\psi_{n^{\prime},{\bf k}^{\prime}}({\bf r})=
\delta_{n,n^{\prime}}\delta_{k,k^{\prime}}$ and 
$1/\Omega_{u.c.}\int_{u.c.}d{\bf r}u^{\ast}_n({\bf r})
u_{n^{\prime}}({\bf r})=\delta_{n,n^{\prime}}$.
Here $V$ is the crystal volume and $\Omega_{u.c.}$ is the unit cell
volume. In GaAs, $\Omega_{u.c.}=a_{lc}^3/4$=0.045~nm$^3$ and the lattice constant $a_{lc}=0.565$~nm.
These wavefunctions can be obtained from  ${\bf k} \cdot {\bf p}$ theory which treats the band Hamiltonian
of the system perturbatively expanding around the $\Gamma$-point \cite{Dietl:2001_b,Abolfath:2001_a}. 
$J_0$ in Eq.~(\ref{j0jr}) corresponds
to the average value of $J({\bf R}_i -{\bf r})$ seen by the ${\bf k}=0$ Bloch state over the $i$-th unit cell.

The slowly
varying envelope function sees an effective zero-range
$p-d$ exchange potential, since
\begin{eqnarray}
& &\langle\psi_{n,{\bf k}}|J({\bf R}_i -{\bf r})|\psi_{n^{\prime},{\bf k^{\prime}}}\rangle
\nonumber \\
&=&
\frac{1}{V}\int d{\bf r}\exp\left[i({\bf k}- {\bf k^{\prime}}){\bf r}\right]
J({\bf R}_i -{\bf r})
u^{\ast}_{n,{\bf k}}({\bf r})
u_{n^{\prime},{\bf k}^{\prime}}({\bf r})\nonumber \\
&\approx&
\frac{1}{V}\sum_{j=1}^{N_{u.c.}}
\exp\left[i({\bf k}- {\bf k^{\prime}}){\bf R}_j\right]
\int_{j-u.c.}^{\,} d{\bf r}
J({\bf R}_i -{\bf r})
u^{\ast}_n({\bf r})
u_n({\bf r})\nonumber \\
&=&
\frac{1}{V}\sum_{j=1}^{N_{u.c.}}
\exp\left[i({\bf k}- {\bf k^{\prime}}){\bf R}_j\right]
J_0\Omega_{u.c.}\delta_{i,j}\nonumber \\
&\approx&
\frac{1}{V}\int d{\bf r}\exp\left[i({\bf k}- {\bf k^{\prime}}){\bf r}\right]
J_{pd}\delta({\bf R}_i-{\bf r})\; ,
\end{eqnarray}
where $N_{u.c.}$ is the number of unit cells in the crystal volume $V$ and
$J_{pd}=J_0\Omega_{u.c.}$.
Note that the assumption of zero-range in the
 ${\bf k} \cdot {\bf p}$ treatment corresponds to the neglect of momentum
 dependence of this interaction for wavevectors close to the $\Gamma$-point in a microscopic picture.
%In the following
%sections of this article we will use the notation, $J_{pd}\equiv
%J_0\Omega_{u.c.}$. $J_0$ has units of energy and $J_{pd}$ has units of 
%energy times volume. 
Positive sign of the $J_{pd}$ (or $J_0$) constant corresponds to an
antiferromagnetic $p-d$ coupling. We also note that in the II-VI DMS literature a
symbol $\beta$ is often used instead of $J_{pd}$ and
$N_0\beta$ instead of $J_0$, where $N_0=1/\Omega_{u.c.}$ is the
number of unit cells (or cation sites) 
per volume \cite{Furdyna:1988_a,Dietl:1994_a}. 

The value of the $J_{pd}$ constant is often
considered to be independent of the host semiconductor \cite{Dietl:2000_a,Dietl:2001_b,Jungwirth:2002_b}.
Indeed the increase of $|V_{pd}|^2\sim a_{lc}^{-5}$ \cite{Harrison:1980_a} in Eq.~(\ref{anderson})
with decreasing lattice constant is partly compensated by smaller $\Omega_{u.c.}$ ($\sim a_{lc}^3$),
and the increase
of $|1/\epsilon_d|$ in a wider gap host is partly compensated by the decrease of the term
$1/(\epsilon_d+U)$. 
Although it may have similar values in many materials,  $J_{pd}$ 
will tend to be larger in larger gap, smaller lattice constant hosts.

In a virtual crystal mean-field approximation, the $p-d$ exchange potential due the Mn impurities
in a Ga$_{1-x}$Mn$_x$As DMS, 
$xN_0\Omega_{u.c.}\sum_{R_{u.c.}}J({\bf R}_{u.c.} -{\bf r})\langle {\bf S}\rangle\cdot {\bf s}$,
has the periodicity of the host crystal. (Here $\langle S\rangle$ is the mean-field Mn spin.)
The valence band states  in this approximation experience 
an effective single-particle kinetic-exchange field, ${\bf h}_{MF}=N_{Mn}J_{pd}\langle {\bf S}\rangle$,
where $N_{Mn}=xN_0$ is the Mn$_{\rm Ga}$ density. 

Finally, we discuss the relationship between the exchange
constant $J_{pd}$ and the exchange constant, $\varepsilon$,
used to provide a ${\bf k} \cdot {\bf p}$ interpretation of 
spectroscopic studies of the neutral $A^0(d^5+hole)$ center.
We emphasize that the use of a ${\bf k} \cdot {\bf p}$ approach 
assumes that the bound hole is spread over at least several lattice constants in each direction.  The fact that it is 
possible to achieve a reasonably consistent interpretation of the very detailed
spectroscopic data in this way is in itself strong support for 
the validity of this assumption.   
The coupling between the weakly bound hole moment $J$ and
the local spin $S$ of the Mn$_{\rm Ga}$ core is expressed in the form $ \varepsilon
{\bf S}\cdot{\bf J}$, where ${\bf J}={\bf j}+{\bf L}$, ${\bf j}$
is the (atomic scale) total angular momentum operator of the band-hole at the
$\Gamma$-point ($j=3/2$ or 1/2 for the As $4p$-orbitals forming the
band states near ${\bf k}=0$), and ${\bf L}$ is the additional
(hole binding radius scale) angular momentum acquired by the hole upon binding to the
Mn$_{\rm Ga}$ impurity. The IR spectroscopy 
data \cite{Linnarsson:1997_a} have been analyzed 
\cite{Bhattacharjee:2000_a} within a spherical
approximation, {\em i.e.}, considering only the $L=0$, $s$-like bound
state. (Note that a sizable anisotropic $d$-like
component in the bound-hole ground state has been identified in the analysis of the
STM data \cite{Tang:2004_a,Yakunin:2004_b}.) 
Further simplification is achieved by the
neglect of the admixture of the two $j=1/2$ ($j_z=\pm 1/2$)
$\Gamma$-point states which is justified by the large  spin-orbit
splitting, $\Delta_{SO}=341$~meV, of these states from the 
two heavy-hole states ($j=3/2$, $j_z=\pm 3/2$) and the two light-hole states
($j=3/2$, $j_z=\pm 1/2$). 

Writing the ground-state wavefunction in the form $\psi_{j_z}({\bf r})=
F_{j_z}({\bf r})u_{j_z}({\bf r})$ with a spherically
 symmetric envelope function $F_{j_z}({\bf r})$ and for
$j_z=\pm3/2,\pm1/2$ ($j=3/2$),
the expectation
value of the exchange potential reads 
\begin{eqnarray}
\langle\psi_{j_z}|J({\bf R}_I -{\bf r}){\bf S}\cdot{\bf s}|
\psi_{j_z}
\rangle&=&\overline{|f({\bf R}_I)|^2}J_{pd}
\langle j_z|{\bf S}\cdot{\bf s}|j_z\rangle
\nonumber \\
&=&\overline{|f({\bf R}_I)|^2}\frac{J_{pd}}{3}
\langle j_z|{\bf S}\cdot{\bf j}|j_z\rangle\; ,
\label{expjpd}
\end{eqnarray}
where $\overline{|f({\bf R}_I)|^2} =<F^{\ast}_{j_z}({\bf
r})F_{j_z}({\bf r})>_{u.c.}$ is the mean value of the square of
the slowly varying envelope function within the unit cell
containing the Mn$_{\rm Ga}$ impurity. Eq.~(\ref{expjpd}) implies
that
\begin{equation}
\varepsilon=\frac{J_{pd}}{3}
\overline{|f({\bf R}_I)|^2}\;,
\label{epsjpd}
\end{equation}
{\em i.e.}, the ratio between $\varepsilon$ and $J_{pd}$ is determined by the
strength of the binding of the hole to the $A^-(d^5)$ Mn$_{\rm Ga}$
core and is larger for more localized holes.

A combination of IR data and theoretical calculations has
been used to analyze this in more detail 
\cite{Linnarsson:1997_a,Bhattacharjee:2000_a}. First of all, the value
of the g-factor $g=2.77$ of the neutral $A^0(d^5+hole)$ Mn$_{\rm
Ga}$ complex obtained from IR spectroscopy measurements is in
agreement with the theoretical value expected for the
total angular momentum state $F=S-J=1$ of the complex, confirming
the antiferromagnetic character of the $p-d$ coupling between the hole and the local
Mn spin. The contribution of the $p-d$ potential to the
binding energy is then given by $\varepsilon{\bf S}\cdot{\bf
J}=\varepsilon[F(F+1)-S(S+1)-J(J+1)]/2$ which for the $F=1$ ground
state gives $-21\varepsilon/4$. The IR spectroscopy measurement of
the splitting $2\varepsilon$ between the $F=1$ and $F=2$ states
gives $\varepsilon\approx 5$~meV, {\em i.e.}, the contribution to the
binding energy from the $p-d$ interaction is
approximately 26.25~meV. The remaining  binding energy,
112.4-26.25=86.15~meV, is due to the central field potential of the
impurity. Bhattacharjee and Benoit \cite{Bhattacharjee:2000_a}
used the hydrogenic-impurity model with a screened Coulomb potential and a central cell
correction, whose strength was tuned to reproduce the value
86.15~meV, to obtain a theoretical estimate for
$\overline{|f({\bf R}_I)|^2}\approx 0.35$~nm$^{-3}$. From
Eq.~(\ref{epsjpd}) they then obtained $J_{pd}\approx
40$~meV~nm$^3$.  
Given the level of approximation used in the
theoretical description of the $A^0(d^5+hole)$ state,
this value is in a reasonably good agreement with
the exchange constant value $J_{pd}=54\pm 9$~meV~nm$^3$ 
($N_0\beta=1.2\pm0.2$~eV) inferred from
photoemission data \cite{Okabayashi:1998_a}. We note here that photoemission spectroscopy
\cite{Okabayashi:1998_a,Okabayashi:1999_a,Okabayashi:2001_a,Okabayashi:2002_a,Rader:2004_a,Hwang:2005_b}
has represented one of the key experimental tools to study the properties of Mn impurity in DMSs, in particular, the position 
of the Mn $d$-level and the strength and sign of the $p-d$ coupling. An indirect measurement of the $J_{pd}$ constant
is performed by fitting the photoemission data to a theoretical spectrum of an isolated MnAs$_4$ 
cluster \cite{Okabayashi:1998_a}. This procedure
is justified by the short-range character of the $p-d$ exchange interaction.  

%%%%%%%%%%%%%%%%%%%%%%%%%%%%%%%%%%%%%%%%%%%%%%%%%%%%%%%%%%%%%%%%%%%%%%%%%%%%%%
\subsection{Other common impurities in (Ga,Mn)As}
\label{micro-other_imp}

Most of the single Mn-impurity spectroscopic studies mentioned
above were performed in samples with doping levels $x<0.1\%$ for
which the Ga$_{1-x}$Mn$_x$As random alloy can be grown under
equilibrium conditions. In these materials Mn can be expected to
occupy almost exclusively the low energy  Ga-substitutional
position. Ferromagnetism, however, is observed only for $x>1\%$ which
is well above the equilibrium Mn solubility limit in GaAs and,
therefore, requires a non-equilibrium growth technique (in practice
low-temperature MBE) to avoid Mn precipitation. The price paid for
this is the occurrence of a large number of metastable impurity
states.  The most important additional defects are 
interstitial Mn ions and As atoms on cation sites (antisite defects).
Both act as donors and can have a severe impact on the electric and magnetic
properties of the DMS epilayers.  More unintended defects form at higher
Mn doping because of the tendency of the material, even under non-equilibrium
growth conditions, toward self-compensation.

\subsubsection{Interstitial Mn}
\label{micro-other_imp-int_mn} Direct experimental evidence for
Mn impurities occupying interstitial  (Mn$_{\rm I}$) rather than substitutional
positions was uncovered by combined channeling
Rutherford backscattering and particle induced x-ray emission
measurements \cite{Yu:2002_a}. 
This technique can distinguish between Mn$_{\rm
I}$ and Mn$_{\rm Ga}$ by counting the relative number of exposed
Mn atoms and the ones shadowed by lattice site host atoms
at different channeling angles. In highly doped as-grown samples,
the experiment identified nearly 20\% of Mn as residing on
interstitial positions. The metastable nature of these impurities
is manifested by the substantial decrease in their density upon
post-growth annealing at temperatures very close to the growth
temperatures \cite{Yu:2002_a,Edmonds:2002_b,Chiba:2003_b,Ku:2003_a,Stone:2003_a}. 
Detailed resistance-monitored annealing studies
combined with Auger surface analysis established  the
out-diffusion of Mn$_{\rm I}$ impurities towards the free DMS
epilayer surface during annealing \cite{Edmonds:2004_a}. 
The characteristic energy barrier of this diffusion process is estimated to 
be 1.4~eV.  (Note that a factor of 2 was omitted in
the original estimate of this energy in \cite{Edmonds:2004_a}.)

Isolated Mn$_{\rm I}$ spectroscopy data are not available
in (Ga,Mn)As, underlying the importance of theoretical work
on the electric and magnetic nature of this impurity. 
Density functional calculations \cite{Ernst:2005_a} suggest, {\em e.g.}, that minority-spin Mn$_{\rm I}$
$d$-states form a weakly dispersive band at $\sim 0.5$~eV below Fermi energy, a feature which is
absent in the theoretical Mn$_{\rm Ga}$ spectra.
{\em Ab initio} total
energy calculations \cite{Maca:2002_a,Masek:2003_b}
showed that Mn can occupy two metastable interstitial
positions, both with a comparable
energy, one surrounded by
four Ga atoms (see Fig.~\ref{Mn_Ga}) and the other surrounded by four As atoms.
The two Mn$_{\rm I}$ states have
similar local magnetic moments and electro-negativity.

The calculations have
confirmed that Mn$_{\rm I}$ acts as a double-donor, as
expected for a divalent metal atom occupying an interstitial
position. Each interstitial Mn therefore compensates two
substitutional Mn acceptors.  
It seems likely that because of the strong Coulombic attraction between 
positively charged Mn$_{\rm I}$ and negatively charged Mn$_{\rm Ga}$ defects,
the mobile interstitials pair-up
with substitutional Mn during the growth, as illustrated in Fig.~\ref{Mn_Ga}
\cite{Blinowski:2003_a}. The
total spin of a Mn$_{\rm Ga}$-Mn$_{\rm I}$ pair, inferred from {\em ab initio}
calculations is much smaller than the local spin $S=5/2$ of the
isolated Mn$_{\rm Ga}$ acceptor. This property, interpreted 
as a consequence of short-range antiferromagnetic interactions between two local moment
defects that have comparable local moments, has been confirmed experimentally \cite{Edmonds:2005_b}.
According to theory, the strength of this magnetic
coupling contribution to the Mn$_{\rm Ga}$-Mn$_{\rm I}$
binding energy is 26~meV \cite{Masek:2003_b}.

{\em Ab-initio} calculations of the spin-splitting of the valence
band indicate that the $J_{pd}$ coupling constants of interstitial
and substitutional Mn are comparable \cite{Masek:2003_b}. 
This would suggest a negligible net $p-d$ coupling between the
antiferromagnetically coupled  Mn$_{\rm Ga}$-Mn$_{\rm I}$ pair and
the valence band holes. We do note, however, that these LSDA calculations give
$J_{pd}\approx 140$~meV~nm$^3$ which is more than twice as
large as the experimental value, so this conclusion must be regarded somewhat cautiously.  
It is generally accepted that this  
discrepancy reflects the general tendency of the DFT-LSDA and similar 
theories to systematically underestimate the splitting between the
occupied Mn $d$ states in the valence band continuum and the empty Mn $d$-states. 
Eq.~(\ref{anderson}) illustrates
how this deficiency of SDF band theories translates into an
overestimated strength of the $p-d$ exchange interaction. Apart from
this quantitative inaccuracy, the relative strength of the
$p-d$ interaction of  Mn$_{\rm I}$ compared to Mn$_{\rm Ga}$ is
still a somewhat controversial issue in the theoretical
literature \cite{Blinowski:2003_a,Masek:2003_b}, 
which has not yet been settled experimentally.  This property, and 
many others related to magnetism are sensitive to details of the 
SDF implementation, to the way in which disorder is accounted
for, and even to technical details associated with the way in which the 
Bloch-Schr\"{o}dinger equation is solved numerically. 
For example, the positive charge of the Mn$_{\rm Ga}$-Mn$_{\rm I}$ 
(single-acceptor double donor) pair, which will tend to reduce its exchange 
coupling with valence band holes due to Coulomb repulsion, is not included in all approaches. 

In addition to the direct hole and local moment 
compensation effects of Mn$_{\rm I}$ defects 
on ferromagnetism in (Ga,Mn)As, the structural changes they
induce in the crystal are indirectly related to important magnetic
properties, particularly to the various magnetic and transport
anisotropies. {\em Ab initio} theory predicts that the separation of the four
nearest As neighbors surrounding the  Mn$_{\rm I}$ in a
relaxed lattice is increased by 1.5\% compared to the clean GaAs
lattice \cite{Masek:2003_a}. Because of the Coulomb repulsion
between Ga cations and Mn$_{\rm I}$ defects, an  
even larger lattice expansion ($\sim 2.5\%$) is found for
the Mn$_{\rm I}$ in the four Ga tetrahedral position. 
When grown on a GaAs substrate, this effect of interstitial Mn leads
to a lattice-matching compressive strain in the (Ga,Mn)As thin
layers that induces a large uniaxial magnetic anisotropy, as we
discuss in detail in Section~\ref{bulk-mag-micromag}.

\subsubsection{As antisites}
\label{micro-other_imp-as_anti}
Low-temperature growth of GaAs is known to lead to the incorporation
of high levels of As antisite defects.  This property is a combined 
consequence of the non-equilibrium growth conditions and the 
As overpressure often used in the MBE process to
assure the 2D growth mode.  These double-donor defects are likely to be also
present  in the (Ga,Mn)As epilayers and may contribute to  hole
compensation.

Unlike Mn$_{\rm I}$ impurities, As antisites are stable up to
$\sim$450~$^o$C \cite{Bliss:1992_a}. 
This is well above the temperature at which Mn
precipitation starts to dominate the properties of (Ga,Mn)As and,
therefore, the As antisites cannot be removed from the epilayer by
a post-growth annealing treatment. Experimental studies
suggest that the degradation of (Ga,Mn)As magnetic properties
due to hole compensation by As antisites can be reduced by using
As$_2$ dimers
instead of As$_4$ tetramers 
and by maintainting a strictly stoichiometric growth mode \cite{Campion:2003_a}.

\subsection{Qualitative picture of ferromagnetism in (Ga,Mn)As and other (III,Mn)V materials} 
\label{micro_picture}
The following elements of the qualitative picture of ferromagnetism in (Ga,Mn)As
emerge from the experimental data and 
theoretical interpretations discussed in this section (see also \cite{Dietl:2002_b}).  The low-energy
degrees of freedom in (Ga,Mn)As materials are the orientations of 
Mn local moments and the occupation numbers of acceptor levels 
near the top of the valence band.  The number of local moments participating in the ordered state and the number
of holes
may differ from the number of
Mn$_{\rm Ga}$ impurities in the III-V host due to the presence of charge and moment compensating defects.
Hybridization between 
Mn $d$-orbitals and valence band orbitals, mainly on neighboring 
As sites, leads to an antiferromagnetic interaction between the 
spins that they carry. 

At low concentrations of substitutional Mn, 
the average distance between Mn impurities (or between holes
bound to Mn ions),
$r_c=(3/4\pi N_{Mn})^{1/3}$, is much larger than the size of the bound hole
characterized approximately by the impurity effective Bohr radius,
$a^{\ast}=\epsilon\hbar/m^{\ast}e^2$. Here $N_{Mn}=4x/a_{lc}^3$ is the number of Mn impurities
per unit volume, $\epsilon$ and $a_{lc}$ are the semiconductor dielectric function and 
lattice constant, respectively, and $m^{\ast}$ is the effective mass near the top of the valence band.
For this very dilute insulating limit, a theoretical concept was introduced in the late 1970's in which
a ferromagnetic exchange interaction between Mn local moments is mediated by thermally 
activated band carriers \cite{Pashitskii:1979_a}. Experimentally, ferromagnetism in (Ga,Mn)As is
observed when Mn doping reaches approximately 1\% \cite{Ohno:1999_a,Campion:2003_b,Potashnik:2002_a}
and the system is near the Mott insulator-to-metal transition, {\em i.e.}, 
$r_c\approx a^{\ast}$ \cite{Marder:1999_a}.
At these larger Mn concentrations,
the localization length of the impurity band states is extended to a degree that allows them to mediate
ferromagnetic exchange interaction between  Mn moments, 
even though the moments are dilute. Several approaches have been used to address ferromagnetism in DMSs near
the metal-insulator transition (for a review see \cite{Dietl:2002_b,Timm:2003_a}, including 
finite-size exact diagonalization studies of
hole-hole and hole-impurity Coulomb interaction effects \cite{Timm:2002_a,Yang:2003_a}, and the picture
of interacting bound magnetic polarons or holes hopping within the impurity band
\cite{Inoue:2000_a,Litvinov:2001_a,Durst:2002_a,Chudnovskiy:2002_a,Kaminski:2002_a,Berciu:2001_a,Bhatt:2002_a,Alvarez:2002_a}.  
Because of the hopping nature of conduction and
the mixed $spd$ character of impurity band states, the regime is sometimes regarded as an 
example of double-exchange ferromagnetism.

At even higher Mn concentrations, the impurity band gradually merges
with the valence band \cite{Krstajic:2004_a} 
and the impurity states delocalize. In these metallic (Ga,Mn)As ferromagnets, which
we focus on in the following sections,
the coupling between Mn local moments is mediated by the 
$p-d$ kinetic-exchange mechanism \cite{Dietl:1997_a,Matsukura:1998_a,Jungwirth:1999_a,Dietl:2000_a,Jain:2001_a}. 
A qualitatively similar picture applies for (In,Mn)As and Mn-doped antimonides.  In the metallic limit
the influence of Coulomb and exchange disorder on perfect crystal valence-band states can be treated 
as perturbatively.  

The crossover from impurity-band mediated to Bloch valence-band mediated 
interactions between Mn moments is a gradual one.  In the middle of the crossover regime, 
it is not obvious which picture to use for a qualitative analysis and quantitative 
calculations are not possible within either picture.  Strongly localized impurity-band states away 
from Fermi energy may play a role in spectroscopic properties \cite{Okabayashi:2001_a}, even when they play 
a weaker role in magnetic and transport properties.   The crossover is controlled not only
by the Mn density but (because of the importance of Coulomb interaction screening) also by the 
carrier density.  There is a stark distinction between the 
compensation dependence predicted by impurity-band and Bloch valence-band pictures.  When the 
impurity-band picture applies, ferromagnetism does not occur 
in the absence of compensation\cite{Kaminski:2003_a,DasSarma:2003_a,Scarpulla:2005_a},
because the impurity band is filled.  Given this, we can conclude from
experiment that the impurity band picture does not apply to optimally annealed 
(weakly-compensated) samples which exhibit robust ferromagnetism.  

\begin{figure}[h]
\ifXHTML\Picture{review/figures/Fig06.png}\else\includegraphics[width=2.0in,angle=0]{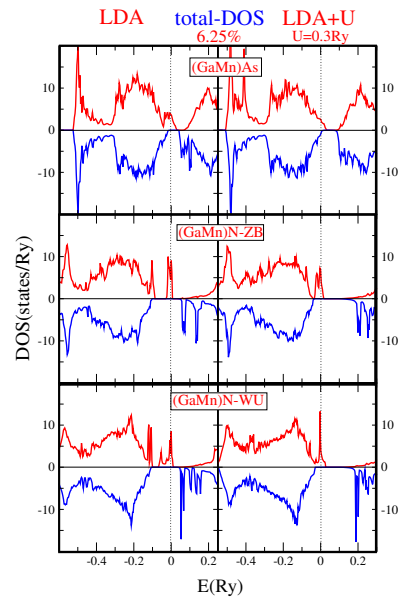}\fi
\caption{The spin-resolved total DOS of
(GaMn)As and (GaMn)N for a Mn concentration of 6.25\%.
Calculations are performed within LSDA and LDA+U approaches.
For (GaMn)N, both zinc-blende and wurzite
 structures are presented. The spin-up/spin-down DOS is
shown above/below the abscissa axis. The total DOS is given
per chemical unit cell of the semiconductor.
From \protect\cite{Sandratskii:2004_a}.
} 
\label{LSDA_total_DOS}
\end{figure}

\begin{figure}[h]
\ifXHTML\Picture{review/figures/Fig07.png}\else\includegraphics[width=2.0in,angle=0]{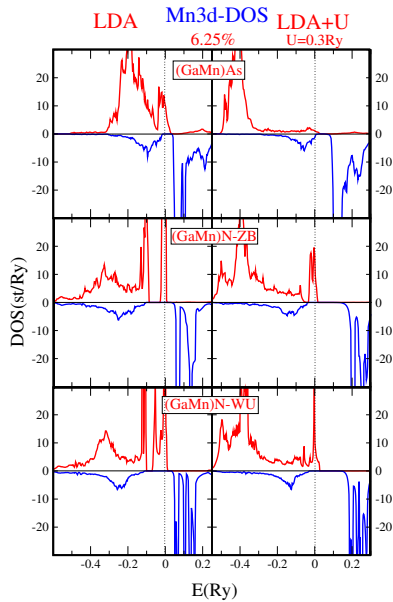}\fi
\caption{Same as in Fig.~\protect\ref{LSDA_total_DOS} but for partial Mn 3$d$ DOS.
In LDA+U the $p-d$ hybridization in (Ga,Mn)As is weak for both majority band and minority band states.} 
\label{LSDA_Mn_DOS}
\end{figure}

The phenomenology of Mn-doped phosphide or nitride DMSs is more complex,
with many aspects that probably cannot be captured by 
the Zener kinetic-exchange model \cite{Dietl:2002_a,Krstajic:2004_a}. Experimentally, the nature of the Mn impurity
is very sensitive to the presence of other impurities or defects in the 
lattice \cite{Korotkov:2001_a,Graf:2002_a,Graf:2003_b,Arkun:2004_a,Edmonds:2004_d,Hwang:2005_b}. This substantially 
complicates the development of a consistent picture of ferromagnetism in these materials. 
Larger $p-d$ coupling in the wide gap 
DMSs (reported {\em e.g.} in the photoemission experiment \cite{Hwang:2005_b}), and stronger bonding of the hole to the Mn ion,  might shift the metal-insulator transition
to higher Mn densities. At typical dopings of several per cent of Mn, the impurity band is still detached from
the valence band \cite{Kronik:2002_a} and ferromagnetic Mn-Mn coupling is mediated by holes hopping within the impurity
band. Recent experiments indicate that this scenario 
may apply to (Ga,Mn)P with 6\% Mn doping \cite{Scarpulla:2005_a}.
LSDA calculations suggest \cite{Sanyal:2003_a,Wierzbowska:2004_a,Sandratskii:2004_a}
that the $p-d$ hybridization can be so strong that the admixture of Mn $3d$ spectral weight at the Fermi energy
reaches a level at which the system effectively turns into a $d$-band metal. To illustrate this trend
we show in Fig.~\ref{LSDA_total_DOS} the LSDA and LDA+U calculations  of the spin-split
total density of states (DOS), and in Fig.~\ref{LSDA_Mn_DOS} the results for 
the Mn $d$-states projected DOS in (Ga,Mn)As
and (Ga,Mn)N \cite{Sandratskii:2004_a}. 
Indeed, in the wider gap (Ga,Mn)N the spectral weight of Mn $d$-orbitals at the Fermi energy is large and is not
significantly suppressed even if  strong on-site correlations are accounted for by introducing the phenomenological 
Hubbard parameter in the LDA+U method (see next section for details on combined Hubbard model and SDF
techniques).

Another possible scenario for these more ionic III-P and III-N semiconductors, supported
by EPR and optical absorption measurements and {\em ab initio} calculations \cite{Graf:2002_a,Luo:2004_a,Kreissl:1996_a,Schulthess:2005_a}, 
involves a transition of the  
substitutional Mn from a divalent ($d^5$) impurity to a trivalent ($d^4$) impurity. This strongly correlated
 $d^4$ center,
with four occupied  $d$-orbitals and a non-degenerate empty $d$-level shifted deep into the host band gap,
may form as a result of a spontaneous (Jahn-Teller) lowering of the cubic symmetry 
near the Mn site. If the energy difference between divalent and trivalent Mn impurity states is small, the DMS will
have a mixed Mn-valence which evokes the conventional double-exchange mechanism. Systems with dominant
$d^4$ character of Mn impurities, reminiscent of a charge transfer insulator, will inevitably require 
additional charge co-doping to provide for ferromagnetic coupling between dilute Mn moments \cite{Schulthess:2005_a}.     

%%%%%%%%%%%%%%%%%%%%%%%%%%%%%%%%%%%%%%%%%%%%%%%%%%%%%%%%%%
\section{Theoretical approaches}
\label{theory}

Our focus is on the theory of (III,Mn)V ferromagnets and
we therefore present in this section an overview of the different approaches that can be used to 
interpret the existing experimental literature and to predict the properties 
of materials that might be realized in the future.  Since the electronic and 
magnetic properties of ferromagnetic semiconductors are extremely sensitive to 
defects that are difficult to control in real materials and may not be 
completely characterized, the ability to make reasonably reliable theoretical predictions that are 
informed by as many relevant considerations as possible can be extremely
valuable to the effort to discover useful new materials.  For example we would like
to be able to make confident predictions of the ferromagnetic transition 
temperature of a (III,Mn)V material as a function of the 
density of substitutional Mn, interstitial Mn, co-dopants, antisite defects, 
and any other defects whose importance might be appreciated in the future.  
This ability is developing, we believe, although there is no simple {\em silver 
bullet} that solves all difficulties for all host materials, and there may still
be some considerations that are important for less studied host 
materials and are not yet part of the discussion.  
In this section we address in a general way the strengths and weaknesses of 
some of the different theoretical approaches.  The following sections of the review
compare predictions made with different types of theoretical approaches with 
experimental data on a variety of important electronic and magnetic properties.  
 
\subsection{First-principles calculations}
\label{theory-first}
In SDF theory \cite{Hohenberg:1964_a,Kohn:1965_a} all many-body effects are buried in a complex 
exchange-correlation energy functional.  Once an approximation is 
made for this functional, predictions for electronic and magnetic 
properties depend only on the particular arrangement of atomic nuclei 
under consideration.  In principle nuclear positions can be relaxed to 
make sure that the spatial distribution of nuclei is metastable and therefore 
realizable.  The exchange-correlation energy functional leads to a 
self-consistently determined spin-dependent exchange-correlation potential that appears 
in an effective independent-particle Hamiltonian.  The main technical 
challenge in DFT theory applications is the development of numerically
efficient methods that provide accurate solutions of single-body 
Schr\"{o}dinger equations (see review articles \cite{Jones:1989_a,Sanvito:2002_b}). 
DFT theory is established  as a flexible and valuable tool for studying
the microscopic origins of magnetism and for predicting electronic, magnetic, and
ground-state structural properties in a wide-variety of materials \cite{Jones:1989_a,Moruzzi:1993_a}.
It has the advantage that
it is a first principles approach 
without any phenomenological parameters.  DFT theory falls short of being a
complete and general solution to the many-electron problem only because 
the exact form of the exchange-correlation energy functional is unknown. 
A simple and widely successful approximation is the LSDA \cite{Barth:1972_a}.

The problem of solving LSDA equations with 
adequate accuracy remains a challenge even in perfectly ordered crystals.
In DMSs 
the degrees of freedom that are important for ferromagnetism,
the orientations of the Mn local moments, typically reside on approximately
1/40 of the atomic sites which further complicates numerical implementation
of the LSDA technique.  Other length scales that are characteristic
for the physics of interest in these materials, like the Fermi wavelengths of 
the valence band carriers, are also longer than the atomic length scale on 
which DFT theory interrogates matter.  This property limits the number of 
independent magnetic degrees of freedom that can be included in a DFT simulation 
of DMS materials.  The problem is exacerbated by the alloy disorder in (Ga,Mn)As.
Even if all Mn atoms substitute for randomly chosen lattice sites, it is 
necessary to find some way to average over microrealizations of the alloy. 
 
Disorder-averaging coherent-potential approximation
(CPA) \cite{Soven:1967_a,Velicky:1968_a}
and supercell approaches, have been used successfully in combination
with DFT calculations to address 
those physical parameters of (III,Mn)V DMSs that are derived from total-energy
calculations, such as the lattice constants, formation and binding
energies of various  defects, and the type
of the magnetic order (see {\em e.g.}
\cite{Park:2000_a,Schilfgaarde:2001_a,Sanvito:2002_b,Maca:2002_a,Erwin:2002_a,Sato:2003_a,Edmonds:2004_a,Sandratskii:2004_a,Wierzbowska:2004_a,Mahadevan:2004_b,Petit:2004_a,Xu:2005_b,Luo:2004_a}). 

Supercell calculations have usually studied interactions 
between Mn moment orientations by comparing the energies of parallel spin and 
opposite spin orientation states in supercells that contain two Mn atoms.
An effective spin Hamiltonian can be extracted from this approach if it is 
assumed that the interactions are pairwise and of Heisenberg form.  Even when
these assumptions are valid, the interaction extracted from these calculations
is the sum of the interactions at a set of separations connected by the supercell
lattice vector.  If the Mn-Mn spin interaction has a range larger than a couple 
of lattice constants, this poses a problem for the supercell approach.
Longer range interactions can, however, be estimated using a 
spin wave approach which allows spin-orientation variations that are 
incommensurate with the supercell.  Effective interactions extracted in this 
way lead to a classical Heisenberg model from which the critical temperature
can usually be calculated without substantial further approximation \cite{Sandratskii:2004_a,Xu:2005_b}.  
The CPA approach can estimate the energy cost of flipping a 
single spin in the ferromagnetic ground state, which is proportional to the
mean-field-approximation for the critical temperature of the effective Heisenberg
model \cite{Sandratskii:2004_a,Sato:2003_a}, 
and in this sense is limited in its predictive powers when mean-field theory
is not reliable.  Alternatively, a more detailed picture of magnetic interactions
is obtained by direct mapping of the CPA total energy to the Heisenberg Hamiltonian
\cite{Kudrnovsky:2004_a,Liechtenstein:1987_a,Bergqvist:2004_a}.

LSDA predictions for spectral properties, like the local DOS, 
are less reliable than predictions for total energy related properties.  This is 
especially true for states above the Fermi energy, and is manifested by a 
notorious inaccuracy in predicted semiconductor band gaps. 
From a DFT theory point of view, this inconsistency arises from attempting
to address the physics of quasiparticle excitations using ground-state DFT.  
In Mn-doped DMSs, the LSDA also fails
to account for strong correlations that suppress fluctuations in the
number of electrons in the $d$-shell.  One generally accepted consequence 
is that the energy splitting between the occupied and empty $d$-states is 
underestimated in SDF theory, leading to an unrealistically large $d$-state local DOS 
near the top of the valence band and to an overestimate of the strength
of the $p-d$ exchange.

LDA+U \cite{Anisimov:1991_a} and self-interaction corrected (SIC) LSDA 
\cite{Perdew:1981_a} schemes have been used to obtain more
realistic energy spectra and help to establish theoretically the microscopic
origins of ferromagnetism in (III,Mn)V semiconductor alloys 
\cite{Park:2000_a,Shick:2004_a,Sandratskii:2004_a,Petit:2004_a,Filippetti:2003_a,Wierzbowska:2004_a,Schulthess:2005_a}. 
LDA+U schemes used in studies of
(III,Mn)V DMSs combine  SDF theory
with the Hubbard description of  strongly correlated localized orbitals. 
Additional parameters from the  Hubbard model are added to the energy functional which are
obtained by fitting to experiment or, in principle, can be calculated self-consistently \cite{Anisimov:1991_a}. 
The SIC-LSDA method is based
on realizing  that spurious self-interactions present in the SDF lead to unphysically large energy penalties
for occupying localized states. Subtracting these interactions of a particle with itself from the density functional
suppresses the tendency of the LSDA to delocalize strongly correlated atomic orbitals.

\subsection{Microscopic tight-binding models}
\label{theory-tb}
A practical approach  that circumvents some of the complexities of this
strongly-correlated many-body problem is based on the Anderson many-body
Hamiltonian theory \cite{Anderson:1961_a,Haldane:1976_a,Fleurov:1976_a,Krstajic:2004_a} and a  tight-binding-approximation (TBA) band-structure
theory \cite{Slater:1954_a,Harrison:1980_a}.  
The TBA Hamiltonian
includes the 8$\times$8 $sp^3$ term with second-neighbor-interaction integrals describing the
host semiconductor \cite{Talwar:1982_a} and terms describing 
hybridization with non-magnetic impurities and Mn. Effective single-particle TBA
theory is obtained from the Anderson Hamiltonian by replacing the density operators in the Hubbard term in Eq.~(\ref{Anderson}) with
their mean values \cite{Masek:1991_a}.
In the TBA model, local changes of the crystal potential at Mn and other impurity 
sites are represented by shifted atomic levels.  The parameters chosen 
for the atomic level shifts and  the hopping amplitudes between atoms
can be inferred from experiment in a manner which  corrects for 
some of the limitations of LSDA theory.  
The parameterization, summarized in \cite{Talwar:1982_a,Masek:1991_a},
provides  the correct band gap
for the host   crystal and  the  appropriate
exchange splitting of  the Mn $d$-states.  
In the calculations, the hole density can be varied independently of
Mn doping by adding non-magnetic donors ({\em e.g.} Si or Se in GaAs) or acceptors ({\em e.g.} C or Be in GaAs).
  
Although the TBA model 
is a semi-phenomenological theory, it shares with first principles
theories the advantage of treating disorder 
microscopically.  A disadvantage of the tight-binding model approach, which is often combined
with the CPA, 
is that it normally neglects Coulomb interaction effects which influence 
the charge and spin densities over several lattice constants surrounding 
the Mn ion positions. Curie temperatures, magnetizations, the lifetimes of Bloch quasiparticle states, 
the effects of doping and 
disorder on the strength of the $p-d$ exchange coupling, and the effective
Mn-Mn magnetic interaction are among the problems
that have been analyzed using this tool 
\cite{Blinowski:2003_a,Jungwirth:2003_c,Timm:2004_b,Jungwirth:2005_b,Jungwirth:2005_a,Sankowski:2005_a,Vurgaftman:2001_b,Tang:2004_a}.

\subsection{${\bf k} \cdot {\bf p}$ effective Hamiltonian theories}
\label{theory-kp}
The highest critical temperatures in (Ga,Mn)As DMSs are achieved in optimally
annealed samples and at Mn
doping levels above 1.5\% for which the band holes are itinerant, as
evidenced by the metallic conductivities   \cite{Campion:2003_b}. 
In this regime, semi-phenomenological
models that are built on  crystal Bloch states rather than localized basis states
for the band quasiparticles
might be expected to provide more useful insights into
magnetic and magneto-transport properties.
A practical approach to this type of
modeling starts from recognizing that
the length scales associated with holes in 
the DMS compounds are still long enough that 
a ${\bf k} \cdot {\bf p}$, envelope function
description of the semiconductor valence bands is appropriate.
Since for many properties it is necessary to incorporate spin-orbit
coupling in a realistic way, six- or eight-band Kohn-Luttinger (KL)
${\bf k} \cdot {\bf p}$ Hamiltonians that include the spin-orbit
split-off band are desirable 
\cite{Luttinger:1955_a,Vurgaftman:2001_a}.  

The kinetic-exchange effective Hamiltonian approach 
\cite{Zener:1951_a,Bhattacharjee:1983_a,Furdyna:1988_b,Dietl:1994_a} 
asserts the localized
character  of  the five  Mn$_{\rm  Ga}$  $d$-orbitals  forming a  moment
$S=5/2$ and describes  hole  states  in the  valence  band using  the
KL Hamiltonian and 
assuming the $p-d$ exchange interaction
between Mn$_{\rm  Ga}$ and hole  spins. 
As discussed in Section~\ref{micro-subMn_pd},
the exchange interaction follows from hybridization between Mn $d$-orbitals and valence band $p$-orbitals.
The approach implicitly assumes that a canonical transformation has been performed which 
eliminated the hybridization \cite{Schrieffer:1996_a,Timm:2003_a}.
The  ${\bf k} \cdot {\bf p}$ approximation applies when  all relevant 
wavevectors are near the Brillouin-zone center and the model also assumes 
from the outset 
that the states near the Fermi energy mainly have the character of the host 
semiconductor valence band, even in the neighborhood of a substitutional Mn.
When these assumptions are valid it follows from symmetry considerations that 
the spin-dependent part of the effective coupling between Mn and band spins
is an isotropic Heisenberg interaction characterized by a single parameter. 
If the KL Hamiltonian parameters are 
taken from the known values for the host III-V compound \cite{Vurgaftman:2001_a}, 
the strength of this exchange interaction $J_{pd}$ can be extracted 
from one set of data, for example from spectroscopic studies
of isolated Mn acceptors as explained in Section~\ref{micro-subMn_pd}, and used to predict all other properties
\cite{Dietl:1997_a,Matsukura:1998_a,Jungwirth:1999_a,Dietl:2000_a,Konig:2000_a}. 
Since the value of  $J_{pd}$
can be obtained from experiments in a paramagnetic state the approach uses no
free parameters to model ferromagnetism in these systems.
In the absence of an external magnetic field the KL kinetic-exchange Hamiltonian has the general form:
\begin{equation}
{\cal H} = {\cal H}_{holes}+ J_{pd} \sum_{i,I} {{\bf S}_I}
\cdot {{\bf s}_i} \delta({\bf r}_i - {\bf R}_I),
\label{coupling}
\end{equation}
where ${\cal H}_{holes}$ includes the ${\bf k} \cdot {\bf p}$  KL Hamiltonian and the interactions of holes
with the random disorder potential and with other holes. The second term in Eq.~(\ref{coupling}) represents
the $p-d$ exchange interaction between
local Mn spins ${\bf S}_I$ and hole  spins ${\bf s}_i$.

The ${\bf k} \cdot {\bf p}$ approach has the advantage that it focuses 
strongly on the magnetic degrees of freedom introduced by the dilute moments,
which can simplify analysis of the model's properties.  Disorder can be treated in 
the model by introducing Born approximation lifetimes for the Bloch states 
or by more sophisticated, exact-diagonalization
or Monte-Carlo methods \cite{Jungwirth:2002_c,Konig:2003_a}. This approach makes it possible to use 
standard electron-gas theory tools to account for
hole-hole Coulomb interactions \cite{Jungwirth:1999_a}.  The envelope function approximation
is simply extended to model magnetic semiconductor 
heterostructures, like superlattices or quantum wells 
\cite{Lee:2002_a,Kechrakos:2004_a,Souma:2005_a,Frustaglia:2003_a,Brey:2000_a,Lee:2000_a}.
This strategy will fail, however, if the $p-d$ exchange
is too strong and the Mn acceptor level is correspondingly
too spatially localized or too deep in the gap.  
For example, Mn-doped GaP and GaN compounds are likely less 
favorable for this approach than (III,Mn)As and (III,Mn)Sb compounds. 
Generally speaking the advantages of a fully microscopic approach have increasing 
importance for more localized acceptors, and hence shorter range Mn-Mn interactions, while the advantages of the 
${\bf k} \cdot {\bf p}$ approach are more clear when the acceptors are more shallow 
and the Mn-Mn interactions have longer range. 

\subsection{Impurity band and polaronic models}
\label{theory-lattice}
There has also been theoretical work on (III,Mn)V DMS materials
based on still simpler models in which holes are assumed to hop between 
Mn acceptor sites, where they interact with the Mn moments via phenomenological
exchange interactions \cite{Berciu:2001_a,Chudnovskiy:2002_a,Bhatt:2002_a,Alvarez:2002_a,Mayr:2002_a,Fiete:2003_a}. 
Hamiltonians used in these studies have a form of (or similar to),
\begin{equation}
{\mathcal H}=-t \sum_{<ij>,\sigma}\hat{c}_{i\sigma}^{\dagger}\hat{c}_{j\sigma}
+J\sum_I {\bf S}_I\cdot\sigmav_I,
\label{spinlatham}
\end{equation}
where $\hat{c}_{i\sigma}^{\dagger}$ creates a hole at site $i$ with spin $\sigma$, the hole spin
operator $\sigmav_I=\hat{c}_{I\alpha}^{\dagger}\sigmav_{\alpha\beta}\hat{c}_{I\beta}$,
and $\sigmav_{\alpha\beta}$ are the Pauli matrices.
The models apply at least qualitatively in the low Mn density limit and 
are able to directly attack the complex and intriguing physics of these unusual insulating ferromagnets.  
Insulating ferromagnetism persists in this limit even when the carrier density is not strongly compensated
\cite{Fiete:2003_a}.  The dependence of ferromagnetism in this regime on the degree of compensation has not yet been systematically studied experimentally and seems certain to pose challenging theoretical questions. 
The models that have been used to study ferromagnetism in this regime can easily be adapted to include holes that 
are localized on ionized defects that may occur in addition to Mn acceptors.

Other related models assume that the Mn acceptors are strongly compensated 
so that the density of localized holes is much smaller than the density 
of Mn ions, leading to a polaronic picture in which a single hole 
polarizes a cloud of Mn spins \cite{Kaminski:2002_a,Durst:2002_a}.  The
free-parameter nature of these phenomenological models 
means that they have only qualitative predictive power. 
They are not appropriate for the high $T_c$ (Ga,Mn)As 
materials which are heavily doped by weakly compensated 
Mn acceptors and are metallic. On the other hand, the impurity band models may represent a useful approach 
to address experimental magnetic and transport properties of ferromagnetic (Ga,Mn)P where holes are more strongly 
localized \cite{Scarpulla:2005_a,Kaminski:2003_a,DasSarma:2003_a}.
%%%%%%%%%%%%%%%%%%%%%%%%%%%%%%%%%%%%%%%%%%%%%%%%%%%%%%%%%%%%

\section{Structural properties}
\label{bulk-struct}

\subsection{Impurity formation energies and partial concentrations}
\label{bulk-struct-formen}
Experimental efforts to increase Mn doping in (Ga,Mn)As DMSs beyond
the solubility limit of 0.1\% have been assisted by modern {\em ab initio}
theoretical studies of impurity formation energies
and effects related to the growth kinetics
\cite{Erwin:2002_a,Masek:2002_a,Masek:2003_b,Masek:2004_a,Mahadevan:2003_a,Edmonds:2004_a}.
Substitution of Ga by Mn
is expected, based on these studies,  to be enhanced
when the Ga chemical potential is kept low
relative to the Mn chemical potential, {\em i.e.}, under Ga-poor, Mn-rich growth
conditions \cite{Mahadevan:2003_a}. Calculations also suggest that one of
the major drawbacks of  bulk growth techniques is  that they
allow  phase separated precipitates, such as MnAs, to attain their most
stable free-standing lattice geometry, leading to  relatively low
formation energies for these unwanted phases \cite{Mahadevan:2003_a}.

In thin film epitaxy, competing phases are forced to adopt
the crystal structure of the substrate, which can significantly increase
their formation energies \cite{Mahadevan:2003_a}. The non-equilibrium 
LT-MBE has been a particularly successful growth
technique which allows a synthesis of  single-phase (Ga,Mn)As
DMSs with Mn concentrations up to $\sim 10\%$. If the growth
temperature is precisely controlled, a 2D growth mode of  uniform
DMSs can be maintained and, at the same time, a large number of Mn
atoms is incorporated in  Mn$_{\rm Ga}$ positions
\cite{Foxon:2004_a}.

\begin{figure}
\ifXHTML\Picture{review/figures/Fig08.png}\else\includegraphics[width=3.4in,angle=0]{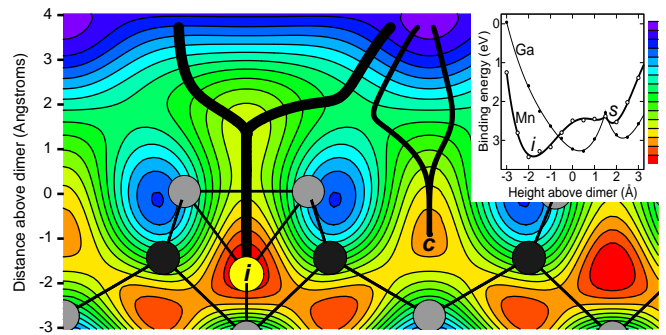}\fi
\caption{
The Potential-energy surface for Mn adsorption on GaAs(001),
plotted in a plane normal to the surface and containing the As surface
dimer.  The minimum energy adsorption site is the subsurface
interstitial site labeled $\textsl{i}$; the corresponding surface
geometry is shown (light gray for As, dark gray for Ga, yellow for
Mn).  Typical adsorption pathways funnel Mn adatoms to this
interstitial site (heavy curves) or to a cave site, $\textsl{c}$
(light curves). Inset: Binding energy of a Mn adatom centered
on the As-dimer; for comparison, results are also
shown for a Ga adatom. When additional As is deposited, the metastable
Mn site, $\textsl{s}$, becomes more favorable and leads to
partial incorporation of substitutional Mn.
From \protect\cite{Erwin:2002_a}.
}
\label{erwin}
\end{figure}

In (Ga,Mn)As DMSs a significant fraction of the Mn atoms is also incorporated  in interstitial
positions \cite{Yu:2002_a}. Adsorption pathways that can funnel Mn to interstitial sites have been
identified theoretically (see Fig.~\ref{erwin}) using  {\em ab initio} calculations of the potential energy surface of Mn
adsorbed on GaAs(001) \cite{Erwin:2002_a}. First principles calculations have also confirmed that
interstitial Mn$_{\rm I}$ impurities are metastable in GaAs, showing that the three distinct 
positions they can
occupy are two tetrahedral T(As$_{4}$ or Ga$_{4}$) positions surrounded by four near-neighbor As or Ga atoms, and one hexagonal position with three Ga and three As nearest neighbors. 
Among the  three interstitial sites the hexagonal position is clearly less favorable, especially so
in an overall p-type (Ga,Mn)As material \cite{Masek:2003_b}.
The typical energy barrier for Mn diffusion between interstitial sites is
approximately 1~eV \cite{Masek:2003_b,Edmonds:2004_a}.
On the other hand, diffusion of Mn between Ga substitutional positions
involves a kick-out mechanism of Mn$_{\rm I}$ + Ga$_{\rm Ga}$~$\rightarrow$
Mn$_{\rm Ga}$ + Ga$_{\rm I}$ for which the typical barrier is
about 3~eV \cite{Erwin:2002_a}.
Interstitial Mn$_{\rm I}$ is therefore much more mobile than 
substitutional Mn$_{\rm Ga}$.

As mentioned in Section~\ref{micro-other_imp-int_mn}, Mn$_{\rm I}$ donors
are likely to form near-neighbor pairs with Mn$_{\rm Ga}$ acceptors
in as-grown materials due to  the strong Coulomb attraction.  The net magnetic
moment of such a pair is
close to zero \cite{Blinowski:2003_a,Masek:2003_b,Edmonds:2004_a,Edmonds:2005_b}.
Although Mn$_{\rm I}$ can be removed by low-temperature annealing, the number of
substitutional  Mn$_{\rm Ga}$ impurities will
remain smaller than the total nominal Mn doping. The Mn$_{\rm Ga}$ doping
efficiency is, therefore, one of the key parameters that may limit the maximum
$T_c$ that can be
achieved in (Ga,Mn)As epilayers, as we discuss in detail in Section~\ref{bulk-mag-tc}.
{\em Ab initio} calculations of the  formation energies can be used to estimate
the dependence of  Mn$_{\rm Ga}$ and Mn$_{\rm I}$ partial concentrations
on  total Mn
doping in as-grown  materials \cite{Masek:2004_a,Jungwirth:2005_b}.
Similarly, correlated doping effects can
be studied for other defects that occur frequently in LT-MBE (Ga,Mn)As materials,
such as the arsenic antisites As$_{\rm Ga}$ \cite{Masek:2002_a}.

Correlations between acceptors (Mn$_{\rm Ga}$) and donors (Mn$_{\rm I}$ or  As$_{\rm Ga}$)
in III-V semiconductors like GaAs are strong due to the nearly covalent
nature of bonding in these crystals.
The cohesion energy of the covalent networks has a maximum if the Fermi energy $E_F$
lies within a band gap. Whenever $E_F$ is shifted to the valence band or the conduction
band the strength of the bonds is reduced because of the
occurrence of unfilled
bonding states or occupied antibonding states,
respectively.

In case of weak doping, small changes in the impurity
concentration can easily move  $E_{F}$ across the
band gap with a negligible influence on the energy spectrum.
The dependence of the formation energy, {\em i.e.} the energy cost for incorporating a
particular impurity in a crystal,  on the number
of electrically active impurities can then be represented by the
corresponding change in $E_{F}$ multiplied by the charge state
of the impurity \cite{Mahadevan:2003_a}.
In case of  strongly doped and
mixed crystals, the redistribution of  electronic states
in the valence band due to the impurities may play a more important
role and should therefore be included in the microscopic calculations.

In general, the
formation energy as a function of impurity
concentrations can be obtained from the composition-dependent
cohesion energy $W_{coh}$ of the crystal. Assuming a sample
consisting of $N$ unit cells of the impure (mixed) crystal, the
formation energy $E_{A}$ of an acceptor A
replacing atom X is defined as the reaction energy of the substitution
process,
\begin{center}
 {\rm sample + A}  $\longrightarrow$ {\rm sample with one extra acceptor + X}.
\end{center}
The corresponding reaction energy is
\begin{eqnarray}
E_{A}(x_{A},x_{D}) &=& N \cdot ( W_{tot}(x_{A} + 1/N,x_{D}) -
W_{tot}(x_{A},x_{D}))\nonumber \\
&+& E_{atom}(X) - E_{atom}(A)\; ,
\label{reaction_en_A}
\end{eqnarray}
where $W_{tot}(x_{A},x_{D})$ is the total energy of the doped crystal
normalized to a unit cell, $x_{A}$ and $x_{D}$ are the acceptor
and donor concentrations, and the last two terms represent total
energies of free-standing atoms X and A.
With increasing size of
the
sample, $N \rightarrow \infty$, the first term in
Eq.~(\ref{reaction_en_A}) approaches the derivative of
$W_{tot}(x_{A},x_{D})$ with respect to $x_{A}$
and the atomic energies can be absorbed by using the relation between the total
and cohesion energy. As a result, the cohesion energy represents a generating functional for the
formation energies, {\em i.e.},
\begin{equation}
E_{A}(x_{A},x_{D}) = \frac{\partial W_{coh}(x_{A},x_{D})}{\partial
x_{A}}.
\label{formation_en_A}
\end{equation}
The formation energy of a donor D substituting for an atom Y 
has the same form with $A\leftrightarrow D$ and $X\leftrightarrow Y$.

Having defined  $E_{A}$ and $E_{D}$, we note that in the low concentration
regime where they depend linearly on $x_A$ and $x_D$,  
\begin{equation}
\frac{\partial E_{A}}{\partial x_{D}} = \frac{\partial
E_{D}}{dx_{A}} \equiv K(x_{A},x_{D}) ,
\label{corr_energy}
\end{equation}
{\em i.e.}, that the mutual influence of the two kinds of impurities is
symmetric. $K(x_{A},x_{D})$ in Eq.~(\ref{corr_energy}) plays the role of a correlation energy
characterizing the co-doping process. For positive $K(x_{A},x_{D})$, the formation
energy of one impurity increases in the presence of the other. In
this case the material tends to be either n-type or p-type rather
than a compensated semiconductor. On the other hand, negative
correlation energy indicates that the presence of impurities of
one kind makes the incorporation of the other dopants easier. In
this case compensation is favored.

\begin{figure}
\begin{center}
\ifXHTML\Picture{review/figures/Fig09.png}\else\includegraphics[width=54mm,height=87mm,angle=270]{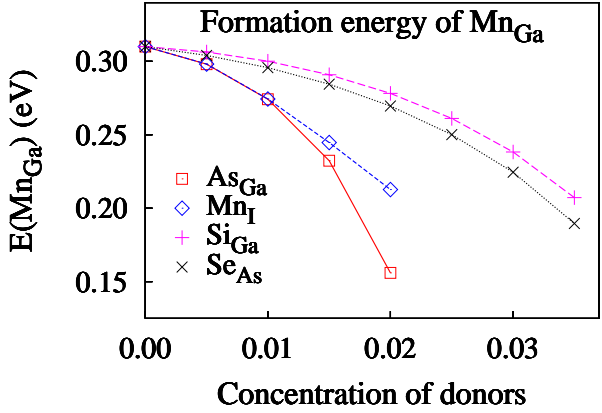}\fi
\caption{Theoretical LMTO formation energy $E({\rm Mn}_{\rm Ga})$ of 
substitutional Mn$_{\rm Ga}$ in Ga$_{0.96}$Mn$_{0.04}$As as a
function of the concentration of various donors.
From \protect\cite{Masek:2004_a}.
}
\label{formen1}
\end{center}
\end{figure}

\begin{figure}
\begin{center}
\ifXHTML\Picture{review/figures/Fig10.png}\else\includegraphics[width=54mm,height=87mm,angle=270]{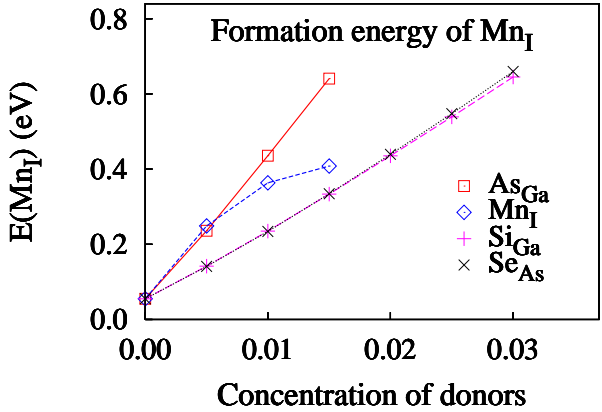}\fi
\caption{Theoretical LMTO formation energy $E({\rm Mn}_{\rm I})$ of the
interstitial Mn impurity in Ga$_{0.96}$Mn$_{0.04}$As a function of
the concentration of various donors.
From \protect\cite{Masek:2004_a}.
}
\label{formen2}
\end{center}
\end{figure}

Doping correlations over a wide
and continuous range of impurity concentrations have been
studied using the CPA,
combined with either the parameterized TBA
model or the {\em ab initio}
linearized-muffin-tin-orbital (LMTO) DFT method \cite{Masek:2004_a,Jungwirth:2005_b}.
Results of these calculations are summarized in Figs.~\ref{formen1} and \ref{formen2}.
The zero of energy is set to correspond to the formation
energy of a reference Ga$_{0.96}$Mn$_{0.04}$As system with all Mn
atoms occupying substitutional Ga positions.
Four representative examples are considered here. Se$_{\rm As}$ and Si$_{\rm
Ga}$ are typical single donors in GaAs which occupy the anion and
cation sublattice, respectively. The As antisite defects As$_{\rm
Ga}$ and the Mn$_{\rm I}$ interstitials, are the most important
native defects in (Ga,Mn)As, both acting as double donors. Fig.
\ref{formen1} shows that the formation energy of Mn$_{\rm Ga}$ decreases
with an increasing number of donors. The curves are grouped into
pairs according to the charge state of the donors, with only a
minor influence of the particular chemical origin of the defect.
The dependence is almost linear for low concentrations and the
slope of the function is roughly proportional to the charge state
of the donor. All this indicates that the variation of the
formation energy of Mn$_{\rm Ga}$ is mostly determined by the
above mentioned Fermi-level effect, and that the redistribution of
the density of states induced by donor defects plays a minor role.
Formation energies of the interstitial Mn$_{\rm I}$ in the
tetrahedral T(As$_{4}$) position are shown in Fig.~\ref{formen2}. For the
donor Mn$_{\rm I}$ impurity, the formation energy increases with
the density of other donors. 
This means that the creation of Mn$_{\rm I}$ is efficiently
inhibited in the presence of As$_{\rm Ga}$.
Analogous results are obtained for
As$_{\rm Ga}$ antisite defects.

\begin{figure}
\begin{center}
\ifXHTML\Picture{review/figures/Fig11.png}\else\includegraphics[width=54mm,height=87mm,angle=270]{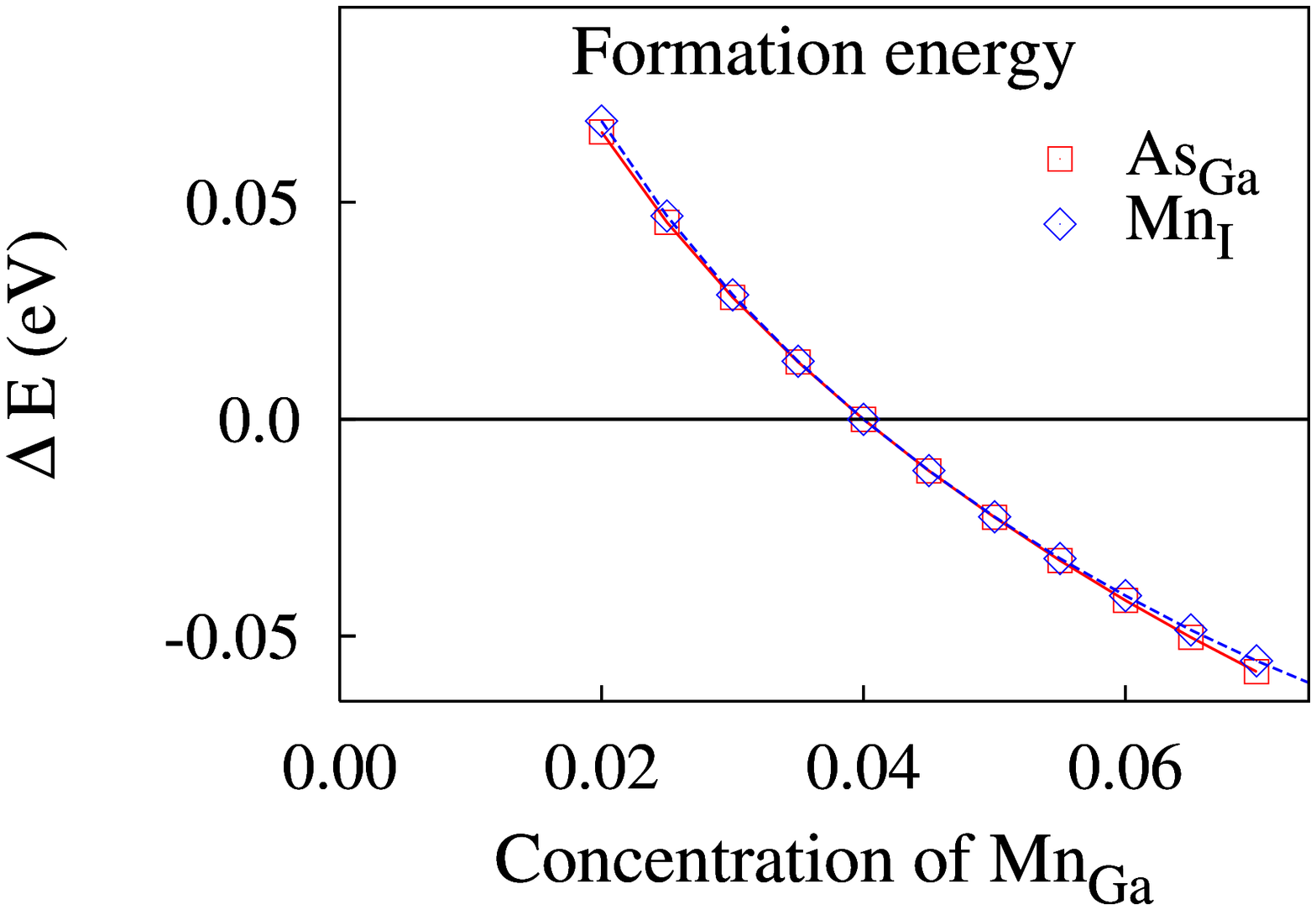}\fi
\caption{Theoretical LMTO variations $\Delta E$ of the formation energies of
the interstitial Mn impurity and As$_{\rm Ga}$ antisite defect in
Ga$_{1-x}$Mn$_{x}$As a function of the concentration $x$ of the Mn
atoms substituted in the Ga sublattice. The variations are
measured from the values for the reference material with 4 \% of
Mn$_{\rm Ga}$
From \protect\cite{Masek:2004_a}.
}
\label{formen4}
\end{center}
\end{figure}

In Fig. \ref{formen4} we show the change of 
formation energies of As$_{\rm Ga}$ and Mn$_{\rm I}$
as a function of the number of Mn$_{\rm Ga}$.
In both cases, the formation energy is a decreasing
function of the density of Mn$_{\rm Ga}$.
This self-compensation tendency is an important mechanism
controlling the properties of as-grown (Ga,Mn)As mixed
crystals.  It explains the observed charge compensation in as-grown
materials and is responsible to a large degree for the 
lattice expansion of highly Mn-doped (Ga,Mn)As DMSs, as we discuss in Section~\ref{bulk-struct-lc}.

The formation
energies can be used to theoretically  estimate partial concentrations
of substitutional Mn$_{\rm Ga}$, $x_s$, and interstitial Mn$_{\rm I}$, $x_i$, in as-grown (Ga,Mn)As
materials.
An assumption is made in these calculations, whose validity is tested by
a comparison with experimental data, that the probabilities of  Mn
atoms to occupy substitutional or interstitial positions are determined by
the respective formation energies, $E_{s}$ and
$E_{i}$, even in the non-equilibrium LT-MBE grown materials.

\begin{figure}
\ifXHTML\Picture{review/figures/Fig12.png}\else\includegraphics[angle=0,width=3.8in]{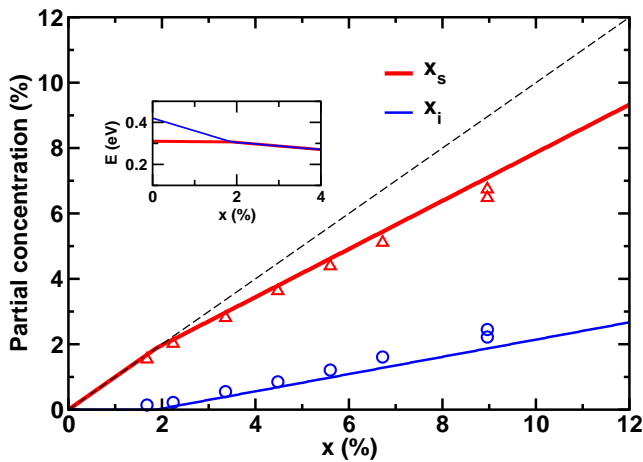}\fi
\caption{
Main panel: Theoretical TBA
equilibrium partial concentrations of substitutional Mn$_{\rm Ga}$ (red line) and interstitial
Mn$_{\rm I}$ (blue line) impurities. Triangles and circles represent experimental Mn$_{\rm Ga}$ and
Mn$_{\rm I}$ partial concentrations, respectively. 
Inset: Formation energies of Mn$_{\rm Ga}$ and Mn$_{\rm I}$ as a function
of total Mn concentration.
From \protect\cite{Jungwirth:2005_b}.
}
\label{partial}
\end{figure}

The balanced distribution  of
Mn$_{\rm Ga}$ and Mn$_{\rm I}$  is reached  when \cite{Masek:2002_a,Jungwirth:2005_b},
\begin{equation}
E_{s}(x_{s},x_{i}) = E_{i}(x_{s},x_{i}),
\label{cohesion_cond}
\end{equation}
as  also expected from  the growth  point  of  view.  Partial
concentrations $x_{s,i}$  of Mn are  obtained by solving  
Eq.~(\ref{cohesion_cond})
together   with   the  condition   $0   \leq   x_{s,i}   \leq   x$,
$x_{s}+x_{i}=x$. In Fig.~\ref{partial} we show results of TBA/CPA calculations \cite{Jungwirth:2005_b};
for $x>1.5\%$  $x_{s} \approx 0.8 x$ and
$x_{i} \approx 0.2 x$. LMTO/CPA 
theory calculations give very similar predictions \cite{Masek:2004_a}.

The  linear relations between  $x_{s}$, $x_{i}$,  and $x$  reflect the
fact that  the difference of  the formation energies of  Mn$_{\rm Ga}$
and Mn$_{\rm I}$ impurities (see inset of Fig.~\ref{partial}) can be, up to $x = 10\%$, approximated by a
linear function of $x_{s}$ and $x_{i}$,
\begin{eqnarray}
\Delta(x_{s},x_{i}) &=& E_{s}(x_{s},x_{i}) - E_{i}(x_{s},x_{i}) \nonumber \\ 
&\approx& -0.1 + 5.9 x_{s} - 15.1 x_{i} ({\rm
eV}).
\end{eqnarray}
This  relation indicates that for $x < 1.5\%$,
Mn$_{\rm  Ga}$  has a  lower  formation  energy than Mn$_{\rm I}$ 
and Mn  atoms  tend to occupy substitutional   positions.  
At and above $x\approx 1.5\%$,
$\Delta(x_{s},x_{i})$ approaches zero and both Mn$_{\rm
Ga}$ and Mn$_{\rm I}$ are formed.

The theoretical results are in a very good agreement with
experimental data, as shown in Fig.~\ref{partial}. The
balance considerations, confirmed experimentally in samples with Mn$_{\rm Ga}$ concentrations up to 6.8\%,
suggest that  there is no fundamental physics barrier to increasing   Mn$_{\rm Ga}$ concentration up to 10\%
and beyond. Very precise control over the growth temperature and stoichiometry is, however,  
required for maintaining the 2D growth mode of the uniform (Ga,Mn)As materials at these high doping levels.

Finally we note that during   growth  the formation   energies  control  
incorporation of Mn  atoms, assuming that the total amount
of  Mn in  the material  is related  to a  sufficiently  high chemical
potential in the Mn source. The annealing processes, on the other hand, do not
depend  on formation  energies but rather on  energy barriers 
surrounding individual metastable positions of Mn in the lattice. 
The barriers are larger for Mn$_{\rm Ga}$ \cite{Erwin:2002_a,Masek:2003_b} 
so that  post-growth low temperature annealing
can be used to remove Mn$_{\rm I}$ without changing the
number of Mn$_{\rm Ga}$ significantly.

\subsection{Lattice constant variation}
\label{bulk-struct-lc} Changes in the lattice constant of
(Ga,Mn)As DMSs, relative to the lattice constant of undoped GaAs,
are too small to significantly suppress or enhance $p-d$ kinetic
exchange or other magnetic coupling mechanisms. Direct effects of
doping-induced lattice distortion on the onset of ferromagnetism
are therefore negligibly small. Nevertheless, variations in the
lattice parameter  provide a measure of impurity concentrations in
the DMS material.  The impurities do, of course, control ferromagnetism 
through their doping properties. Also, because thin film
(Ga,Mn)As epilayers are not relaxed, lattice constant mismatch
between the DMS layer and the substrate induces strains that in
many cases determine magnetocrystalline and magnetotransport
anisotropies, as we discuss in detail in Sections~\ref{bulk-mag-micromag-aniso} and 
\ref{bulk-magtransp-amr}. Theoretical
calculations \cite{Masek:2003_a} of the dependence of the lattice
constant on the density of the most common impurities in DMSs
represent, therefore, another piece to the mosaic of our
understanding of ferromagnetism in these complex systems.

Considering the values of the atomic radii of Mn
($R_{\rm Mn}$ = 1.17~\AA) and Ga ($R_{\rm
Ga}$ = 1.25~\AA), the substitutional
Mn$_{\rm Ga}$ impurity may be expected to lead to only very small
changes (reductions)
in the lattice constant. This expectation is consistent with the calculated \cite{Zhao:2002_a}
lattice constant of a hypothetical zinc-blende MnAs
crystal  whose value is comparable to that of GaAs.
On the other hand, As antisites produce an expansion of the GaAs lattice
\cite{Liu:1995_a,Staab:2001_a} and a similar trend can be expected for interstitial
Mn.

Modern density-functional techniques allow one to move beyond 
intuitive theoretical considerations and discuss the dependence of
the lattice constant on impurity concentrations on a more
quantitative level \cite{Masek:2003_a}. The CPA is again a useful
tool here for studying (Ga,Mn)As properties over a wide range of
impurity concentrations. 
Some quantitative inaccuracies in  theoretical results due to the
limitations of the LMTO/CPA approach ({\em e.g.} neglect of local
lattice relaxations) have been
corrected by using the full-potential linearized-
augmented-plane-wave supercell method
\cite{Masek:2005_a}.
Starting with an ideal (Ga,Mn)As mixed crystal with all Mn
atoms occupying substitutional Ga positions, these calculations give
the following Vegard's law expression for the doping dependence of
the lattice constant:
\begin{equation}
a_{lc}(x_{s})=a_0+a_s x_{s} (\AA), 
\label{a_Mn_Ga}
\end{equation}
with the expansion coefficient $a_{s}$ ranging from -0.05 to 0.02
depending on the method used in the calculation \cite{Masek:2003_a,Masek:2005_a}.
As expected, $a$ changes only weakly with the Mn$_{\rm Ga}$
density $x_{s}$. 

\begin{figure}
\begin{center}
\ifXHTML\Picture{review/figures/Fig13.png}\else\includegraphics[width=67mm,height=86mm,angle=270]{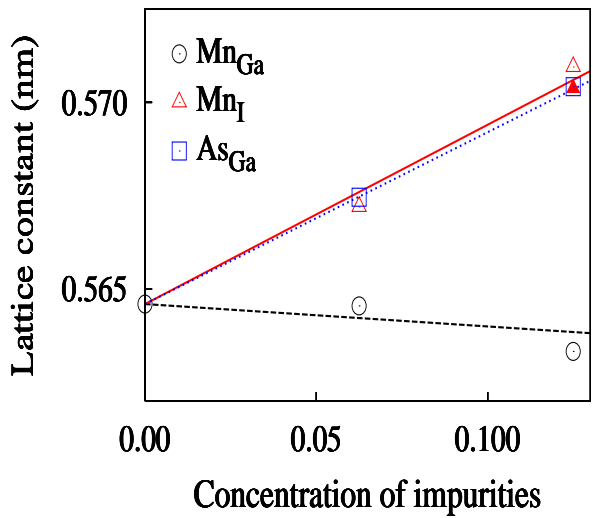}\fi
\caption{Theoretical LMTO/CPA  lattice constant of (Ga,Mn)As as a function of
the concentration of the most important impurities Mn$_{\rm Ga}$
(circles), Mn$_{\rm I}$ (full triangles), and As$_{\rm Ga}$
(empty triangles).
From \protect\cite{Masek:2005_a}.}
\label{alatt1}
\end{center}
\end{figure}

A similar linear dependence is obtained for hypothetical crystals where As$_{\rm Ga}$ (or
Mn$_{\rm I}$) is the only impurity present in the material, as
shown in Fig.~\ref{alatt1}. According to the more reliable full-potential supercell calculations
\cite{Masek:2005_a}, the composition-dependent
lattice constant is found to obey:
\begin{equation}
a_{lc}(x_{s},x_{i},y)=a_0-0.05x_{s}+0.48x_{i}+0.46y (\AA),
\label{a_all}
\end{equation}
where $x_{i}$ and $y$ are the densities of Mn$_{\rm I}$ and
As$_{\rm Ga}$, respectively, and $a_0$ is the lattice constant of
pure GaAs. 

\begin{figure}
\ifXHTML\Picture{review/figures/Fig14.png}\else\includegraphics[angle=-90,width=3.2in]{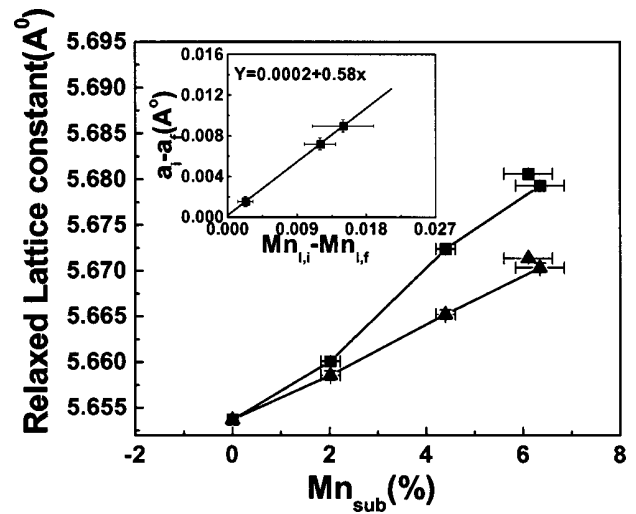}\fi
\caption{
Relaxed lattice constants for the 50-nm-thick (Ga,Mn)As films before
annealing (squares) and after annealing (triangles) as a function of substitutional
Mn$_{\rm Ga}$ content. The inset shows the change of the relaxed lattice constant
as a function of the change in the fraction of interstitial Mn 
 due to the out-diffusion of interstitial Mn during annealing ({\em i.e.}, Mn$_{\rm I,i}$-Mn$_{\rm I,f}$).
From \protect\cite{Zhao:2005_a}.
}
\label{lattice_const_nott}
\end{figure}

Recently, several experimental works
\cite{Potashnik:2002_a,Kuryliszyn-Kudelska:2004_a,Sadowski:2004_a,Zhao:2005_a}
have studied the dependence of lattice constants in (Ga,Mn)As
materials on disorder, based on the comparison between as-grown
and annealed samples. The measurements confirmed that both
As$_{\rm Ga}$ and Mn$_{\rm I}$ defects lead to a significant
expansion of the lattice. In samples grown under As-rich
condition, which is expected to inhibit formation of Mn$_{\rm I}$
impurities, annealing has virtually no effect on the measured
lattice constant. This is consistent with the stability of
As$_{\rm Ga}$ defects up to temperatures that are far above the
annealing temperatures. Mn$_{\rm I}$ impurities, on the other
hand, can be efficiently removed by low-temperature annealing.
Consistently, annealing leads to a significant reduction of the
lattice constant in materials which contain a large number of
these defects in the as-grown form, as shown in Fig.~\ref{lattice_const_nott}.
On a quantitative level,
experimental data suggest a stronger lattice expansion due to
Mn$_{\rm Ga}$ and a weaker expansion due to As$_{\rm Ga}$ and
Mn$_{\rm I}$,  compared to the theoretical predictions of
Eq.~(\ref{a_all}).  The quantitative
disagreement can be attributed, in part, to the simplified description
of the system within the theoretical model. Also, the presence
of other lattice imperfections or inaccuracies in the
determination of experimental doping values may have  partly obscured the
direct quantitative comparison between experiment and theory.

%%%%%%%%%%%%%%%%%%%%%%%%%%%%%%%%%%%%%%%%%%%%%%%%%%%%%%%%%%
\section{Magnetic properties}
\label{bulk-mag}
\subsection{Ferromagnetic critical temperature}
\label{bulk-mag-tc}
Curie   temperatures   in  metallic   (Ga,Mn)As   have  been   studied
theoretically        starting        from        the {\bf k}$\cdot${\bf p} kinetic-exchange effective Hamiltonian
\cite{Jungwirth:1999_a,Dietl:2000_a,Jungwirth:2002_b,Brey:2003_a,DasSarma:2004_a,Jungwirth:2005_b}
and                                   from                      microscopic TBA or SDF
band structure calculations
\cite{Sandratskii:2002_a,Sato:2003_a,Sandratskii:2004_a,Hilbert:2005_a,Jungwirth:2003_c,Timm:2004_b,Jungwirth:2005_b,Xu:2005_b,Bouzerar:2004_a,Bouzerar:2005_b,Bergqvist:2004_a,Bergqvist:2005_a}.
For a more detail description of these theoretical approaches see Sections~\ref{theory-first}-\ref{theory-kp}. 
The advantage
of     the    {\bf k}$\cdot${\bf p} kinetic-exchange  model     is     that     it    uses     the experimental
value    for    the $p-d$ coupling constant  $J_{pd}$,  {\em i.e.},  it  correctly  captures  the  strength  of  the
magnetic interaction that has been established to play the key role in
ferromagnetism  in  (Ga,Mn)As.    The  model   also  accounts  for   strong  spin-orbit
interaction present  in the host  valence band which splits  the three
$p$-bands  into a  heavy-hole,  light-hole, and  a  split-off band  with
different   dispersions.   The   spin-orbit  coupling    was shown \cite{Konig:2001_a,Brey:2003_a} to play 
an  important role  in suppressing  magnetization  fluctuation effects
and, therefore, stabilizing the ferromagnetic state up to high temperatures. On the other hand,
describing the potentially complex behavior
of Mn$_{\rm  Ga}$ in GaAs by a  single parameter may oversimplify the problem. The
calculations omit, for example, the suppression of $T_c$ in low hole-density (Ga,Mn)As 
materials due to the direct antiferromagnetic superexchange
contribution to the coupling of near-neighbor Mn pairs. The whole model inevitably breaks down
in DMS systems with holes strongly bound to Mn acceptors or with
large charge fluctuations on Mn$_{\rm  Ga}$ $d$-shells.

The advantage of  microscopic approaches to Curie temperature calculations
is that they make no  assumption about  the character of  Mn$_{\rm Ga}$  impurities in
GaAs and their magnetic coupling. They are therefore useful  for studying material trends in $T_c$
as a function  of Mn doping or the density  of
other intentional or unintentional  impurities and defects present  in real systems.  
Because spin-orbit interactions add to the numerical complexity of calculations 
that are already challenging, they have normally been neglected in this approach.  
Another   shortcoming, discussed already in Section~\ref{theory-first},  of   the  LSDA
approaches is   an overestimated strength
of the $p-d$ exchange as compared to experiment.
Within the mean-field approximation, which considers thermodynamics of
an isolated  Mn moment in  an effective field and  neglects correlated
Mn-Mn  fluctuations, microscopic calculations 
typically  yield larger   $T_c$'s  than   the effective Hamiltonian
model  that  uses the  experimental
value for $J_{pd}$. 
Stronger  $p-d$  exchange and the omission of spin-orbit coupling effects in  the DFT  theories, 
however, also leads to a  larger suppression of the Curie temperature
due   to   fluctuation   effects. (A closer agreement on the character
of  the  $T_c$ versus  Mn-doping  curves,  calculated  within the  two
formalisms, is  obtained when  the deficiencies of  LSDA
theories  are partly  eliminated by using, {\em e.g.}, the LDA+U.) Despite
the   above  weaknesses   of  semi-phenomenological   and  microscopic
calculations, a qualitatively consistent picture is clearly
emerging  from  these  complementary  theoretical approaches  that,  as  we
discuss  below, provides  a  useful framework  for analyzing  measured
$T_c$'s.

\subsubsection{Mean-field theory in (Ga,Mn)As}
\label{bulk-mag-tc-mf}
Our review of theoretical $T_c$ trends in (Ga,Mn)As  starts
with the results of the KL kinetic-exchange Hamiltonian  and mean-field  approximation
to  set  up   a  scale  of  expected  Curie
temperatures. These  estimates which are not accurate in all regimes, are 
simplified by assuming a homogeneous distribution of Mn$_{\rm Ga}$ ions
and neglecting the role of other defects,   
apart  from their  potential contribution to hole or  moment compensation
\cite{Dietl:2000_a,Dietl:1997_a,Jungwirth:1999_a,Konig:2003_a}.
In the case of microscopic models this assumption is equivalent to the virtual crystal approximation.    
Microscopic TBA calculations have shown  very little
effect of positional disorder on the strength of magnetic couplings in
(Ga,Mn)As  epilayers with  metallic conductivities,  
partly justifying the  virtual-crystal approach \cite{Jungwirth:2003_c}.  In addition,
 detailed
theoretical studies confirm the  absence   of  any   significant  magnetic  
frustration   associated with  the random  positions of
Mn$_{\rm Ga}$ moments in the lattice in  the more metallic
ferromagnetic  semiconductors \cite{Timm:2004_b,Fiete:2004_a}.
In the very dilute limit, however, $T_{c}$ becomes sensitive to
the distribution of the Mn moments in the lattice 
\cite{Bergqvist:2004_a}.

In  the  mean-field approximation \cite{Jungwirth:1999_a,Dietl:2000_a},
each local Mn$_{\rm  Ga}$  moment is described by a  Hamiltonian ${\bf S}_{I}
\cdot  {\bf H}_{MF}$  where ${\bf S}_{I}$ is the  Mn$_{\rm  Ga}$ local spin operator,
${\bf H}_{MF}  = J_{pd}  \langle {\bf s}\rangle$,  and $\langle
{\bf  s}\rangle$  is  the  mean  spin density  of  the  valence  band
holes (for the definition of the $J_{pd}$ field see Section~\ref{micro-subMn_pd}). 
$H_{MF}$ is an effective field seen by the local moments due to
spin-polarization of  the band holes, analogous to  the nuclear Knight
shift.   Similarly   ${\bf  h}_{MF}   =  J_{pd}   N_{Mn}   
\langle {\bf S}\rangle $  is an effective  magnetic field experienced  by the
valence band  holes which is proportional to the density and the  mean spin
polarization of the Mn$_{\rm  Ga}$ local moments. 
The dependence  of $\langle {\bf S}\rangle$ on
temperature  and  field $H_{MF}$  is  given \cite{Konig:2003_a} by  the
Brillouin function \cite{Ashcroft:1976_a}:
\begin{equation}
\langle {\bf S}\rangle=       \frac{{\bf H}_{MF}}{|H_{MF}|}S
B_s(S|H_{MF}|/k_BT)\; . 
\label{bs}
\end{equation}
The  Curie temperature  is  found by  linearizing  $H_{MF}$ and  $B_s$
around $\langle {\bf S}\rangle =0$:
\begin{eqnarray}
{\bf H}_{MF}&\approx& J_{pd}^2N_{Mn} \langle {\bf S}\rangle\chi_f
\nonumber \\
B_s&\approx&\frac{S+1}{3}\frac{S|H_{MF}|}{k_BT_c}\; . 
\label{linear}
\end{eqnarray}
Here $\chi_f$ is the itinerant hole spin susceptibility given by
\begin{equation}
\chi_f=\frac{d\langle s\rangle}{dh_{MF}} =-\frac{d^2e_T}{dh_{MF}^2}\; , \label{chi}
\end{equation}
and $e_T$ is the total energy per volume of the holes. Eqs.~(\ref{bs}) 
and (\ref{linear}) give
\begin{equation}
  k_B T_c = \frac{N_{\rm Mn} S (S+1)}{3}
  J_{\rm pd}^2\,\chi_f \; .
\label{tc}
\end{equation}

The qualitative    implications     of    this
$T_c$-equation can be understood within  a model
itinerant hole system  with a single spin-split band  and an effective
mass $m^{\ast}$. The kinetic  energy contribution, $e_k$, to the total
energy of the band holes gives a susceptibility:
\begin{equation}
\chi_{f,k}=-\frac{d^2e_k}{dh_{MF}^2}= \frac{m^{\ast}k_F}{4\pi^2\hbar^2}
  \; ,
\label{chik}
\end{equation}
where $k_F$  is the Fermi wavevector. Within  this approximation $T_c$
 is  proportional to  the Mn$_{\rm  Ga}$  local moment density, to  the hole  Fermi
 wavevector, {\em i.e.} to  $p^{1/3}$ where $p$ is the  hole density, and to
 the hole effective mass $m^{\ast}$.

\begin{figure}
\ifXHTML\Picture{review/figures/Fig15.png}\else\includegraphics[width=3.3in]{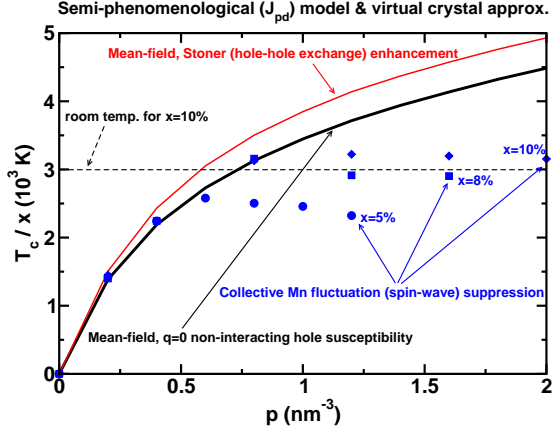}\fi
\caption{Ferromagnetic transition temperatures of (Ga,Mn)As calculated within the
effective Hamiltonian and virtual crystal approximation: 
mean-field (thick black line), Stoner enhancement of $T_c$ (thin red line), 
$T_c$ after correcting for correlated Mn orientation fluctuations using a spin-wave approximation
(blue symbols).
From \protect\cite{Jungwirth:2005_b}.
} 
\label{tc_mf_sw}
\end{figure}

A more quantitative prediction for the Curie temperature is obtained by
evaluating  the  itinerant hole  susceptibility  using  a
realistic band Hamiltonian, 
\begin{equation}
{\mathcal H}={\mathcal H}_{KL}+ {\bf s}\cdot {\bf h}_{MF},
\label{kl-pd-ham}
\end{equation}
where  ${\mathcal H}_{KL}$ is the  six-band KL  Hamiltonian of  the  GaAs host
band \cite{Vurgaftman:2001_a}   and   $\vec  s$   is   the  hole   spin
operator \cite{Dietl:2000_a,Dietl:2001_b,Abolfath:2001_a}. The results,
represented  by  the  solid  black line  in  Fig.~\ref{tc_mf_sw},  are
consistent with  the qualitative analysis based on  the parabolic band
model, {\em i.e.},  $T_c$ roughly follows  the $\sim xp^{1/3}$  dependence. Based on
these  calculations, room temperature  ferromagnetism in  (Ga,Mn)As is
expected for 10\% Mn$_{\rm Ga}$ doping in weakly compensated samples \cite{Jungwirth:2005_b}.

Hole-hole Coulomb interaction effects can be included, in the lowest order of  perturbation
theory by adding the  hole exchange contribution to the total energy  \cite{Mahan_1981_a}.
The  red   line  in
Fig.~\ref{tc_mf_sw}  shows  this  Stoner $T_c$  enhancement  calculated
numerically using the kinetic-exchange effective model with the six-band KL Hamiltonian.  
$T_c$ stays roughly
proportional to $xp^{1/3}$ even  if hole-hole exchange interactions are
included,  and  the  enhancement  of  $T_c$  due  to
interactions is of the order $\sim 10-20$\% \cite{Jungwirth:1999_a,Dietl:2001_b,Jungwirth:2005_b}.

The mean-field effective Hamiltonian analysis above neglects   discreteness in
the random Mn$_{\rm  Ga}$ positions in the lattice and magnetic coupling mechanisms 
additional to the kinetic exchange contribution, particularly the near neighbor superexchange. 
The former point can be expected to influence $T_c$ at large hole densities, {\em i.e.}, when the hole Fermi
wavelength approaches inter-atomic distances. Of course, the entire 
phenomenological scheme fails on many fronts when the Fermi wavelength 
approaches atomic length scales since it is motivated by the assumption that all relevant 
length scales are long; the {\bf k}$\cdot${\bf p} band structure, 
the use of the host material band parameters, and the neglect of momentum dependence in the 
$J_{pd}$ parameter all become less reliable as the hole density 
increases to very large values. The approximations are apparently not fatal, however, even for $x\sim 10$\% 
and any degree of compensation.

In the opposite limit of strongly compensated systems, where the overall
magnitude
of the hole-mediated exchange is weaker, 
antiferromagnetic superexchange
can dominate  the near-neighbor Mn$_{\rm  Ga}$-Mn$_{\rm  Ga}$ coupling \cite{Kudrnovsky:2004_a},  
leading  to a reduced
Curie temperature \cite{Jungwirth:2005_b,Sandratskii:2004_a}. 
We emphasize that the  {\bf k}$\cdot${\bf p} kinetic-exchange model cannot be applied consistently 
when nearest neighbor interactions dominate, 
since it implicitly assumes that all length scales are longer than a lattice constant.  
We also note that net antiferromagnetic coupling of near-neighbor Mn$_{\rm Ga}$-Mn$_{\rm Ga}$ 
pairs is expected only in systems with large charge compensations. In weakly compensated (Ga,Mn)As the ferromagnetic contribution takes over \cite{Kudrnovsky:2004_a,Mahadevan:2004_c,Dietl:2001_b}.

In addition to   the above effects related to random Mn distribution, Mn positional disorder can directly modify  $p-d$ 
interactions when the coherence of  Bloch states becomes significantly
disturbed. Microscopic theories, such as the TBA/CPA 
\cite{Jungwirth:2005_b} or {\em ab initio} approaches based on either
CPA or supercell band structures \cite{Sandratskii:2004_a,Sato:2003_a}, 
capture all these effects on an equal footing and can be used 
to estimate trends in mean-field $T_c$ beyond the virtual crystal
approximation.   

\begin{figure}[h]
\ifXHTML\Picture{review/figures/Fig16.png}\else\hspace*{-.5cm}
\includegraphics[width=2.6in,angle=-90]{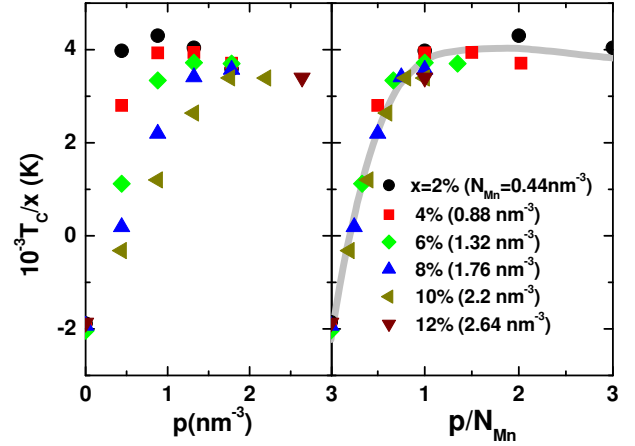}\fi
\caption{$T_c$ calculations within the microscopic TBA/CPA model: $T_c$ versus hole density 
(left panel), $T_c$ versus number of holes per  Mn$_{\rm Ga}$ (right panel). The overall theoretical
$T_c$ trend is highlighted in grey.
From \protect\cite{Jungwirth:2005_b}.} 
\label{tc_tba}
\end{figure} 

The mean-field CPA Curie temperatures are obtained   by evaluating
the energy cost of flipping one Mn$_{\rm Ga}$ moment with all 
other moments held fixed in the ferromagnetic ground state.
It can be evaluated for any given chemical composition
\cite{Liechtenstein:1987_a,Masek:1991_a} and  defines an
effective exchange field $H_{eff}$ acting on the local moment. This energy change
corresponds to $H_{MF}$ in the 
kinetic-exchange model used in the previous section, {\em i.e.},
\begin{equation}
k_BT_c=\frac{S+1}{3}H_{eff} \; .
\label{tc_heisenberg}
\end{equation}
Results based on microscopic TBA band structure calculations are shown in Fig.~\ref{tc_tba} as a function
of hole density  for several Mn$_{\rm Ga}$ local moment concentrations \cite{Jungwirth:2005_b}.
(The hole density is varied independently of 
Mn$_{\rm Ga}$ doping in these calculations
by adding non-magnetic donors or acceptors.)
Comparison with Fig.~\ref{tc_mf_sw} identifies the main
physical origins of the deviations from the $T_c\sim xp^{1/3}$ trend.
Black dots in the left panel of Fig.~\ref{tc_tba}
correspond to a relatively low local Mn$_{\rm Ga}$ moment 
concentration ($x=2\%$) and hole densities ranging up to $p=4N_{Mn}$ and show
the expected suppression of $T_c$ at large $p$. The effect of superexchange
in the opposite limit is clearly seen when inspecting, {\em e.g.}, the $x=10\%$ data
for $p<1$~nm$^{-3}$. The mean-field TBA/CPA 
Curie temperature is largely suppressed
here or even negative, meaning that the ferromagnetic state becomes unstable
due to the short-range antiferromagnetic coupling. Note, that the neglect of Coulomb interactions in these
TBA/CPA calculations likely leads to an overestimated strength of the antiferromagnetic superexchange.
The 
inhomogeneity of the carrier distribution in the disordered mixed 
crystal may also contribute to the steep decrease of $T_c$ with increasing 
compensation seen in Fig.~\ref{tc_tba}.

Although
the Curie temperatures in the left panel of Fig.~\ref{tc_tba} appear to
depart strongly for the $T_c\sim xp^{1/3}$ dependence, the linearity
with $x$ is almost fully recovered when
$T_c$ is plotted as a function of the number of holes per Mn$_{\rm Ga}$ local moment,  $p/N_{Mn}$
(see right panel of Fig.~\ref{tc_tba}). Note that for compensations ($1-p/N_{Mn}$) reaching 100\%
this property of the superexchange coupling is reminiscent of the behavior
of (II,Mn)VI diluted magnetic semiconductors \cite{Furdyna:1988_b} in which Mn acts as an isovalent
magnetic impurity. The dependence on $p$ in (Ga,Mn)As is expected
to become very weak, however, when approaching the uncompensated state.
Similarly, the prospects for substantial increases in $T_c$ by 
non-magnetic acceptor co-doping of weakly compensated material appear to be quite limited. 

\begin{figure}[h]
\ifXHTML\Picture{review/figures/Fig17.png}\else\hspace*{-.5cm}
\includegraphics[width=1.6in,angle=-90]{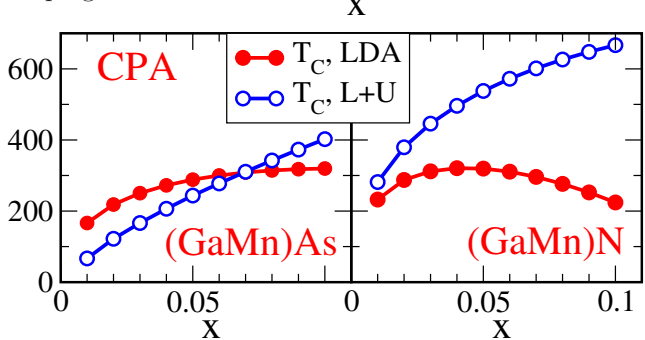}\fi
\caption{LDA/CPA (red) and LDA+U/CPA (blue) calculations of the Curie temperature as a
function of Mn doping in uncompensated (one hole
per Mn) (Ga,Mn)As and (Ga,Mn)N.  
From \protect\cite{Sandratskii:2004_a}.
} 
\label{lda_tc}
\end{figure}

In the left panel of Fig.~\ref{lda_tc} we show mean-field CPA Curie temperatures in uncompensated
($p/N_{Mn}=1$) (Ga,Mn)As DMSs as a function of Mn doping calculated using LDA and LDA+U {\em ab initio}
methods. The LDA+U calculations, which give more realistic values of the $p-d$ exchange
coupling, confirm the linear dependence of $T_c$ on $x$,  showing no signs of saturation even at the largest doping
$x=10$\% considered in these calculations.

%%%%%%%%%%%%%%%%%%%%%
\subsubsection{Role of collective Mn-moments fluctuations and different III-V hosts}
\label{bulk-mag-tc-sw} 
The potential influence of correlated Mn-moment fluctuations (corrections to mean field theory)
on ferromagnetic ordering in (Ga,Mn)As can be recognized by considering, 
within a simple parabolic band model, the RKKY  oscillations effect which
occurs as a consequence
of the $2k_F$ anomaly in the wavevector dependent susceptibility of the 
hole system \cite{Dietl:1997_a,Brey:2003_a}. 
In this theory, which treats the hole system perturbatively around the paramagnetic state,
the sign of the hole-mediated 
Mn$_{\rm  Ga}$-Mn$_{\rm  Ga}$ coupling varies
as $\cos(2k_Fd)$, where $d$ is the distance between Mn$_{\rm  Ga}$ moments, and its
amplitude decays as $d^3$. 
Estimating the average Mn$_{\rm  Ga}$-Mn$_{\rm  Ga}$ separation in
a (Ga,Mn)As random alloy as
$\bar{d}=2(3/4\pi N_{Mn})^{1/3}$ for uncompensated
(Ga,Mn)As systems  and neglecting spin-orbit coupling and band warping,  
$\cos(2k_F\bar{d})\approx -1$, which means that the role of the RKKY oscillations cannot be generally discarded. 
In realistic valence bands the  oscillations are suppressed due to  non-parabolic and
anisotropic dispersions of the heavy- and light-hole bands and due to  strong 
spin-orbit coupling \cite{Brey:2003_a,Konig:2001_a}.

More quantitatively, the   range   of   reliability   and corrections
to  the   
mean-field   approximation in  (Ga,Mn)As can be estimated by accounting for the
suppression
of   the   Curie   temperature using the
quantum theory of long-wavelength   spin-waves or using Monte Carlo simulations which treat 
Mn-moments as classical variables. For weak $p-d$ exchange coupling, $SN_{Mn}J_{pd}/E_F\ll 1$ where
$E_F$ is the hole Fermi energy, the spin-polarization of the hole system is small and
the RKKY and spin-wave approximations treat collective Mn-moment
fluctuations on a similar level. The advantage of the spin-wave theory is that it can be used to explore
the robustness of ferromagnetic states over a wider range of $p-d$ couplings, including the 
more strongly exchange-coupled as-grown materials with large Mn density and large hole compensation
(See Section~\ref{discussion-general} for a general discussion of magnetic interactions in the two coupling-strength
limits).

Calculations in metallic systems have been performed starting from the {\bf k}$\cdot${\bf p} kinetic-exchange effective  
Hamiltonian  \cite{Schliemann:2001_b,Brey:2003_a,Jungwirth:2005_b}
or from the SDF band structure calculations \cite{Xu:2005_b,Hilbert:2005_a,Bouzerar:2004_a,Bergqvist:2004_a}.
(Monte Carlo studies of $T_c$ in the regime near the metal-insulator are reported in
\cite{Mayr:2002_a}.)
Within a non-interacting spin-wave approximation,
the  magnetization   vanishes  at  the
temperature where  the number of  excited spin waves equals  the total
spin of the ground state \cite{Jungwirth:2002_b,Jungwirth:2005_b}:
\begin{equation}
   k_BT_c = \frac{2S+1}{6}  k_D^2 D(T_c) \; ,
\label{collective_tc}
\end{equation}
where 
$k_D=(6\pi^2N_{\rm Mn})^{1/3}$ is the Debye cutoff and $D(T)=D_0\, \langle S\rangle(T)/S$ 
is proportional to the zero-temperature  spin-wave 
stiffness  parameter $A$ ($D_0=2A/SN_{Mn}$) 
and the mean-field temperature-dependent  average spin on Mn, $\langle S\rangle(T)$
\cite{Konig:2001_a,Schliemann:2001_a}.
(For a detail discussion of the micromagnetic parameter $A$ see Section~\ref{bulk-mag-micromag}). 

Comparing the spin-stiffness  results obtained using the {\bf k}$\cdot${\bf p} kinetic-exchange  model
with a  simple parabolic band and with the more
realistic, spin-orbit coupled KL Hamiltonian,  
the spin stiffness is observed to always be much larger in
the KL model \cite{Konig:2000_a,Konig:2001_a}. 
For (Ga,Mn)As, the  parabolic-band model underestimates
$D$  by a factor  of $\sim$10-30  for typical hole densities.   
This larger spin stiffness in the spin-orbit coupled valence bands 
is due to the heavy-hole -- light-hole  mixing.  
Crudely, the large mass heavy-hole
band  dominates the  spin  susceptibility and  enables local (mean-field)  magnetic
order  at  high temperatures,  while  the  dispersive light-hole  band
dominates   the  spin  stiffness   and  enables   long-range  magnetic
order.  The analysis highlights that the
 multi-band  character  of the  semiconductor valence  band
plays an important role in the ferromagnetism of (Ga,Mn)As.

Critical temperature estimates based on Eq.~(\ref{collective_tc}), the KL kinetic-exchange model,
and including also the  Stoner enhancement of $T_c$  are summarized in
 Fig.~\ref{tc_mf_sw}  by the blue symbols. These  $T_c$ estimates  indicate
 that  $T_c$ will remain  roughly proportional to $x$  even at
large dopings.  The suppression of  $T_c$ due to  spin-waves increases
with   increasing  hole   density   relative  to   the  local   moment
concentration, resulting in saturation of the critical temperature
with increasing $p$ at about 50\% compensation.

\begin{figure}[h]
\ifXHTML\Picture{review/figures/Fig18.png}\else\hspace*{-.5cm}
\includegraphics[width=2.6in,angle=-90]{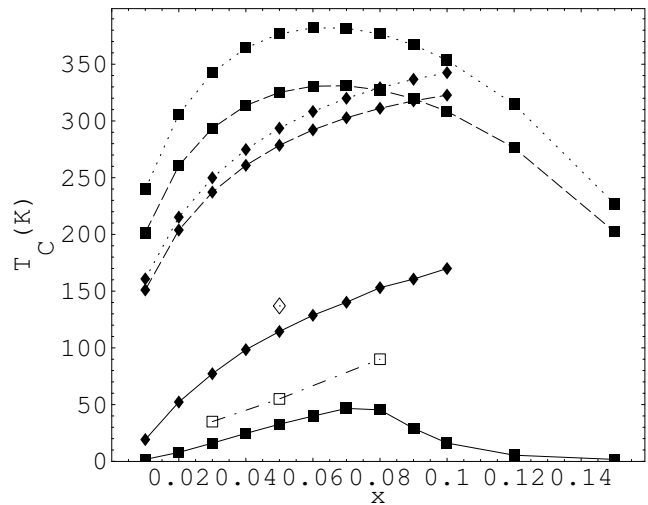}\fi
\caption{ Comparison of the {\em ab initio} Curie temperatures  of
Ga$_{1-x}$Mn$_{x}$As (diamonds) and Ga$_{1-x}$Mn$_{x}$N (squares) 
obtained by the spin-wave approach (solid line, filled symbols),
virtual-crystal-approximation-RPA (dashed line, filled symbols), mean-field-approximation (dotted line,
filled symbols) and Monte Carlo (dash-dotted line, open symbols). 
From \protect\cite{Hilbert:2005_a}.
} 
\label{lda_tc_sw}
\end{figure}

The suppression of $T_c$ due to  correlated Mn-moment fluctuations is also observed  in the 
LSDA calculations \cite{Xu:2005_b,Hilbert:2005_a,Bergqvist:2004_a,Bergqvist:2005_a,Bouzerar:2004_a,Bouzerar:2005_b}. 
The trend is illustrated in Fig.~\ref{lda_tc_sw} where  collective 
fluctuations are accounted for using the spin-wave theory or the Monte Carlo approach 
\cite{Hilbert:2005_a,Bergqvist:2004_a}; similar trends of suppressed mean-field $T_c$ due to collective
Mn-moment fluctuations have been predicted by a spin-wave
theory using a more elaborate, self-consistent RPA technique
on random lattice \cite{Bouzerar:2004_a}. A larger suppression of the mean-field $T_c$  seen
in the {\em ab initio} calculations, compared to the KL kinetic-exchange  model results,
can be attributed partly to the simpler, three-fold degenerate LSDA valence band structure 
in theories that neglect spin-orbit coupling.
Also, the stronger $p-d$ exchange in the LSDA theories may result in a weaker spin stiffness
of the magnetic system, as the holes are more strongly bound to the Mn acceptors and the hole mediated
Mn-Mn coupling has a more short-range character. The enhancement of fluctuation effects in stronger $p-d$ coupled
systems is clearly seen in  Fig.~\ref{lda_tc_sw} when comparing the LSDA results for narrower gap
(weaker $p-d$ exchange) (Ga,Mn)As and wider gap (stronger $p-d$ exchange) (Ga,Mn)N.

The quantitative discrepancies between KL kinetic-exchange model and LSDA results for
the mean-field $T_c$ and for the suppression of ferromagnetism 
due to collective Mn-moment fluctuations partly cancel out, leading
to similar overall predictions of the Curie temperatures in (Ga,Mn)As. Based on the $T_c$ analysis alone
it is therefore difficult to  determine whether the magnetic interactions have a 
short-range or long-range character in (Ga,Mn)As DMSs. 

Theoretically, the localization of the hole around the Mn impurity and the range of
magnetic Mn-Mn interactions can be studied using microscopic TBA or {\em ab initio}
calculations of the charge and moment distribution in the lattice or by mapping 
the total energy of the DMS crystal to the Heisenberg 
Hamiltonian 
\cite{Schilfgaarde:2001_a,Sandratskii:2002_a,Sato:2003_a,Wierzbowska:2004_a,Mahadevan:2004_c,Kudrnovsky:2004_a,Hilbert:2005_a,Schulthess:2005_a,Timm:2004_b}.
Moving down the anion column in the periodic table from the nitride DMSs to antimonides, the holes become more
delocalized \cite{Mahadevan:2004_c} and, consequently, 
the Mn-Mn interactions are  more long-range in these microscopic calculations.
In (Ga,Mn)As the LSDA theory predicts short-range magnetic coupling 
while the LDA+U or SIC-LSDA
results suggest  that the holes which
mediate the Mn-Mn exchange interaction are more delocalized \cite{Schilfgaarde:2001_a,Wierzbowska:2004_a,Schulthess:2005_a}.
 
Combined theoretical and experimental 
studies of remanent magnetization, micromagnetic parameters, and magneto-transport coefficients, discussed in 
detail in Sections~\ref{bulk-mag-magnet}-\ref{bulk-ac_magtransp}, indicate that in high quality (Ga,Mn)As
ferromagnets with metallic conductivities (conductivity increases with decreasing temperature) the holes
are sufficiently delocalized to make the kinetic-exchange model approach applicable.  It is natural to expect that
the free-carrier-mediated ferromagnetism picture will also apply in narrower gap antimonide DMSs, such as
(In,Mn)Sb, with even larger conductivities due to the smaller hole effective mass \cite{Wojtowicz:2003_a}.
Smaller effective mass and larger unit cell volume, as compared to arsenide DMSs, explain 
the smaller Curie temperatures in (III,Mn)Sb. This  trend is illustrated 
in Table~\ref{tc_mf_as_sb} by comparing the respective mean-field KL kinetic-exchange model $T_c$'s. 
In Mn-doped phosphides and nitrides the suppression of $T_c$ due to effects beyond the mean-field virtual-crystal
approximation, seen in the LSDA calculations and
related to the short-range nature of
magnetic interactions, may prevail \cite{Scarpulla:2005_a}. We note also that
the above LSDA Curie temperature studies do not capture the possible transition of the Mn impurity state in
the wide gap III-V's to the
highly correlated tri-valent ($d^4$) center with four strongly localized $d$-electrons
and an empty $d$-state deep in the gap, in which case the hole mediated ferromagnetism picture these
calculations imply is not applicable \cite{Luo:2004_a,Kreissl:1996_a,Schulthess:2005_a}.

\begin{figure}
\ifXHTML\Picture{review/figures/Fig19.png}\else%
%\vspace*{20cm}
\includegraphics[angle=-90,width=3.4in]{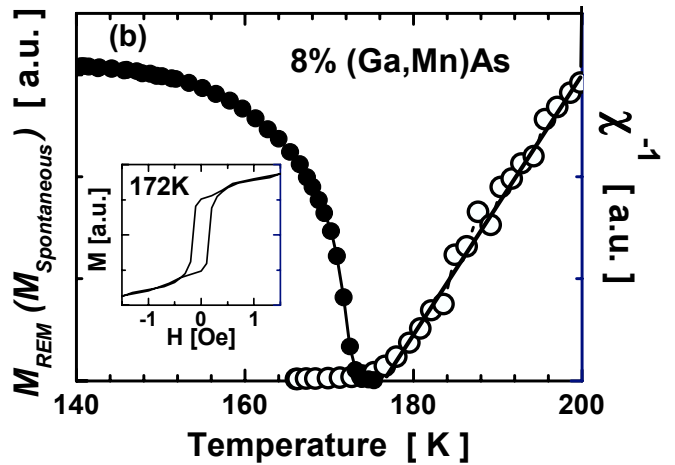}%
%\vspace{-10cm}
\fi
\caption{ Temperature dependence
of remanent magnetization and inverse
paramagnetic susceptibility for Ga$_{0.91}$Mn$_{0.09}$As sample with $T_c=173$~K; inset:
hysteresis loop for the same sample at 172~K.
From \cite{Wang:2004_c}.
}
\label{nott_173_sample}
\end{figure}

\begin{figure}
\ifXHTML\Picture{review/figures/Fig20.png}\else%
%\vspace*{20cm}
\hspace*{-0.5cm}%
\includegraphics[angle=-90,width=4.0in]{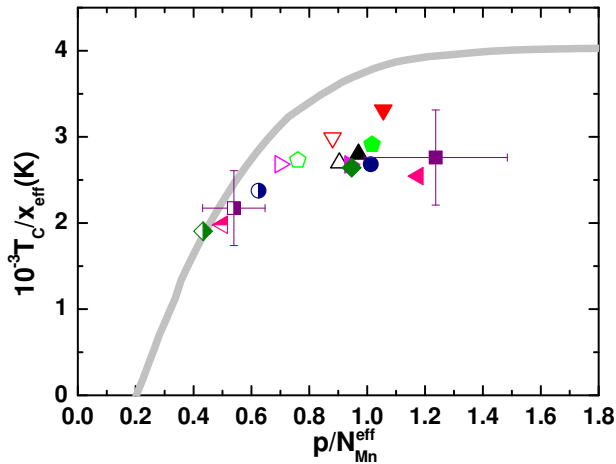}%
%\vspace{-10cm} 
\fi
\caption{ Experimental $T_c/x_{eff}$ vs. hole density  relative to  effective concentration
of Mn$_{\rm Ga}$ moments. Deviations from linear dependence on $x_{eff}$ are seen only for
high compensations  ($1-pa_{lc}^3/4x_{eff}=1-p/N_{Mn}^{eff}>40\%$) in agreement with  theory. For weakly
compensated samples $T_c$ shows  no signs of saturation with increasing
$x_{eff}$. Theoretical (grey) $T_c$ trend from Fig.~\protect\ref{tc_tba} is plotted for comparison.  
From \cite{Jungwirth:2005_b}.}
\label{tc_p_Mn_eff_exp}
\end{figure}

\begin{table}
\begin{center}   
\begin{tabular}{cccccc}
host & $p$ (nm$^{-3}$)
& $T_c^{\rm MF}$  & $T_c^{\rm ex}$ & $T_c^{\rm coll}$ & $T_c^{\rm est}$ (K)
\\ \hline
AlAs & 0.1 & 45 & 53 & 41 & 47\\
     & 0.5 & 134 & 158 & 105 & 119\\ \hline
GaAs & 0.1 & 40 & 43 & 38 & 41\\
     & 0.5 & 124 & 138 & 106 & 115\\ \hline
InAs & 0.1 & 14 & 15 & 14 & 15\\
     & 0.5 & 41 & 44 & 40 & 41\\ \hline
AlSb & 0.1 & 19 & 22 & 18 & 20\\
     & 0.5 & 58 & 64 & 49 & 53\\ \hline
GaSb & 0.1 & 18 & 19 & 18 & 19\\
     & 0.5 & 85 & 88 & 82 & 85 \\ \hline
InSb & 0.1 & 11 & 12 & 11 & 11 \\
     & 0.5 & 37 & 38 & 35 & 36 \\ 
\end{tabular}
\end{center}
\caption{
Mean-field ($T_c^{\rm MF}$), Stoner enhanced ($T_c^{\rm ex}$),
spin-wave suppressed ($T_c^{\rm coll}$), and estimated including both Stoner
enhancement and spin-wave suppression ($T_c^{\rm est}$) ferromagnetic
transition temperatures in III-V host semiconductors
doped with 5\% of Mn and with itinerant hole densities
$p=0.1$ and $p=0.5$~nm$^{-3}$.
From \protect\cite{Jungwirth:2002_b}.
}
\label{tc_mf_as_sb}
\end{table}
Experimentally,  $T_c$ trends have been most extensively studied in 
(Ga,Mn)As
\cite{Ohno:1998_a,Potashnik:2002_a,Edmonds:2002_b,Yu:2003_a,Chiba:2003_b,Ku:2003_a,Stone:2003_a,Jungwirth:2005_b}.  
In Fig.~\ref{nott_173_sample} we show the temperature-dependent magnetization and inverse susceptibility
of the current record $T_c$ material \cite{Wang:2004_c}. 
The Brillouin-function character of the magnetization curve confirms that
the mean-field theory is appropriate in these high quality DMS materials with metallic conductivities.
Curie temperatures for a series of as-grown and 
annealed (Ga,Mn)As samples with experimentally characterized charge and moment   
compensations are plotted  in Fig.~\ref{tc_p_Mn_eff_exp} \cite{Jungwirth:2005_b}.  
The concentration of uncompensated Mn moments in the plot
is $x_{eff}=x_s-x_i$, where it is assumed that the
Mn$_{\rm I}$ donors present in the system are attracted
to Mn$_{\rm Ga}$ acceptors  and that these pairs couples antiferromagnetically, as discussed
in Sections~\ref{micro-other_imp-int_mn} and \ref{bulk-struct-formen}.
(The consistency of this assumption is confirmed by independent magnetization studies reviewed 
in Section~\ref{bulk-mag-magnet}).
The experimental $T_c/x_{eff}$ plotted
against $p/N_{Mn}^{eff}$ in Fig.~\ref{tc_p_Mn_eff_exp}, where $ N_{Mn}^{eff}=4x_{eff}/a_{lc}^3$,
show a common $T_c$ trend which is consistent 
with  theoretical expectations \cite{Jungwirth:2005_b,Bouzerar:2004_a,Bergqvist:2005_a}. 
In particular, theory and experiment agree on the
very weak dependence of $T_c/x_{eff}$ on $p/N_{Mn}^{eff}$ for low compensation and 
the relatively rapid fall of $T_c/x_{eff}$ with decreasing  $p/N_{Mn}^{eff}$
for compensations of $\sim 40\%$ or larger. It should be noted that the 
maximum experimental $x_{eff}$ is only 4.6\%  in the as grown 
sample and 6.8\%  after annealing for a total Mn concentration 
$x=9\%$. Hence the modest $T_c$'s observed so far. Achieving 
$T_c$ values close to room temperature in (Ga,Mn)As, which is expected to occur for  $x_{eff}\approx 10\%$, 
appears to be essentially a 
material growth issue, albeit a very challenging one \cite{Jungwirth:2005_b}. 

Only a few experimental studies of LT-MBE grown (III,Mn)Sb DMSs have been reported to date
\cite{Abe:2000_a,Wojtowicz:2003_a}. Curie temperatures measured in these materials are lower 
than $T_c$'s in Mn-doped arsenides, consistent with the kinetic-exchange model predictions in  
Table~\ref{tc_mf_as_sb}.
The nature of ferromagnetism and, therefore, the interpretation of experimental Curie temperatures 
observed in phosphide and nitride DMSs are not established yet, as we have already pointed out 
in Section~\ref{transtemp}.

%%%%%%%%%%%%%%%%%%%%%%%%%%%%%%%%%%%%%%%%%%%%%%%%%%%%%%%%%

\subsection{Magnetization}
\label{bulk-mag-magnet}
In this section we focus on low-temperature ferromagnetic moments in (Ga,Mn)As DMSs with
metallic conductivities.
Early experimental studies, reporting large apparent 
magnetization deficits in (Ga,Mn)As \cite{Ohno:2001_a,Korzhavyi:2002_a,Potashnik:2002_a},
motivated a theoretical search for possible intrinsic 
origins of frustrating magnetic interactions in this material. Using a wide spectrum
of computational techniques, ranging from {\em ab initio} 
methods \cite{Korzhavyi:2002_a,Mahadevan:2004_b,Kudrnovsky:2004_a} and microscopic 
TBA \cite{Timm:2004_b} to  
{\bf k}$\cdot${\bf p}
kinetic-exchange 
models \cite{Schliemann:2002_a,Schliemann:2003_a,Brey:2003_a,Zarand:2002_a,Fiete:2004_a}
the theoretical studies have identified several mechanisms that can 
lead to non-collinear ground states.  The observation that long wavelength 
spin-waves with negative energies frequently occur within
the parabolic-band  kinetic-exchange model illustrates
that randomness in the distribution
of Mn moments  can result in an instability of the collinear ferromagnetic state \cite{Schliemann:2002_a}.
Frustration can be further enhanced when positional disorder is combined with anisotropies in 
Mn-Mn interactions. The {\em pd} character of electronic states forming the magnetic moment leads
to magnetic interaction anisotropies with respect to the crystallographic orientation of the 
vector connecting two Mn moments \cite{Mahadevan:2004_b,Kudrnovsky:2004_a,Timm:2004_b,Brey:2003_a}. 
When spin-orbit coupling is taken into 
account, magnetic interactions
also become anisotropic with respect to the relative orientation of the Mn-Mn connecting vector and the magnetic moment
\cite{Zarand:2002_a,Fiete:2004_a,Schliemann:2003_a,Timm:2004_b}. 

Some degree of non-collinearity is inevitable as a combined consequence of positional disorder and 
spin-orbit coupling.  Nevertheless it was argued theoretically that
a large suppression of the ferromagnetic moment is not expected 
in metallic (Ga,Mn)As samples with Mn concentrations
above $\sim 1$\% \cite{Timm:2004_b}.  The minor role of non-collinearity is due 
largely to the long-range character of magnetic
interactions, which tends to average out the frustrating 
effect of anisotropic coupling between randomly distributed Mn impurities \cite{Timm:2004_b,Zhou:2003_a}. 
Indeed, {\em ab initio}, microscopic TBA, and {\bf k}$\cdot${\bf p} kinetic-exchange  model
calculations of zero-temperature magnetic moments in
(Ga,Mn)As ferromagnets which neglect 
effects that would lead to non-collinearity 
\cite{Dietl:2001_b,Wierzbowska:2004_a,Schulthess:2005_a,Jungwirth:2005_a}
are consistent 
with experiments reported in a series  of high-quality (Ga,Mn)As ferromagnets 
\cite{Edmonds:2005_b,Wang:2004_d,Jungwirth:2005_a}. It
rules out any marked intrinsic  frustrations in the ground
state of these DMSs. 
The substantial magnetization suppression seen in many early (Ga,Mn)As samples 
can be attributed primarily to the role played in those samples by 
interstitial Mn atoms and other unintentional defects. 
 
\subsubsection{Magnetization of an isolated  Mn($d^5$+hole) complex}
\label{bulk-mag-magnet-singleMn}
We start the discussion by identifying  the key physical considerations
that influence  the ground-state magnetization of (Ga,Mn)As ferromagnets by
focusing first on a single Mn($d^5$+hole) complex and approximating the total magnetization in the 
collinear state by a simple sum
of individual (identical)  Mn($d^5$+hole) complex contributions. This crude model is used only  to 
qualitatively  clarify
the connection between $p-d$ hybridization
and  antiferromagnetic kinetic-exchange coupling, the sign of the hole contribution to
total moment per Mn, and the expected mean-field contribution
to magnetization per Mn from the Mn local moments and from the antiferromagnetically
coupled holes. We also explain in this section that quantum fluctuations around the mean-field
ground state are generically present because of antiferromagnetic character of the
$p-d$ kinetic exchange interaction. 

\begin{figure}[h]
\ifXHTML\Picture{review/figures/Fig21.png}\else\includegraphics[width=3.3in,angle=0]{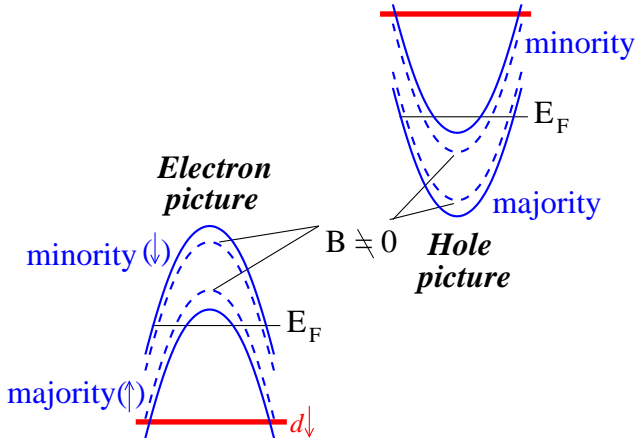}\fi
\caption{Cartoon of Zeeman coupling with an external magnetic field assuming $g>0$, in the electron and in the hole 
picture. In the ferromagnetic state the valence band is spin-split at zero magnetic field (solid lines). 
The majority band in both electron and hole pictures
move up in energy when the field is applied (dashed lines), 
resulting in a negative band-contribution to the magnetization.} 
\label{el-hole-pict}
\end{figure}

Magnetization at $T=0$ is defined thermodynamically by the 
dependence of the ground-state energy
$E$ on external magnetic field $B$:
\begin{equation}
m=\left. -\frac{\partial E}{\partial B}\right|_{B=0}
\label{magdef}
\end{equation}
To avoid confusion that may result from using the hole picture to describe magnetization
of carriers in p-type (Ga,Mn)As materials, we recall first the relationship between magnetizations
evaluated using the physically direct electron-picture 
and magnetizations evaluated using the indirect but computationally more convenient hole-picture.
In  mean-field theory the magnetization is related to the
change of single-particle energy with field, summed over all occupied orbitals.
Orbitals that decrease in energy with field make a positive contribution to the 
magnetization.  For ${\bf B}\parallel +\,\hat{\bf z}$, the $d$-electron spins are
aligned along $(-z)$-direction (down-spins) and the majority spin band electrons have spin-up due
to antiferromagnetic $p-d$ exchange coupling. Then, if the majority  band 
moves up in energy with $B$ and the minority band moves down, as illustrated
in the left part of Fig.~\ref{el-hole-pict}, the band kinetic energy increases with $B$ and, according to
Eq.~(\ref{magdef}), the corresponding contribution to the magnetization is negative.
In the hole-picture, we obtain 
the same respective sense of shifts of the majority {\em hole} and minority {\em hole}  
bands, as illustrated in the right part of Fig.~\ref{el-hole-pict}, and therefore the correct (negative in our case) sign
of the magnetization.
The cartoon shows that in order to 
circumvent the potentially confusing notion of the spin of a hole 
in magnetization calculations, it is safer to start from the full Hamiltonian ${\mathcal H}(B)$ in the physically 
direct picture of electron states, where the sign of the coupling of electron spin to the field 
and the exchange energy are unambiguously defined. The electron picture $\rightarrow$ hole picture transformation (${\mathcal H}(B)\rightarrow
-{\mathcal H}(B)$)
and the clearly defined notion of majority 
and minority bands in either picture guarantees the sign consistency of the calculated magnetization.
Note that the language used here neglects spin-orbit interactions which lead to single-particle
orbitals that do not have definite spin character.  Although spin-orbit interactions are important
they can be neglected in most qualitative considerations \cite{Jungwirth:2005_a}. 

The electron-electron exchange energy has a negative sign and its magnitude increases monotonically when
moving from the paramagnetic
to the half-metallic (empty minority band) state. This together with Eq.~(\ref{magdef}) implies
that the magnetization contribution from the electron-electron
exchange energy has the same sign as the contribution from the kinetic energy. Using the same arguments as above
we see that in the electron-electron exchange energy case the sign of magnetization is also treated consistently by 
the electron picture $\rightarrow$  hole picture transformation.

The mean-field ground state wavefunction of the Mn($d^5$+hole) complex is $|S_z=-S\rangle|j_z=+j\rangle$ and
the magnetization per Mn equals $m_{MF}=(g_SS-g_jj)\mu_B$, 
where $S$ and $j$ are local $d$-electrons and hole
moments and $g_S$ and $g_j$ are the respective Land\'e g-factors. The five $d$-electrons have
zero total orbital angular momentum, {\em i.e.} $g_S=2$, and for spin $j=1/2$ hole ($g_j=2$) we get $m_{MF}=4\mu_B$. 
Hole states near the valence band edge have  $p$-character, however, so more realistically
we should consider $g_j j=4/3 * 3/2=2$ which
gives $m_{MF}=3\mu_B$. We show below that this basic picture of a 
suppressed $m_{MF}$ 
due to holes applies also to highly Mn-doped (Ga,Mn)As materials, although the magnitude of the 
mean-field hole
contribution is weaker because
of the occupation of both majority and minority hole bands and, partly, because of spin-orbit coupling effects.   

The two-spin, $S$ and $j$ model can be also used to demonstrate the presence of
quantum fluctuations around the mean-field ground state, which is a consequence of the 
antiferromagnetic sign of the ${\bf S}\cdot {\bf j}$ coupling \cite{Jungwirth:2005_a}. 
In the limit of $B\rightarrow 0$ the two-spin Hamiltonian is given by,
\begin{equation}
{\mathcal H}_{Sj}=J \;{\hat {\bf S}}\cdot {\hat {\bf j}}=\frac{J}{2} \, (\hat{S}_{tot}^2-\hat{S}^2-\hat{j}^2)\, .
\label{two-spin-Ham}
\end{equation}
For antiferromagnetic coupling ($J>0$), $S_{tot}=S-j$ and
the corresponding ground-state energy $E_{AF}=\frac{|J|}{2}[(S-j)(S-j+1)-
S(S+1)-j(j+1)]=-|J|(Sj+j)$ is lower than the mean-field energy, $-|J|Sj$.
The mean-field ground
state is not exact here and quantum fluctuation corrections to the
magnetization will be non-zero in general.
The difference between magnetizations of the exact and mean-field state is obtained from
respective expectation
values of the Zeeman Hamiltonian, $g_S\mu_BB\hat{S}_z+g_j\mu_BB\hat{j}_z$, 
and from Eq.~(\ref{magdef}) \cite{Jungwirth:2005_a}:
\begin{equation}
m-m_{MF}\equiv m_{QF}=-\mu_B
\frac{j}{S+j}(g_S-g_j)
\; .
\label{qf_correction}
\end{equation}
When $j=1/2$ and $g_S=g_j=2$ the quantum fluctuation correction to the magnetization vanishes
even though the mean-field ground state is not exact.  The correction
 remains relatively weak also in the case of $j=3/2$ and $g_j=4/3$), for which 
$m_{QF}=-0.25\mu_B$. 

\subsubsection{Magnetization of (Ga,Mn)As ferromagnets}
\label{bulk-mag-magnet-manyMn}
\begin{figure}[h]
\ifXHTML\Picture{review/figures/Fig22.png}\else\includegraphics[width=3.6in,angle=0]{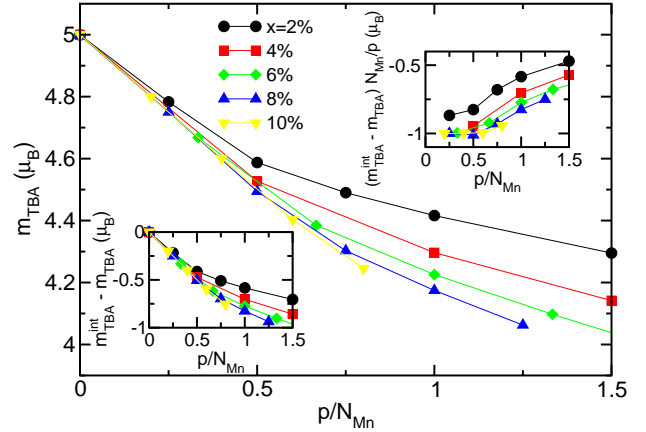}\fi
\caption{Main panel: mean-field total magnetization per Mn as a function of  hole density relative to the local
Mn moment density. Lower inset: hole contribution
to magnetization (see definition in the text) per Mn. Upper inset: hole contribution to magnetization per hole. 
Results are obtained using the TBA/CPA model.
From \cite{Jungwirth:2005_a}.
} 
\label{tba}
\end{figure}

As in the Mn($d^5$+hole) complex, the 
magnetization of coupled Mn moment systems  can be decomposed into mean-field contributions
from Mn local moments and valence band holes and a quantum fluctuations correction.
At a mean-field level, the TBA description of (Ga,Mn)As mixed crystals is particularly useful 
for explaining the complementary role of local and
itinerant moments in this p-type magnetic semiconductor and we therefore start by reviewing results of this 
approach \cite{Jungwirth:2005_a}.
In Fig.~\ref{tba} the microscopic TBA/CPA magnetic moments per Mn, $m_{TBA}$, in
(Ga,Mn)As ferromagnets are plotted as a function of $p/N_{Mn}$.
The value of $m_{TBA}$ is obtained here using the electron picture by
integrating over occupied states up to the Fermi energy. 
Spin-orbit coupling is neglected
in these TBA calculations and only the spin-polarization contribution
to magnetization is considered in $m_{TBA}$, which simplifies the qualitative discussion below.

A common way of microscopically separating
contributions from local atomic and itinerant moments is by projecting the occupied electron states
onto Mn $d$-orbitals and $sp$-orbitals, respectively. In this decomposition, the resulting local Mn moments are
smaller than 5$\mu_B$ per Mn  due to the admixture of $d$-character in empty states near the valence band edge.
The effective kinetic-exchange model corresponds, however, to a
different decomposition of contributions, in effect associating one spectral region with local Mn moments and
a different spectral region with itinerant hole moments.
The kinetic-exchange model, in which local moments have $S=5/2$, is obtained from the 
microscopic models, {\em e.g.} from the TBA/CPA,  by expressing  $m_{TBA}$  
as the difference between a contribution $m^{int}_{TBA}$ calculated by
integrating over all electronic states up to mid-gap, {\em i.e.} including the entire valence band, and
a contribution corresponding to the integral from Fermi energy to mid-gap. As long as the valence-conduction
band gap is non-zero, the former contribution is independent of valence band filling and equals 
the moment of an isolated Mn atom, 5$\mu_B$.
The latter term, plotted in the lower inset of  Fig.~\ref{tba},  represents magnetization of itinerant holes.

\begin{figure}[h]
\ifXHTML\Picture{review/figures/Fig23.png}\else\includegraphics[width=3.6in,angle=0]{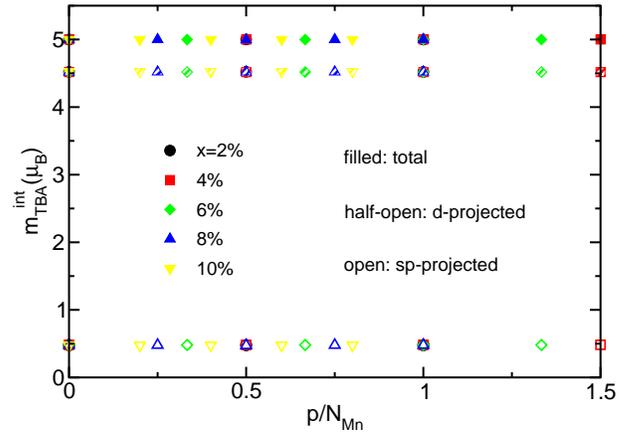}\fi
\caption{Integrated total and $d$- and $sp$-projected magnetizations 
per Mn as a function of  hole density relative to the local
Mn moment density. See text for definition of $m^{int}_{TBA}$. 
From \cite{Jungwirth:2005_a}.} 
\label{tba2}
\end{figure}

The applicability of the kinetic-exchange model relies implicitly on the 
perturbative character of the microscopic
$p-d$ hybridization. The level of  $p-d$ hybridization over a typical doping range
is illustrated in Fig.~\ref{tba2}, which shows the
orbital composition of $m^{int}_{TBA}$. The filled symbols are calculated
including spectral weights from all $spd$ orbitals while the
half-open and open symbols are obtained after projecting onto the
$d$ and $sp$ orbitals, respectively.  If there was no hybridization,
$m^{int}_{TBA}$ projected on the $d$-orbitals would
equal the total $m^{int}_{TBA}$ and the $sp$-orbital projected
$m^{int}_{TBA}$ would vanish. In the TBA/CPA calculations, the
$d$-orbital projected  $m^{int}_{TBA}$ is reduced by only 10\% as
compared to the total $m^{int}_{TBA}$ and, therefore, the $p-d$
hybridization can be regarded as a weak perturbation. The nearly constant
value of the $d$-orbital projected $m^{int}_{TBA}$ also suggests
that the kinetic-exchange coupling parameter $J_{pd}$ in the effective kinetic-exchange
Hamiltonian is nearly independent of doping over the typical range of
Mn and hole densities.

The decrease of
$m_{TBA}$ in Fig.~\ref{tba} with increasing $p/N_{Mn}$ clearly demonstrates the
antiferromagnetic $p-d$ exchange. The initial common slope for data corresponding to different 
Mn concentrations reflects the half-metallic nature
of the hole system (only majority hole band occupied) when spin-orbit interactions
are neglected. Here the hole contribution
to magnetization per volume is proportional to $p$, {\em i.e.}, magnetization per Mn is
proportional to $p/N_{Mn}$. The change in  slope of $m_{TBA}$ at larger hole densities, which now becomes
Mn-density dependent,  
reflects the population of the minority-spin hole band and, therefore, the additional dependence of hole magnetization
on exchange splitting between majority- and minority-hole bands. 
Note that the maximum absolute value of the hole contribution to magnetization per hole (see upper inset
of Fig.~\ref{tba}) observed in the half-metallic state is 1$\mu_B$ in these TBA calculations which assume
$j=1/2$ and $g_j=2$ holes.

Similar conclusions concerning the character of contributions to  the magnetization of (Ga,Mn)As
have been inferred from LDA+U and SIC-LSDA
supercell calculations \cite{Wierzbowska:2004_a,Schulthess:2005_a}. (These microscopic calculations
also neglect spin-orbit coupling.) The half-metallic LDA+U 
band structure in the case of zero charge compensation  ($p/N_{Mn}=1$)
results in a  total
magnetization per Mn  of 4$\mu_B$ \cite{Wierzbowska:2004_a}, in agreement with the 
corresponding $m_{TBA}$ values. In the SIC-LSDA
calculations \cite{Schulthess:2005_a}, the system is not completely half-metallic and, consistently, the total
moment per Mn is larger than 4$\mu_B$. The LDA+U and SIC-LSDA local moments on Mn are 4.7$\mu_B$
and 4.5$\mu_B$, respectively, in good agreement again with the $d$-projected $m^{int}_{TBA}$ values.
In both {\em ab initio}
calculations the oppositely aligned moment on the As sublattice extends over the entire supercell,
confirming the delocalized character of the holes and the antiferromagnetic sign of the $p-d$ exchange.
 
The KL kinetic-exchange model calculations 
\cite{Dietl:2001_b,Jungwirth:2005_a} have been used to refine, quantitatively, predictions 
for the total magnetization based on the above microscopic theories. In particular
the number of minority holes at a given total hole density
is underestimated in these TBA and {\em ab initio} approaches. This is caused in part by
the quantitative value of the exchange spin-splitting of the valence band which, {\em e.g.}, in the TBA/CPA calculations
is a factor of 1.5-2 larger than
value inferred from experiment. Also 
neglecting the spin-orbit interaction  results in  three majority bands that are degenerate
at the $\Gamma$-point, instead of only two bands (heavy-hole and light-hole) in the
more realistic spin-orbit-coupled band structure.   (This deficiency is common
to all calculations that neglect spin-orbit coupling.)  In addition to having more states
available in the majority band which leads to underestimating
the minority hole density, these microscopic calculations also omit the reduction
of the mean spin-density
in the majority band caused by the spin-orbit coupling. 
The total 
magnetization values  would be underestimated due to these effects. On the other hand, assuming only the
spin contribution to the hole magnetization leads to an overestimated total magnetization, as already illustrated
in Section~\ref{bulk-mag-magnet-singleMn}.

\begin{figure}[h]
\ifXHTML\Picture{review/figures/Fig24.png}\else\includegraphics[width=2.0in,angle=-90]{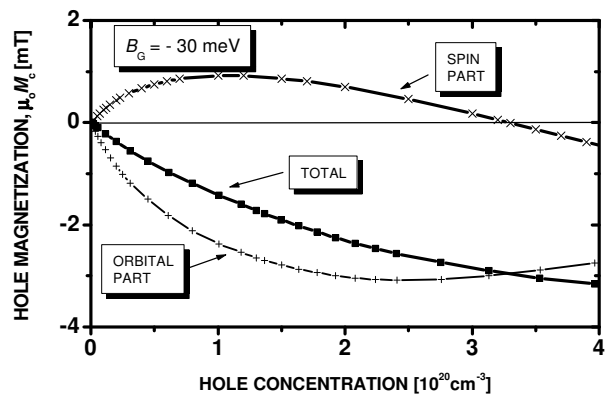}\fi
\caption{Magnetization of the hole liquid (squares) in
Ga$_{1-x}$Mn$_{x}$As computed as a function of the hole
concentration for the spin splitting parameter $B_G = -30$ meV (
corresponding to $x=0.05$ at $T=0$). The crosses show spin and orbital contribution to
the hole magnetization. 
The results were obtained using the six-band Kohn-Luttinger parameterization of the valence band
and the kinetic-exchange model.
From \protect\cite{Dietl:2001_b}.
} 
\label{mag_dietl}
\end{figure}

\begin{figure}[h]
\ifXHTML\Picture{review/figures/Fig25.png}\else\includegraphics[width=3.3in,angle=0]{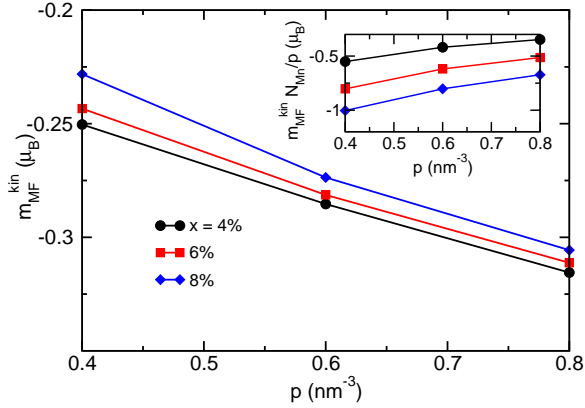}\fi
\caption{Mean-field kinetic energy contribution to the hole magnetization per Mn as a function of hole density.
The results were obtained using the six-band Kohn-Luttinger parameterization of the valence band
and the kinetic-exchange model. 
From \cite{Jungwirth:2005_a}.} 
\label{kin_mf}
\end{figure}

In the kinetic-exchange effective model the 
$T=0$ local moment contribution to the magnetization
per Mn is 5$\mu_B$. As emphasized above, this is not in contradiction with the smaller $d$-electron
contribution to the magnetic moment in microscopic calculations.
The kinetic band energy contribution to the mean-field magnetization per Mn
 is obtained by numerically integrating 
over all occupied hole eigenstates of the Hamiltonian (\ref{kl-pd-ham}) and by finding the 
coefficient linear in $B$ of this kinetic energy contribution to the total energy 
\cite{Dietl:2001_b,Jungwirth:2005_a}. Results of such calculations 
are summarized in Fig.~\ref{mag_dietl} which
shows the spin and orbital contributions to the magnetization of holes, 
and in Fig.~\ref{kin_mf} showing hole moment per Mn, $m^{kin}_{MF}$, for several
local Mn moment and hole densities. Note that the decoupling of the hole magnetization into spin and
orbital terms is partly ambiguous in the spin-orbit coupled valence bands and that only the total
moment, $m^{kin}_{MF}$, has a clear physical meaning \cite{Dietl:2001_b,Jungwirth:2005_a}.
As expected holes give a negative contribution to magnetization, {\em i.e.} they
 suppress the total magnetic moment. 
The magnitude of the  mean-field magnetization per hole,
$|m^{kin}_{MF}|N_{Mn}/p$, is smaller than 2$\mu_B$ obtained in Section~\ref{bulk-mag-magnet-singleMn}
for the isolated spin-orbit coupled hole bound to the Mn impurity.
It is due to occupation of both majority and minority heavy- and
light-hole bands at these typical (Ga,Mn)As hole densities (see inset of
Fig.~\ref{kin_mf}). 
Data shown in Fig.~\ref{mag_dietl} and in the main panel of Fig.~\ref{kin_mf}  indicate
a ~$\sim$0.2 to 0.4$\mu_B$ suppression of the mean-field moment per Mn due to the hole kinetic
energy contribution to magnetization. 

The hole exchange energy contribution
to the total mean-field magnetization was found to be negative and nearly independent of $x$ and $p$ in the
typical doping
range, and its magnitude is about a factor of 5 smaller than the magnitude of
the term originating from the hole kinetic band energy \cite{Jungwirth:2005_a}. 
Quantum fluctuation corrections lead to a ~1\% suppression of the mean-field moment per 
Mn \cite{Jungwirth:2005_a}.
(Details of these calculations, using the imaginary time path-integral formulation of quantum many body theory combined with the Holstein-Primakoff
bosonic representation of the Mn local moments can be found in \cite{Konig:2001_a} and 
\cite{Jungwirth:2005_a}.)
Combining all these considerations  the $T=0$ magnetization per Mn in the
effective kinetic-exchange model has a positive mean-field contribution equal to 5$\mu_B$ from the Mn local 
moments and a negative 
contribution from band holes and quantum fluctuations which suppress the moment per Mn by $\sim$5-10\%.

The total magnetization per nominal Mn density and per effective density of uncompensated Mn$_{\rm Ga}$ local moments 
has been measured by a superconducting quantum interference device (SQUID) 
in a series of as-grown and annealed (Ga,Mn)As samples \cite{Jungwirth:2005_a}. 
The characterization of these materials has already been
discussed in the previous section (see Fig.~\ref{tc_p_Mn_eff_exp} and the related text). 
Within experimental uncertainty, the SQUID magnetization was found to be independent
of the magnetization orientation, in agreement with theoretical expectations \cite{Jungwirth:2005_a}. 
The moment decreases with increasing nominal Mn concentration, and increases on annealing
\cite{Potashnik:2002_a,Jungwirth:2005_a}. This is consistent with the anticipated formation of interstitial Mn for doping above $\sim$2\% \cite{Jungwirth:2005_b},
given the  antiferromagnetic coupling between  Mn$_{\rm I}$
and Mn$_{\rm Ga}$ \cite{Blinowski:2003_a,Edmonds:2005_b}, and with breaking of this coupling by
low-temperature annealing \cite{Yu:2002_a,Edmonds:2004_a}.
In agreement with the above theoretical calculations,  the  total 
magnetization per effective density of uncompensated Mn$_{\rm Ga}$, $m^{eff}_{SQUID}$, falls
within the range 4-5$\mu_B$ for all samples studied. Furthermore, although there
is appreciable scatter, it can be seen that samples with lower hole densities tend to show higher
$m^{eff}_{SQUID}$, consistent with a negative contribution to magnetization from antiferromagnetically
coupled band holes.

The local and $d$-state projected contribution from Mn to the magnetic moment
in (Ga,Mn)As has been probed experimentally by measuring 
the  x-ray magnetic circular dichroism (XMCD) 
\cite{Edmonds:2004_c,Jungwirth:2005_a,Edmonds:2005_b}. 
In agreement with the SQUID measurements and theoretical
expectations, the XMCD data are independent, within experimental uncertainty, of the
direction of magnetization \cite{Jungwirth:2005_a}.
The data are listed in  Table~\ref{mag_xmcd} for two annealed samples with low
 and high Mn doping. In both cases, magnetic moments of around 4.5$\mu_B$ were obtained,
 showing a negligible dependence on the hole density. Similar results were found for samples
 with intermediate Mn doping \cite{Jungwirth:2005_a}. The experimental XMCD
 results are in  good agreement
 with the corresponding TBA values indicated by half-open symbols
in Fig.~\ref{tba2} and with the LDA+U and SIC-LSDA Mn local moments \cite{Wierzbowska:2004_a,Schulthess:2005_a}.
(Note that all these microscopic calculations
 account only  for the spin angular momentum contribution to
the local Mn 3$d$ moment since spin-orbit coupling effects were neglected.)

\begin{figure}[h]
\ifXHTML\Picture{review/figures/Fig26.png}\else\hspace*{-1cm}\includegraphics[height=3.6in,angle=-90]{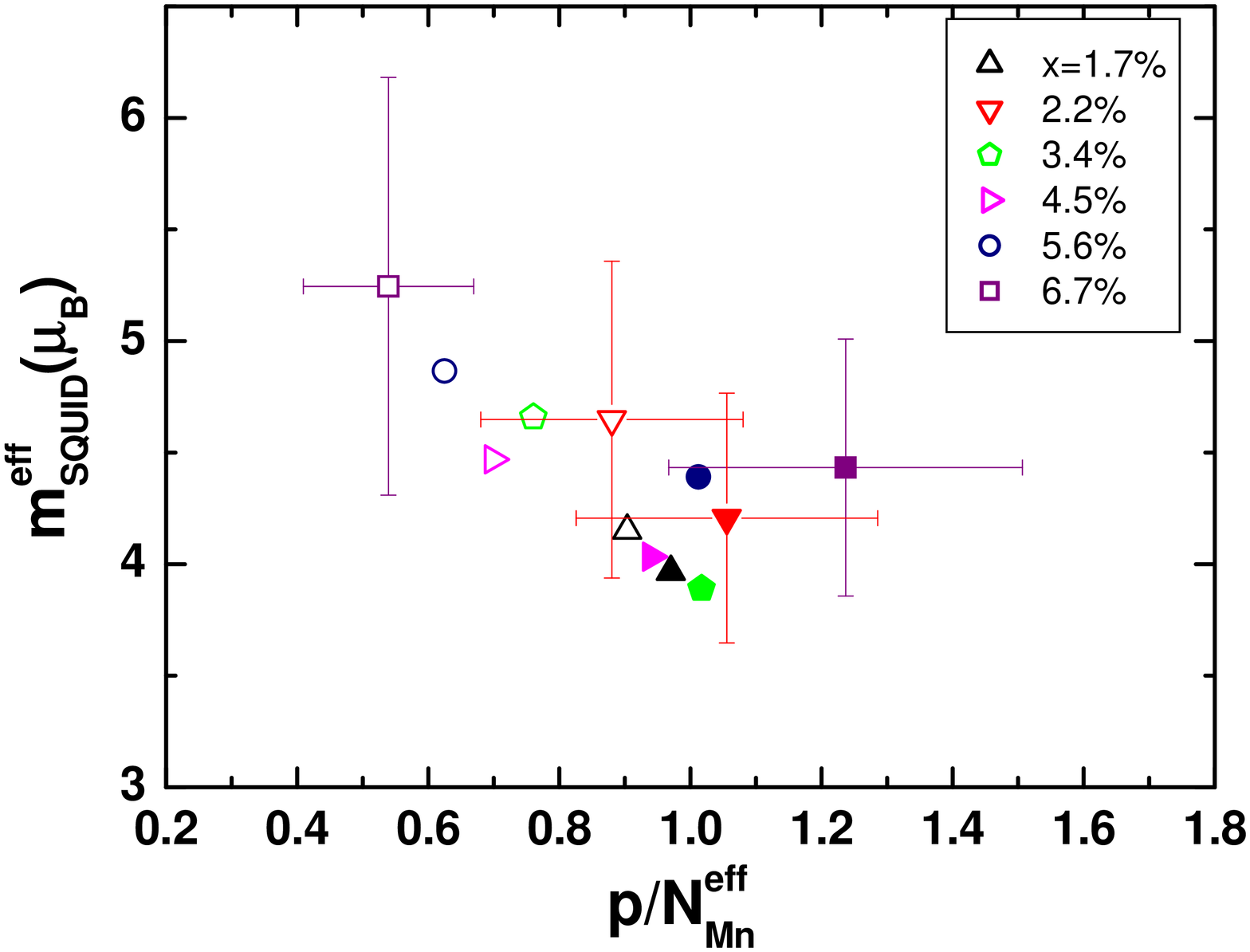}\fi
\caption{SQUID magnetization per effective density of uncompensated
Mn$_{\rm Ga}$ local moments in as-grown (open symbols) and
annealed (filled symbols) (Ga,Mn)As materials plotted as a function of hole density per
effective density of uncompensated
Mn$_{\rm Ga}$ local moments. 
From \cite{Jungwirth:2005_a}. }
\label{Meff-p}
\end{figure}

 \begin{table}[h]
%\vspace*{-0.5cm}
%\begin{center}
\begin{tabular}{cccc}
 \hline
$x$ &   $m^{spin}_{XMCD}$      & $m^{orb}_{XMCD}$        &   $m^{spin}_{XMCD}+ m^{orb}_{XMCD}$   \\

 (\%)& ($\pm0.3\mu_B$)     &  ($\pm0.03\mu_B$)   &  ($\pm0.3\mu_B$)  \\
\hline
2.2& 4.3  &  0.15        &  4.5  \\
\hline
8.4& 4.3 & 0.16 & 4.5 \\

\hline
\end{tabular}

\caption{Mn 3$d$ moments obtained from XMCD and decomposed into the spin and orbital contributions
in annealed samples with nominal Mn dopings 2.2 and 8.4\%. 
From \cite{Jungwirth:2005_a}.}
%\vspace*{-0.25cm}
 \label{table1}

%\end{center}
\label{mag_xmcd}
\end{table}

\subsection{Micromagnetic parameters}
\label{bulk-mag-micromag}
A small set of parameters is often sufficient to phenomenologically describe 
the long-wavelength properties of ferromagnets. 
The description, which usually captures all properties that 
are relevant for applications of  magnetic materials, starts from the micromagnetic  
energy functional $e[{\bf n}]$ of the spatially
dependent magnetization orientation ${\bf n}={\bf M}/M$ \cite{Aharoni:2001_a}.  The micromagnetic energy functional 
treats the long-ranged magnetic dipole interactions explicitly and uses a gradient 
expansion for other terms in the energy. The magnitude $M$ of the magnetization
is one of the micromagnetic parameters characterizing the material. Zeroth order terms in
the energy functional are the
magnetic anisotropy energy, $e_{ani}[{\bf n}]$, and the Zeeman coupling to an external
magnetic field if present, $-\mu_0{\bf H}\cdot {\bf n}M$.
The leading gradient term, $e_{ex}[{\bf n}]$, referred to as exchange energy in  micromagnetic
theory, represents the 
reduction in magnetic condensation energy when the magnetization 
orientation is not spatially constant. 
Micromagnetic 
parameters used to characterize the magnetic anisotropy energy depend on the symmetry of the system. 
For example, in a ferromagnet which possesses  uniaxial anisotropy with the
easy-axis aligned along the $z$-direction,  $e_{ani}=K_un_z^2,$ where $K_u<0$
is the uniaxial anisotropy constant. As we explain below, 
the magnetic anisotropy of  (Ga,Mn)As ferromagnets is a combination of the
cubic term, the in-plane uniaxial anisotropy, and  the
uniaxial term induced by the growth-direction lattice-matching strain which often  dominates
in (Ga,Mn)As epilayers.   The large anisotropies and relatively small magnetic moments
of these dilute magnetic systems rank (Ga,Mn)As DMSs among  hard ferromagnets (magnetic
hardness parameter $\kappa\sim|K_u/\mu_0M^2|^{1/2}>1$) with outstanding micromagnetic properties, including
frequently observed single-domain-like characteristics of field-induced magnetization reversals 
\cite{Ohno:1998_a,Abolfath:2001_b,Dietl:2001_b,Potashnik:2003_a,Wang:2005_e,Goennenwein:2004_a}.

The anisotropy of the exchange term in the micromagnetic functional is often  neglected in which
case it can be written as,  $e_{ex}=A(\nabla{\bf n})^2$, where $A$ is the 
spin stiffness constant.
Collective magnetization dynamics is described by the 
Landau-Lifshitz-Gilbert equation,
\begin{equation}
\frac{d{\bf M}}{dt}=-\frac{g\mu_B}{\hbar}{\bf M}\times\mu_0{\bf H}_{e}
+\frac{\alpha}{M}{\bf M}\times\frac{d{\bf M}}{dt}\; ,
\label{llg}
\end{equation}
where $\mu_0{\bf H}_{e}=-\partial
e/\partial{\bf M}$ and $\alpha$  is  the
Gilbert   damping   micromagnetic parameter. In this section we review 
KL kinetic-exchange model calculations
of the micromagnetic parameters of metallic (Ga,Mn)As ferromagnets 
\cite{Dietl:2001_b,Abolfath:2001_a,Dietl:2001_c,Konig:2001_a,Brey:2003_a,Sinova:2004_b}
and predictions based on the microscopic values of these parameters for anisotropy fields, the characteristic
size of domains, and  the critical current in current-induced magnetization switching
\cite{Abolfath:2001_a,Dietl:2001_c,Dietl:2001_c,Sinova:2004_b}.

\subsubsection{Magnetocrystalline anisotropy}
\label{bulk-mag-micromag-aniso}
Magnetocrystalline  anisotropy, which is the dependence of the energy of
a ferromagnet on the magnetization orientation with respect to crystallographic axes,
is a spin-orbit coupling effect often  associated with localized electrons in magnetic $d$ or $f$-shells.
Local Mn moments in (Ga,Mn)As, however, are treated in the KL kinetic exchange model 
as pure spins $S=5/2$ with angular momentum
$L = 0$ and, therefore do not contribute to the anisotropy. 
The physical origin of the anisotropy energy in 
this model is spin-orbit coupling in the valence band \cite{Dietl:2001_a,Abolfath:2001_a}. 
Even within the  mean-field approximation to the KL kinetic-exchange model,  magnetic anisotropy 
has a rich phenomenology which has explained a number of experimental observations
in the (III,Mn)V DMSs, including  easy axis reorientations   as a 
function of hole density, temperature, or 
strains in the lattice
\cite{Ohno:1996_b,Liu:2003_a,Sawicki:2004_b,Sawicki:2004_a,Masmanidis:2005_a,Liu:2004_a}.

The remarkable tunability of the magnetic properties of (Ga,Mn)As DMSs through lattice matching
strains is an important byproduct of 
LT-MBE growth of ferromagnetic films with lattices  locked to 
those of their substrates in the plane perpendicular to the growth axis. 
X-ray diffraction studies have established that
the resulting strains are not relaxed by dislocations or other defects even
for $\sim 1\mu$m thick epilayers  \cite{Shen:1999_a,Zhao:2005_a}.
We 
mentioned in Section~\ref{bulk-struct-lc} that the lattice constant of  relaxed (Ga,Mn)As is larger
than the lattice constant of GaAs, especially so if interstitial Mn$_{\rm I}$ or As$_{\rm Ga}$ antisites
are present in the DMS crystal. (Ga,Mn)As grown on the GaAs substrate is therefore under compressive strain.
Tensile strained (Ga,Mn)As DMSs have been produced by using (Ga,In)As substrates
\cite{Ohno:1996_b,Shono:2000_a,Liu:2003_a}.

\begin{figure}[h]
\ifXHTML\Picture{review/figures/Fig27.png}\else\includegraphics[width=3.3in,angle=0]{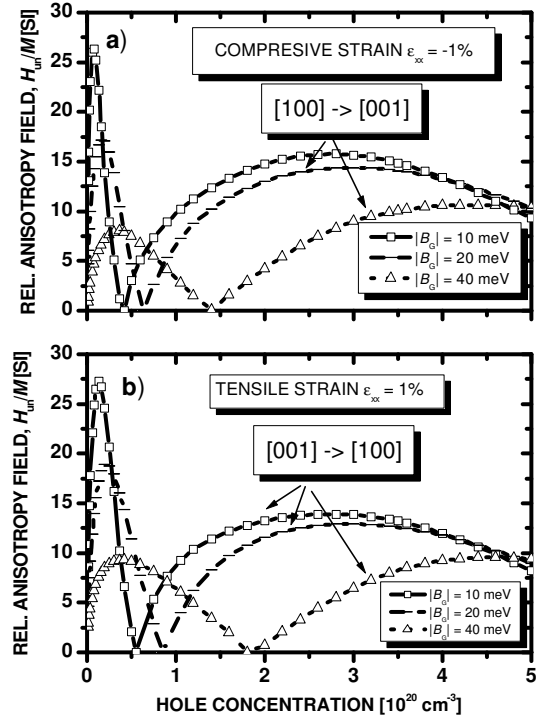}\fi
\caption{Computed minimum magnetic field $H_{un}$ (divided by
$M$) necessary to align magnetization $M$ along the hard axis for
compressive (a) and tensile (b) biaxial strain in
Ga$_{1-x}$Mn$_{x}$As film for various values of the spin-splitting
parameter $B_G$. The easy axis is along [001] direction and in the
(001) plane at low and high hole concentrations for compressive
strain, respectively (a). The opposite behavior is observed for
tensile strain (b). The symbol [100] $\rightarrow$ [001] means that
the easy axis is along [100], so that $H_{un}$ is applied along
[001] ($B_G = -30$ meV corresponds to the saturation value of $M$
for Ga$_{0.95}$Mn$_{0.05}$As).
From \cite{Dietl:2001_b}
} 
\label{aniso_dietl}
\end{figure}

[001] growth direction strain breaks
the cubic symmetry of the (Ga,Mn)As resulting in a combined cubic and uniaxial anisotropy form of the energy
functional,
\begin{eqnarray}
\label{e_ani}
 e_{ani}[{\bf \hat n}]& =& 
K_{c1} \left( n_x^2 n_y^2 + n_x^2 n_z^2 + n_y^2 n_z^2 \right) \nonumber \\
&+& K_{c2} \left( n_x n_y n_z \right) ^2+K_un_z^2\; .
\end{eqnarray}
Here $K_{c1}$ and $K_{c2}$ are the two lowest order cubic anisotropy constants \cite{Dietl:2001_b,Abolfath:2001_a}.
Strain must be included in the  {\bf k}$\cdot${\bf p} description in order to evaluate
the uniaxial anisotropy constant $K_u$. For small strains this is done
by expressing the positional vector  ${\bf r^{\prime}}$ in the
strained lattice in terms of ${\bf r}$ in the unstrained lattice as $r^{\prime}_{\alpha}= r_{\alpha}+\sum_{\beta}e_{\alpha\beta}r_{\beta}$, and expanding the KL Hamiltonian in lowest order of
the strain constants $e_{\alpha\beta}$ \cite{Jones:1973_a,Chow:1999_a}. In (Ga,Mn)As epilayers
grown along the [001] direction, the strain constant $e_{xx}=e_{yy}$ can be tuned from approximately
$-1$\% to $+1$\%. For the larger strain values, 
the uniaxial term dominates the total anisotropy energy \cite{Dietl:2001_b,Abolfath:2001_a}.
In Fig.~\ref{aniso_dietl} we show  mean-field KL kinetic-exchange model calculations of the uniaxial anisotropy field $\mu_0H_u=|2K_u/M|$ relative to the $T=0$ magnetization
$M=g\mu_BN_{Mn}S$.  $\mu_0H_u$ corresponds to the minimum external magnetic field
necessary to align magnetization $M$ along the hard axis. The figure illustrates the dependence of the easy axis 
orientation on the hole density and on the sign of $e_{xx}$. In particular, the easy-axis is in-plane for
compressive strain
($e_{xx}<0$) and out-of-plane for  tensile strain ($e_{xx}>0$), consistent with experiment \cite{Ohno:1996_b,Liu:2003_a}. The ability to manipulate the easy axis orientation from in-plane to
out-of-plane has many implications for  fundamental research on DMS materials and is very attractive
also from the point of view of potential applications in the magnetic recording technologies.

(Ga,Mn)As epilayers have also a relatively strong in-plane uniaxial anisotropy component \cite{Liu:2003_a,Sawicki:2004_b,Sawicki:2004_a,Tang:2003_a}.  
The
in-plane easy axis is consistently associated with particular crystallographic directions and
can be rotated from the [$\overline{1}$10] direction to the [110] direction by low-temperature annealing. Although
the origin of the in-plane uniaxial anisotropy has not been established,
its dependence on  hole concentration (varied by annealing) and temperature was modeled successfully within
the KL kinetic-exchange model  by assuming that it
is associated with a small shear strain $e_{xy}\approx 0.05$\% \cite{Sawicki:2004_a}.

\subsubsection{Spin stiffness}
\begin{figure}[h]
\ifXHTML\Picture{review/figures/Fig28.png}\else\includegraphics[width=3.3in,angle=0]{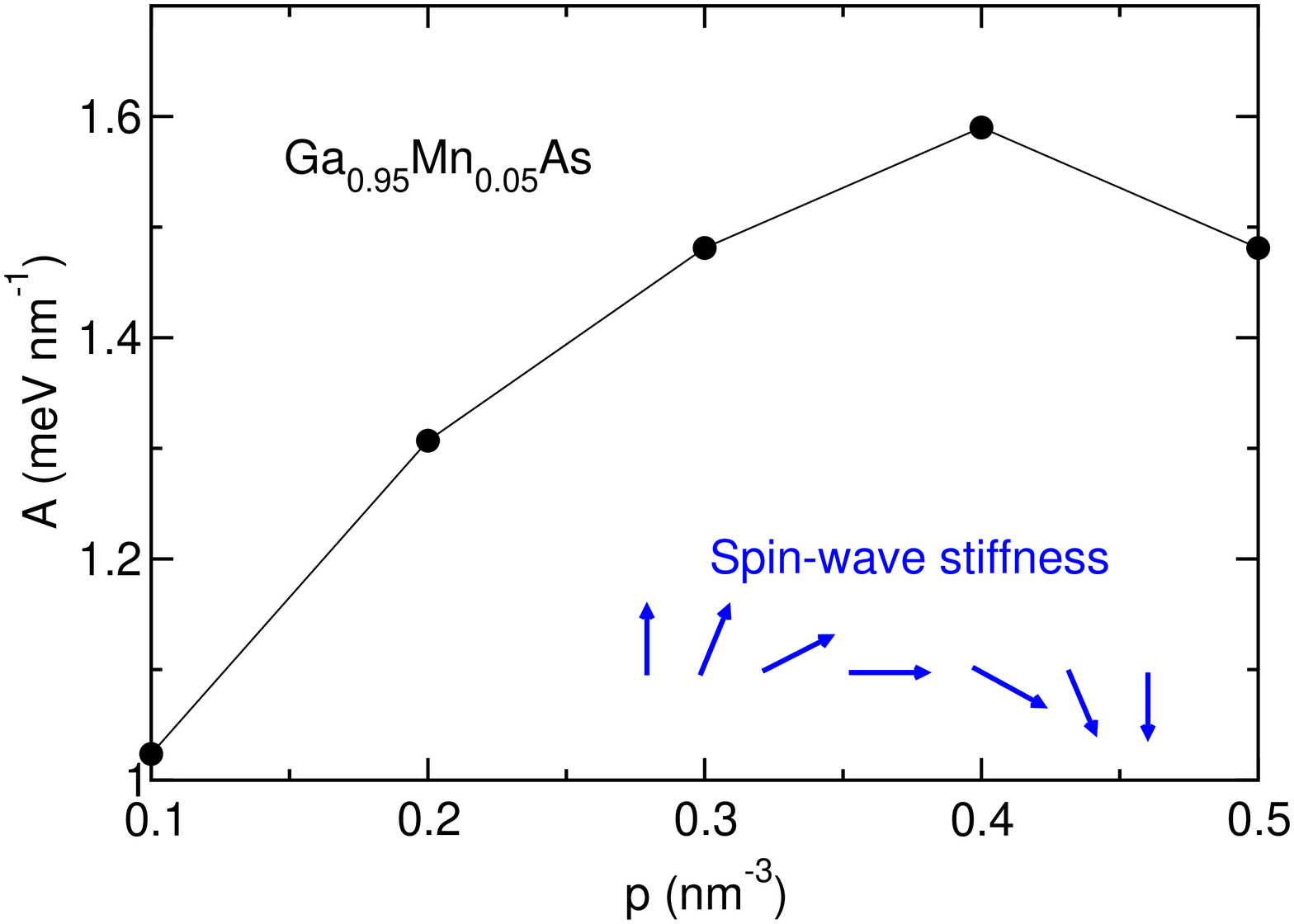}\fi
\caption{Theoretical spin stiffness parameter in Ga$_{0.95}$Mn$_{0.05}$As as a function of the hole density
calculated using the Holstein-Primakoff representation of fluctuating Mn local spins and the KL kinetic-exchange
description of the hole bands.
} 
\label{stiffness}
\end{figure}

Thermal and quantum magnetization fluctuation effects in metallic (Ga,Mn)As DMSs have been described by 
Holstein-Primakoff boson representation of the Mn local spin operators.
Assuming that fluctuations around the mean-field orientation 
are small, the relationship between spin-raising and lowering operators 
and boson creation and annihilation operators is,
$S^{+}=b\sqrt{2N_{Mn}S}$ and $ S^{-}=b^{\dagger}\sqrt{2N_{Mn}S}$
($S_x=(S^++S^-)/2$, $S_y=(S^+-S^-)/2i)$ \cite{Auerbach:1994_a,Konig:2000_a,Konig:2001_a}. 
After integrating out the 
itinerant hole degrees of freedom in the 
coherent-state path-integral formalism of  many-body problem,
the partition function, 
\begin{equation}
Z=\int {\cal D}[\bar z z] \exp (-S_{eff}[\bar z z]),
\label{part_func} 
\end{equation}
depends only on the bosonic
degrees of freedom (represented by the complex numbers $z$ and $\bar z$  in the effective action $S_{eff}$).
The independent spin-wave theory
is obtained by expanding  $S_{eff}[\bar z z]$ up to quadratic order in $z$ and $\bar z$, {\em i.e.},
spin excitations are treated as noninteracting bosons.
The spin stiffness parameter $A$ is then calculated by fitting the microscopic spin-wave dispersion at
long wavelength to the form,
\begin{equation}
   \Omega_{k} = {2K_u\over N_{\rm Mn}S} 
        +  {2A\over N_{\rm Mn}S} \, k^2 + {\cal O}(k^4) \, 
\label{gap+stiffness}
\end{equation}
to match the conventions used for exchange and anisotropy constants in  micromagnetic theory.  
Typical values of $A$ in (Ga,Mn)As derived from the above many-body formalism and the KL kinetic-exchange
description of the hole bands are shown in Fig.~\ref{stiffness}. The values are consistent with the experimental
spin stiffness parameter in a Ga$_{0.949}$Mn$_{0.051}$As measured by ferromagnetic resonance \cite{Goennenwein:2003_a}. 

\subsubsection{Gilbert damping of magnetization precession}
The damping of
small cone-angle magnetization precession in a ferromagnet is parameterized by the Gilbert coefficient. 
For small
fluctuations of the Mn magnetization orientation in (Ga,Mn)As around the easy axis,
Eq.~(\ref{llg})  can be  used  to derive  an expression for the linear response 
of a magnetic system to weak transverse fields in terms of the phenomenological 
constants of micromagnetic theory.   For
zero external static magnetic field and zero wavevector (uniform rotation) the
corresponding inverse susceptibility reads:
\begin{equation}
\chi^{-1}=\frac{\hbar}{(g\mu_B)^2N_{Mn}S}
\left(\begin{array}{cc}
\tilde{K_u}-i\alpha\omega & -i\omega\\
i\omega & \tilde{K_u}-i\alpha\omega\\
\end{array}\right)\; ,
\label{classicalsusc}
\end{equation} 
where  $\tilde{K_u}=K_u/(\hbar N_{Mn}S)$ and $\omega$ is the  frequency of the external rf field
perturbation.          
     
\begin{figure}[h]
\ifXHTML\Picture{review/figures/Fig29.png}\else\includegraphics[width=3.3in,angle=0]{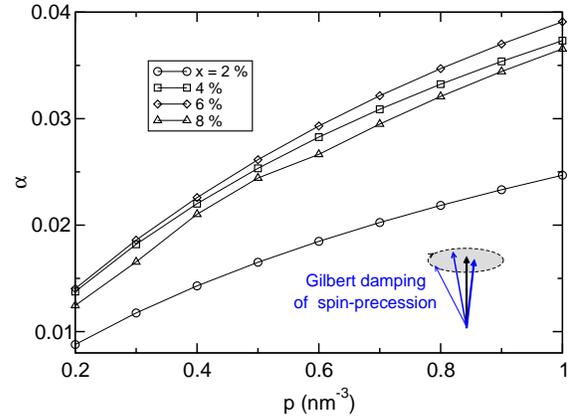}\fi
\caption{Theoretical Gilbert damping coefficient in (Ga,Mn)As.
From \cite{Sinova:2004_b}.
} 
\label{gilbert}
\end{figure}

\begin{figure}[h]
\ifXHTML\Picture{review/figures/Fig30.png}\else\includegraphics[width=3.3in,angle=0]{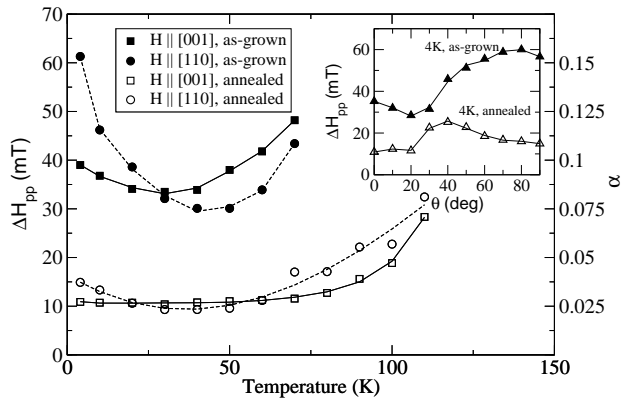}\fi
\caption{Experimental peak-to-peak ferromagnetic resonance linewidth and corresponding
Gilbert coefficient in as-grown
(filled symbols) and annealed (open symbols) Ga$_{0.92}$Mn$_{0.08}$As
samples measured as a function of temperature for [001] and [110]
dc magnetic-field orientations (main plot) and as a function of the field angle at 4K (inset).
From \cite{Liu:2003_a,Sinova:2004_b}.
} 
\label{gilbert_exp}
\end{figure}

Microscopically, Gilbert
damping in (Ga,Mn)As DMSs was attributed to the $p-d$ exchange-coupling between local
Mn moments and  itinerant holes \cite{Sinova:2004_b}. The elementary
process  for this  damping mechanism  is one  in which  a local-moment
magnon  is  annihilated  by  exchange interaction  with  a  band
hole  that suffers  a spin-flip.   This process  cannot  by itself
change  the  total  magnetic  moment since  the  exchange  Hamiltonian
commutes with the total spin ${\bf S}+{\bf s}$.  Net relaxation of the
magnetization  requires  another  independent  process  in  which  the
itinerant hole spin relaxes through spin-orbit interactions.
A fully microscopic theory of
the  kinetic-exchange  contribution  to  the  Gilbert  coefficient  was
derived by  comparing Eq.~(\ref{classicalsusc})  with
microscopic   linear  response   theory and by identifying   the  Gilbert
coefficient with the dissipative part of the quantum-mechanical susceptibility \cite{Sinova:2004_b,Tserkovnyak:2004_a},
\begin{equation}
\chi^R_{i,j}({\bf r},t|{\bf r}^{\prime},t^{\prime})=(g\mu_B)^2\frac{i}{\hbar}
\langle[S_i ({\bf r},t),S_j ({\bf r}^{\prime},t^{\prime})]
\rangle\theta(t-t^{\prime}). 
\label{kubo_susc}
\end{equation}
Here $i=x,y$ and $S_i({\bf r},t)=M_i({\bf r},t)/(g\mu_B)$ are the Mn 
transverse spin-density operators. As in the microscopic theory of the
spin-stiffness, the correlation function (\ref{kubo_susc}) for the uniform (${\bf k}=0$)
precession mode was evaluated using the 
the  long-wavelength non-interacting spin-wave form of the partition function (\ref{part_func}) 
\cite{Sinova:2004_b}. 
Gilbert coefficients in (Ga,Mn)As DMSs obtained
within this formalism agree quantitatively with experimental values of $\alpha$ in homogeneous 
(annealed) systems
measured from the width of the ferromagnetic resonance curves, as shown in Figs.~\ref{gilbert} and
\ref{gilbert_exp}.

\subsubsection{Domains and spin-transfer magnetization switching}
Theoretical values of magnetic anisotropy and spin stiffness have been combined to 
estimate the typical domain size in tensile strained
(Ga,Mn)As epilayers with out-of-plane easy axis \cite{Dietl:2001_c}. The calculated low-temperature 
width of a single domain stripe of 1.1$\mu$m
compares favorably with the experimental value of  1.5$\mu$m seen in the micro-Hall probe experiments
\cite{Shono:2000_a}. Near $T_c$ the discrepancy between theoretical and experimental domain sizes becomes large, 
however, and has been attributed to critical fluctuations effects not included in the mean-field theory.

The calculated Gilbert damping coefficient and magnetic anisotropy constants were used to predict critical
currents for spin-transfer magnetization switching \cite{Slonczewski:1996_a,Berger:1996_a} 
in (Ga,Mn)As based tunneling structures
\cite{Sinova:2004_b}. Spin-polarized perpendicular-to-plane currents 
in magnetic multilayers with non-parallel spin configurations can transfer spin between 
magnetic layers and exert current-dependent torques \cite{Slonczewski:1996_a}.
The critical current for magnetization switching is obtained by adding the torque term to
the Landau-Lifshitz-Gilbert Eq.~(\ref{llg}) \cite{Slonczewski:1996_a,Sinova:2004_b}. 
Critical currents
$\sim 10^5$ A~cm$^{-2}$ obtained in these calculations and confirmed in experiment \cite{Chiba:2004_b} are 
two orders of magnitude smaller than those observed typically in metals. The small moment densities explain a 
large part of the orders of magnitude reduction in the critical current. This finding suggests that DMS materials 
have the potential to be particularly useful for exploiting the current-induced magnetization reversal
effect in magnetic tunnel junctions. 
%%%%%%%%%%%%%%%%%%%%%%%%%%%%%%%%%%%%%%%%%%%%%%%%%%%%%%%%%%%

\section{Magneto-transport}
\label{bulk-dc_magtransp}

Studies of the temperature dependent resistivity, anisotropic
magnetoresistance,  and anomalous   and  ordinary  Hall effects
have been used  to  characterize DMS materials 
and  to test  different  theoretical models describing these ferromagnets.   
In this  section we review dc magneto-transport properties of (Ga,Mn)As focusing mainly
on the metallic regime.

\subsection{Low temperature conductivity}
\label{bulk-magtransp-dc_cond}
(Ga,Mn)As materials  can exhibit insulating or metallic  behavior  depending  on the
doping  and  post-growth  annealing  procedures.  (Strictly speaking, a material is 
defined as being metallic if its resistivity is finite in the limit $T \rightarrow 0$, 
although for practical reasons this adjective is often used to describe a material 
whose resistivity decreases with temperature over most or all of the range 
of temperatures studied in a particular series of experiments.) 
In  optimally annealed samples with low density of unintentional defects, metallic behavior  is 
observed for Mn doping larger than approximately 1.5\% \cite{Campion:2003_b,Potashnik:2002_a}. 
In Section~\ref{micro_picture} we already discussed this observation as a 
consequence of the Mott metal-insulator transition due to doping with substitutional Mn$_{\rm Ga}$ 
acceptors. We also
introduced in Sections~\ref{micro_picture}
and \ref{theory-lattice} some of the theoretical work that has qualitatively addressed magnetic and transport
properties of DMS systems near the metal-insulator transition. To our knowledge, a systematic experimental
analysis has not yet been performed which would allow a reliable
assessment of the theory predictions in this complex and intriguing regime.
On the other hand, 
the dc transport in metallic (Ga,Mn)As DMSs has a rich phenomenology which has been explored in a number
of experimental works and  microscopic understanding of many of these effects is now well established.  

\begin{figure}
\ifXHTML\Picture{review/figures/Fig31.png}\else\includegraphics[width=2.3in,angle=-90]{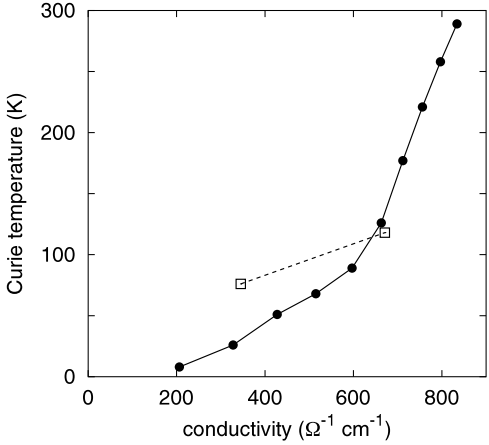}\fi
\caption{Mean-field LSDA/CPA Curie temperatures versus $T=0$ dc conductivities calculated for
(Ga$_{0.95−y}$Mn$_{0.05}$As$_{y}$ )As alloys with varying As antisite content $y$ (full dots) and the experimental
values obtained in as-grown and annealed (Ga,Mn)As thin films with the corresponding
level of compensation \cite{Campion:2003_b} 
(open squares). Note that these are illustrative calculations since in the experiment the change in the unintentional
impurity concentration upon annealing is mostly due to the out-diffusion of interstitial Mn. 
From \cite{Turek:2004_a}.
}
\label{dc_cond_turek}
\end{figure}

\begin{figure}
\ifXHTML\Picture{review/figures/Fig32.png}\else\includegraphics[width=1.8in,angle=-90]{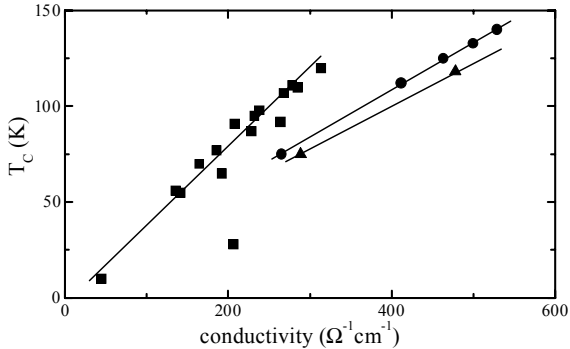}\fi
\caption{Curie temperature versus room temperature
conductivity for (Ga,Mn)As films with x=0.08 (squares), x=0.06
(circles), x=0.05 (triangles). Straight lines are to guide the eye. The data points correspond
to samples annealed at different temperatures and times. Note that similar trends are obtained for
Curie temperatures plotted versus low temperature conductivities, whose values are systematically larger by
$\sim 20-40$\% compared to the room temperature conductivities.
From \cite{Edmonds:2002_b}.
}
\label{dc_cond_tc_nott}
\end{figure}

LSDA/CPA band structure calculations combined with Kubo linear response
theory  were used to study
correlations between the low temperature conductivity and the density of various defects in the lattice, the hole density,
and $T_c$ in metallic (Ga,Mn)As \cite{Turek:2004_a}.  The theory tends to overestimate
the conductivity in low-compensation materials but the overall range of values $\sim 100-1000$~$\Omega^{-1}$cm$^{-1}$ for typical material parameters is 
consistent with measured data. As illustrated in Fig.~\ref{dc_cond_turek} and \ref{dc_cond_tc_nott}, 
the theory models capture at a qualitative level the 
correlation between large conductivities and high Curie temperatures seen in 
experiment \cite{Campion:2003_b,Potashnik:2002_a,Edmonds:2002_b}. 
Comparable absolute values of the $T=0$ conductance and similar trends in the dependence of the
conductance on the density of impurities were obtained using the KL kinetic-exchange model and
the semiclassical Boltzmann description of  the dc 
transport \cite{Jungwirth:2002_c,Lopez-Sancho:2003_b,Hwang:2005_a}. 
We note, however, that there are differences in the detailed microscopic mechanisms
limiting the conductivity in the two theoretical approaches. Scattering in the {\em ab initio} theory is
dominated by the $p-d$ exchange potential on randomly distributed Mn atoms,  and by the local changes
of the crystal potential on the impurity sites.  Long-range Coulomb potentials produced by Mn$_{\rm Ga}$
acceptors
and other charged defects are omitted in the CPA approach.

In the kinetic-exchange effective Hamiltonian model the elastic scattering effects were included using the
first order Born approximation 
\cite{Jungwirth:2002_c}. The corresponding  transport
weighted  scattering  rate from the Mn$_{\rm Ga}$ impurity has a contribution from both the 
$p-d$ exchange potential and the long-range Coulomb potential,
\begin{eqnarray}
\Gamma_{i,\vec k}&=&\frac{2\pi}{\hbar} N_{Mn}\sum_{i^{\prime}}
\int\frac{d\vec k^{\prime}}{(2\pi)^3} 
|M_{i,i^{\prime}}^{\vec k,\vec k ^{\prime}}|^2
\nonumber \\ &\times &\delta(E_{i,\vec k}
-E_{i^{\prime}\vec k ^{\prime}})
(1-\cos \theta_{\vec k, \vec k ^{\prime}})\; ,
\label{gamma}
\end{eqnarray}
with the scattering matrix elements,  
\begin{eqnarray}
M_{i,i^{\prime}}^{\vec k,\vec k ^{\prime}}&=&
J_{pd}S
\langle z_{i \vec k}|{\bf n}\cdot {\bf s}|
z_{i^{\prime}\vec k ^{\prime}}\rangle\nonumber \\
&-&
\frac{e^2}{\epsilon_{host}\epsilon_0(|\vec k -\vec k ^{\prime}|^2
+q_{TF}^2)}\langle z_{i \vec k}|
z_{i^{\prime}\vec k ^{\prime}}\rangle .
\label{mnelement}
\end{eqnarray}
Here $\epsilon_{host}$ is  the host semiconductor dielectric constant,
$|z_{i   \vec   k}\rangle$   is  the   multi-component
eigenspinor  of  the    KL Hamiltonian (\ref{kl-pd-ham})
and $E_{i,\vec k}$ is the corresponding eigenenergy, 
and  the  Thomas-Fermi
screening   wavevector   
$q_{TF}=\sqrt{e^2 {\rm DOS}(E_F)/(\epsilon_{host}\epsilon_0)}$,
where ${\rm DOS}(E_F)$ is the density of states at the  Fermi
energy \cite{Jungwirth:2002_c}. (Analogous expressions apply to scattering rates due to other defects.)

The relative strengths of scattering  off the $p-d$  exchange and Coulomb potentials can be estimated by  
assuming a  simple parabolic-band model of the valence band characterized by the heavy-hole
effective mass $m^{\ast}=0.5m_e$.
The $p-d$  kinetic-exchange  contribution  in this approximation
is $\Gamma_{pd}=N_{Mn} J_{pd}^2 S^2 m^{\ast}\sqrt{2m^{\ast}E_F}/(4   \pi   \hbar^4)$ 
and the scattering rate due to the screened 
Coulomb potential, $\Gamma_{C}$,      is   given        by         the  standard      Brooks-Herring
formula  \cite{Brooks:1955_a}. In (Ga,Mn)As  with 
 $p=0.4$~nm$^{-3}$ and $x=5$\%,  these  estimates   give  $\hbar\Gamma_{pd}\sim  20$~meV  and
$\hbar\Gamma_{C}\sim 150$~meV.  For lower compensation ($p\approx1$~nm$^{-3}$), 
the screening of the Coulomb potential is
more efficient resulting in values  of $\Gamma_{C}$ below 100~meV,  but still several times larger
than $\Gamma_{pd}$. (Note  that these  elastic scattering rates
are  smaller, although by less than a factor of 10,  
than other characteristic energy scales such as the Fermi energy and the
spin-orbit coupling strength in the GaAs valence band which 
partly establishes the consistency of this theoretical approach.)

The dominance of the Coulomb potential in  Born approximation
scattering rates for typical chemical compositions is confirmed by 
calculations based on the 
six-band KL Hamiltonian \cite{Jungwirth:2002_c}. 
The good agreement between $T=0$ conductivity values obtained using
the {\em ab initio}  and the kinetic-exchange model theories should therefore be taken with 
caution  as it may originate, to some extent, from the
stronger local $p-d$ exchange in the LSDA/CPA 
theory which partly compensates the neglect of long-range Coulomb potentials in this {\em ab initio}
approach.

\subsection{Anisotropic magnetoresistance}
\label{bulk-magtransp-amr}
Boltzmann transport theory combined with the KL kinetic-exchange model of the
(Ga,Mn)As band structure is a practical approach for studying magneto-transport effects
that originate from the spin-orbit coupling. In Section~\ref{bulk-mag-micromag} we reviewed the
predictions of the model for 
the role of spin-orbit interaction in magnetic properties of (Ga,Mn)As. Here we focus on
the anisotropic magnetoresistance (AMR) effect which is the  transport analogue of
the magnetocrystalline anisotropy
\cite{Wang:2002_a,Baxter:2002_a,Jungwirth:2002_c,Tang:2003_a,Jungwirth:2003_b,Matsukura:2004_a,Wang:2005_c,Goennenwein:2004_a}.

The AMR effect can be regarded as the first spintronic functionality implemented in microelectronic devices. 
AMR magnetic sensors  replaced  simple horse-shoe magnets in hard-drive read heads in the
early 1990's.  With the introduction of giant magnetoresistance based devices in 1997, 
magnetoresistive sensing launched a new era era in the magnetic memory industry. 
In ferromagnetic metals, AMR has been known for well over a century.  However, the role of the various mechanisms
held responsible for the effect has not been fully sorted out despite  renewed 
interest motivated by practical applications
\cite{Jaoul:1977_a,Malozemoff:1985_a}. 
The difficulty in metal ferromagnets partly stems from the relatively weak
spin-orbit coupling compared to other relevant energy scales and complex band structure. In (Ga,Mn)As ferromagnets
with strongly spin-orbit coupled holes occupying the top of the valence band, modeling of the AMR effect
can be accomplished on a semiquantitative level without using free parameters, as we review below.
  
\begin{figure}[h]
\ifXHTML\Picture{review/figures/Fig33.png}\else\includegraphics[width=1.0in,angle=-90]{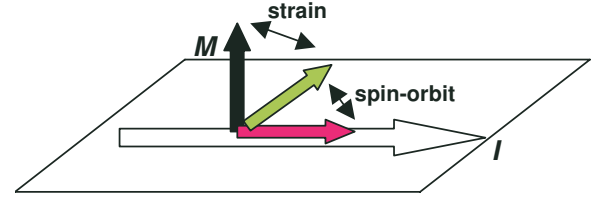}\fi
\caption{Schematic illustration of AMR configurations: electrical
current ${\bf I}$ flows along $\hat{x}$-direction and magnetization is aligned by an 
external magnetic field in either of the two in plane directions,
${\bf M} \parallel {\bf I}$ (${\bf M} \parallel \hat{x}$) and 
in-plane ${\bf M} \perp {\bf I}$ (${\bf M} \parallel \hat{y}$), 
or in the  out-of-plane direction perpendicular
to the current,  ${\bf M} \perp {\rm plane}$ (${\bf M} \parallel \hat{z}$). 
The non-zero coefficient $AMR_{ip}$, defined as the relative difference in resistivities in the
in-plane ${\bf M} \perp {\bf I}$ and  ${\bf M} \parallel {\bf I}$ configurations, originates from
spin-orbit coupling. In strained (Ga,Mn)As materials, $AMR_{ip}$ is in general not
equal to 
$AMR_{op}$,  defined as the relative difference in resistivities in the
${\bf M} \perp {\rm plane}$ and  ${\bf M} \parallel {\bf I}$ configurations.
From \cite{Matsukura:2004_a}.
}
\label{amr_schem}
\end{figure}

(Ga,Mn)As epilayers with broken cubic symmetry due to 
growth-direction lattice-matching strains are usually characterized by two AMR coefficients, 
\begin{eqnarray}
AMR_{op}&=&\frac{\rho_{xx}({\bf   M}\parallel  \hat{x})-  \rho_{xx}({\bf
M}\parallel  \hat{z})}{\rho_{xx}({\bf  M}\parallel  \hat{z})} \nonumber \\   
AMR_{ip}&=&\frac{\rho_{xx}({\bf   M}\parallel  \hat{x})-  \rho_{xx}({\bf
M}\parallel \hat{y})}{\rho_{xx}({\bf M}\parallel \hat{y})}, 
\end{eqnarray}
where
$\hat{z}$  is the  growth direction and the electrical current $I\parallel\hat{x}$.
The three different experimental configurations  used to determine $AMR_{op}$ 
and $AMR_{ip}$
are illustrated in Fig.~\ref{amr_schem}. The coefficient $AMR_{ip}$ is given by 
combined effects of spin-orbit coupling and the current induced broken symmetry between the two
in-plane cubic axes. The difference between $AMR_{ip}$ and $AMR_{op}$ is a consequence
of the lowered symmetry in strained samples.
Theoretical predictions for $AMR_{ip}$ and $AMR_{op}$ 
were compared with measurements in a series of (Ga,Mn)As
epilayers grown under compressive strain \cite{Jungwirth:2003_b}. The calculated sign of the AMR coefficients
and the magnitude which varies form $\sim 1$\% in  highly Mn doped materials to $\sim 10$\%
in samples with low Mn concentration  are consistent with experiment. 

\begin{figure}[h]
\ifXHTML\Picture{review/figures/Fig34.png}\else\includegraphics[width=3.3in]{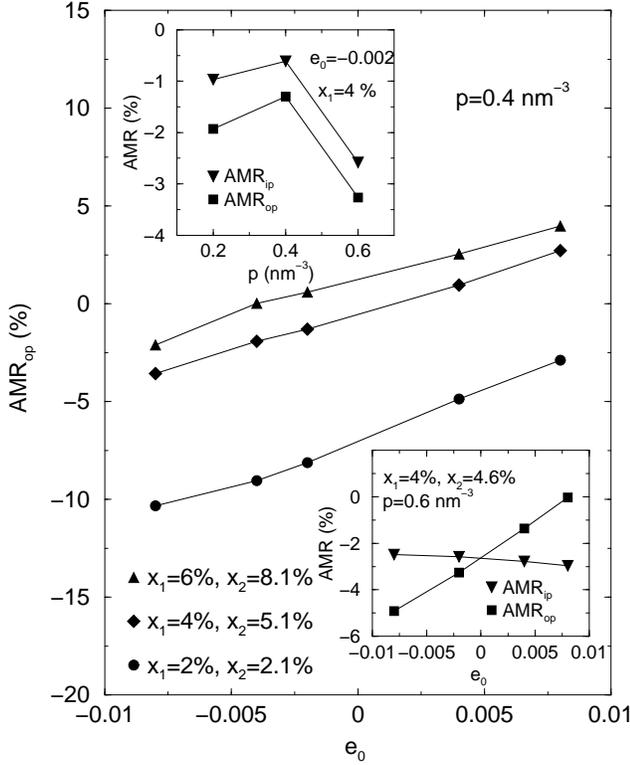}\fi
\caption{Theoretical AMR effects in (Ga,Mn)As obtained from the
KL kinetic-exchange model and Boltzmann transport theory. The curves correspond
to total Mn doping $x_1$ and compensation due to As-antisites
or total  Mn doping $x_2$ and compensation due to Mn-interstitials. Main panel:
out-of-plane AMR coefficient
as a function of strain for several Mn dopings. Lower inset: 
out-of-plane and in-plane AMR coefficients
as a function of strain. Upper inset: out-of-plane AMR coefficient
as a function of the hole density.
From \cite{Jungwirth:2002_c}.
}
\label{amr}
\end{figure}

\begin{figure}[h]
\ifXHTML\Picture{review/figures/Fig35.png}\else\includegraphics[width=3.3in]{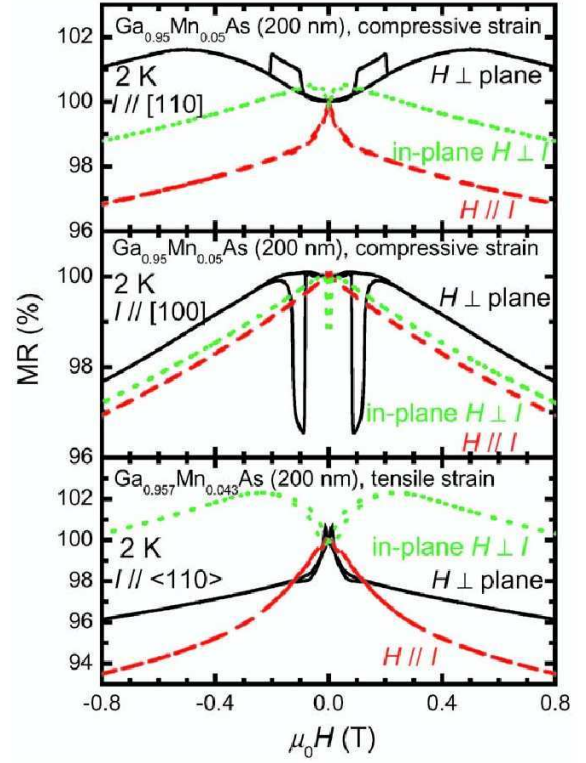}\fi
\caption{Upper and middle panel: experimental 
field-induced changes in the resistance of Ga$_{0.95}$Mn$_{0.05}$As
grown on 
GaAs substrate under compressive strain for current along the [110] crystal direction (upper panel)
and [100] crystal direction (middle panel) for three different orientations of the magnetization 
as indicated  in Fig.~\ref{amr_schem}. Lower panel:
experimental magnetoresistance curves in Ga$_{0.957}$Mn$_{0.043}$As
grown on
(In,Ga)As substrate under tensile strain (current along the [110] crystal direction).
From \cite{Matsukura:2004_a}.
}
\label{amr_dietl}
\end{figure}

Reminiscent of the 
magnetocrystalline anisotropy behavior, the theory predicts rotations as a function of strain of the magnetization direction corresponding
to the high (or low) resistance state.  Calculations illustrating this effect are shown in Fig.~\ref{amr} and the
experimental demonstration in (Ga,Mn)As epilayers with compressive and tensile strains is presented in
Fig.~\ref{amr_dietl}. In the top panel of Fig.~\ref{amr_dietl}, corresponding to  (Ga,Mn)As material with
compressive strain, 
${\bf M} \perp {\rm plane}$ (${\bf M} \parallel \hat{z}$) is the high resistance configuration,
the intermediate resistance state is 
realized for in-plane ${\bf M} \perp {\bf I}$ (${\bf M} \parallel \hat{y}$), and the low resistance
state is measured when ${\bf M} \parallel {\bf I}$ (${\bf M} \parallel \hat{x}$). 
In the sample with tensile strain,
the  in-plane ${\bf M} \perp {\bf I}$ and ${\bf M} \perp {\rm plane}$  curves switch places,
as seen in the bottom panel of Fig.~\ref{amr_dietl}. Consistent with these experimental observations
the theoretical   $AMR_{ip}$ and $AMR_{op}$  coefficients are negative and $|AMR_{ip}|<|AMR_{op}|$
for compressive strain and $|AMR_{ip}|>|AMR_{op}|$ for tensile strain.

\subsection{Anomalous and ordinary Hall effects}
\label{bulk-magtransp-he}
The anomalous Hall effect (AHE) is another  transport phenomenon originating from
spin-orbit coupling which has been used to study and characterize ferromagnetic films for more
than one hundred years \cite{Chien:1980_a}. 
The difficulties that have  accompanied attempts at accurate microscopic description of the effect in
metal ferromagnets are 
reminiscent of those in the AMR. 
The success of AHE modeling in (Ga,Mn)As materials
\cite{Jungwirth:2002_a,Jungwirth:2003_b,Edmonds:2003_a,Dietl:2003_c}, 
reviewed in this section, 
has had implications also beyond the field of DMSs. It 
helped to motivate a reexamination of the AHE in transition metals and
and in
a series of more complex ferromagnetic compounds  which  has led to a significant progress in resolving
the microscopic mechanism responsible for the effect in these 
materials 
\cite{Yao:2004_a,Fang:2003_a,Lee:2004_a,Dugaev:2005_a,Kotzler:2005_a,Sinitsyn:2005_a,Haldane:2004_a}.

 The  Hall resistance
$R_{xy}\equiv \rho_{xy}/d$ 
of a  magnetic thin film of  thickness $d$ is empirically
observed to contain two  distinct contributions, $R_{xy}=R_O B+R_A M$ \cite{Chien:1980_a}. 
The first contribution
arises from
the  ordinary Hall  effect (OHE)  which  is  proportional to  the
applied magnetic field $B$, and the second term is the AHE 
which may remain finite at $B=0$ and depends instead on the magnetization.
 $R_O$ and  $R_A$ are the OHE  and AHE coefficients
respectively. To put the AHE studies in DMSs in a broader perspective, the following paragraphs 
offer a brief  excursion through the
history of AHE theory and an overview of microscopic mechanisms discussed in this context in the 
literature (a more detailed survey is given, {\em e.g.}, in \cite{Sinova:2004_c}).

The  first detailed  theoretical  analysis of  the AHE 
was  given by  Karplus  and Luttinger  \cite{Karplus:1954_a},
where they  considered the problem  from a perturbative point  of view
(with  respect   to  an  applied   electric  field)  and   obtained  a
contribution  to  the anomalous Hall  conductivity  in  systems with  spin-orbit
coupled Bloch  states given  by the expression:  
\begin{equation}
\sigma_{AH}=-\frac{2e^2}{\hbar V}\sum_{n,\vec{k}} f_{n,\vec{k}}
{\rm Im}\left\langle \frac{\partial u_{n,\vec{k}}}{\partial k_x}\right.
\left|\frac{\partial u_{n,\vec{k}}}{\partial k_y}\right\rangle,
\label{sigAHEberry}
\end{equation}
where  $f_{n,\vec{k}}$  is  the  Fermi  occupation number  of  the  Bloch  state
$|u_{n,\vec{k}}\rangle$.  This contribution  
is purely a property of the perfect crystal and has become known in recent years as 
the intrinsic AHE.  It leads to an AHE coefficient $R_A$  proportional to 
$\rho_{xx}^2$ and therefore can dominate in metallic ferromagnets that have a relatively 
large resistivity.  The intrinsic AHE is 
related to Bloch state Berry phases in momentum space and  depends non-perturbatively 
on spin-orbit interaction strength when degeneracies in momentum space are lifted by 
spin-orbit coupling
\cite{Sundaram:1999_a,Jungwirth:2002_a,Onoda:2002_a,Burkov:2003_a,Haldane:2004_a,Dugaev:2005_a}. 
This point is particularly relevant for ferromagnetic semiconductors 
because all carriers that contribute to transport are located near particular points in the Brillouin zone, 
often high symmetry points at which degeneracies occur. 
 
Shortly after the seminal work of Karplus  and Luttinger, Smit proposed 
a different  interpretation of the AHE 
based on a picture
of asymmetric spin-dependent skew scattering off impurity potential  involving spin-orbit 
coupling \cite{Smit:1955_a,Nozieres:1973_a,Leroux-Hugon:1972_a}.
Analytically, the skew scattering appears in the second order
Born approximation applied to  the collision term of  the Boltzmann
transport equation. This mechanism  gives a contribution to  $R_A\propto
\rho_{xx}$, {\em i.e.}, proportional to the density of scatterers  and dependent on the type and range
of the impurity  potential. 

The AHE conductivity has a number of contributions in addition to the skew scattering and intrinsic contributions, that can originate either from spin-orbit coupling in the disorder scattering or spin-orbit coupling in the Bloch bands.  Among these, side jump scattering has been identified has an important contribution \cite{Berger:1970_a}.  
Side jump due to spin-orbit coupling in the Bloch bands appears as a ladder diagram vertex correction to the intrinsic anomalous Hall effect and its importance depends on the nature of that coupling \cite{Dugaev:2005_a,Sinitsyn:2005_a}.

The ratio of intrinsic and skew-scattering contributions to the AHE conductivity
can be approximated, assuming a single spin-orbit coupled band and
scattering off ionized impurities, by the expression
\cite{Leroux-Hugon:1972_a,Chazalviel:1975_a}:
\begin{equation}
\left|\frac{\sigma_{AH}^{int}}{\sigma_{AH}^{sk}}\right|= 
\frac{c N}{p r_s k_F l},
\label{sjtoss_ion}
\end{equation} 
where $N/p$ is  the ratio of the density of ionized impurities
and the carrier density, $r_s$ is  the average distance between carriers in
units of Bohr radius, $l$ is  the mean free path, and $c\sim 10$, varying
slightly with scattering length.  For the short range scattering potential
considered by Luttinger  \cite{Luttinger:1958_a} and Nozieres and Lewiner
\cite{Nozieres:1973_a}, $V(\vec{ r})=V_0\delta(\vec{r}-\vec{r_i})$:
\begin{equation}
\left|\frac{\sigma_{AH}^{int}}{\sigma_{AH}^{sk}}\right|=\frac{3}
{\pi |V_0| {\rm DOS}(E_F) k_F l}.
\label{sjtoss_short}
\end{equation}
The  estimate is a useful first guess
at which mechanism dominates in  different materials but one must keep
in mind the simplicity of the models used to derive these expressions.

In (Ga,Mn)As, Eq.~(\ref{sjtoss_ion}) gives a ratio  of intrinsic to
skew scattering contribution  of the order of 50 and the intrinsic
AHE    is   therefore   likely     to
dominate \cite{Dietl:2003_c}.  Consistently,
the experimental $R_A$  in metallic (Ga,Mn)As DMSs  
is proportional to  $\rho_{xx}^2$ \cite{Edmonds:2002_a}.
Microscopic calculations of the intrinsic low-temperature AHE conductivity 
in (Ga,Mn)As were performed by taking into account
the Berry phase anomalous velocity term in the semiclassical
Boltzmann equation leading to Eq.~(\ref{sigAHEberry})
\cite{Jungwirth:2002_a}, and by applying the fully quantum mechanical 
Kubo formalism \cite{Jungwirth:2003_b}. In both approaches,
the KL kinetic-exchange model was used to obtain the hole band structure.

In  the  Kubo   formula,  the  dc  Hall  conductivity  for
non-interacting quasiparticles at zero external magnetic field is given by \cite{Onoda:2002_a,Jungwirth:2003_b}
\begin{eqnarray}
\sigma_{\rm AH}&=&\frac{i e^2\hbar}{m^2 }\int \frac{d\vec{k}}{(2\pi)^3}
\sum_{n\ne n'}
\frac{f_{n',\vec{k}}-f_{n,\vec{k}}}{E_{n\vec{k}}-E_{n'\vec{k}}} 
\nonumber\\&&\times
\frac{\langle n \vec{k}|\hat{p}_x|n'\vec{k}\rangle\langle 
n'\vec{k}| \hat{p}_y|n \vec{k}\rangle}
{E_{n'\vec{k}}-E_{n\vec{k}}+i\hbar\Gamma}.
\label{sigAHEKubo}
\end{eqnarray}
The real part of Eq.~(\ref{sigAHEKubo}) in the limit of zero scattering rate 
($\Gamma\rightarrow 0$) can be written as, 
\begin{eqnarray}
\sigma_{\rm AH}&=&\frac{e^2\hbar}{m^2 }\int \frac{d\vec{k}}{(2\pi)^3}
\sum_{n\ne n'}
(f_{n',\vec{k}}-f_{n,\vec{k}}) \nonumber\\&&\times
\frac{{\rm Im}[\langle n' \vec{k}|\hat{p}_x|n\vec{k}\rangle\langle 
n\vec{k}| \hat{p}_y|n' \vec{k}\rangle]}
{(E_{n\vec{k}}-E_{n'\vec{k}})^2}.
\label{sigAHEKubo2}
\end{eqnarray}
Realizing  that the  momentum matrix  elements  
$\langle   n'   \vec{k}|\hat{p}_{\alpha}|n\vec{k}\rangle=
(m/\hbar)\langle                  n'                  \vec{k}|\partial
{\mathcal H}(\vec{k})/\partial_\alpha|n\vec{k}\rangle$, Eq.~(\ref{sigAHEKubo2}) can
be shown  to be equivalent  to Eq.~(\ref{sigAHEberry}).  The advantage of 
the Kubo formalism is that it makes it possible to  include finite
lifetime  broadening  of the  quasiparticles  in the simulations. (See 
Section~\ref{bulk-magtransp-amr} for the discussion of  quasiparticle scattering rates in (Ga,Mn)As.
Whether  or not  the lifetime broadening is
included, the  theoretical anomalous Hall conductivities  are of order
10~$\Omega^{-1}$~cm$^{-1}$  for typical  (Ga,Mn)As  DMS  parameters.  
On a quantitative level, non-zero $\Gamma$ tends to enhance $\sigma_{AH}$
at low  Mn doping and suppresses   $\sigma_{AH}$ at high  Mn concentrations where
quasiparticle broadening due to disorder becomes comparable to the
strength of the kinetic-exchange  field. 

\begin{figure}
\ifXHTML\Picture{review/figures/Fig36.png}\else\includegraphics[width=3.3in]{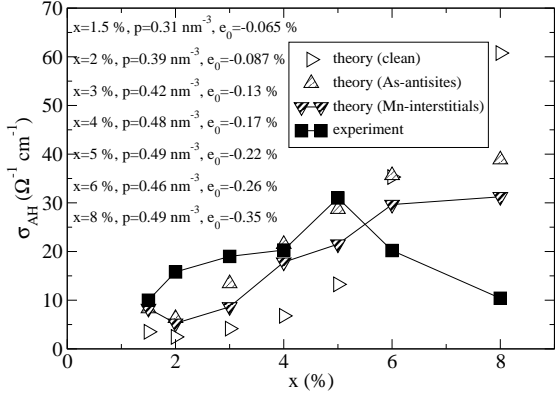}\fi
\caption{Comparison between experimental and theoretical anomalous Hall
conductivities. From \cite{Jungwirth:2003_b}.}
\label{AHE_teor_exp}
\end{figure}

A systematic comparison between  theoretical and experimental AHE data
is   shown  in  Fig.~\ref{AHE_teor_exp}  \cite{Jungwirth:2003_b}.  The
results  are plotted  versus  nominal Mn  concentration  $x$ while  other
parameters  of the  measured samples  are listed  in the  figure
legend.   Experimental  $\sigma_{AH}$ values  are indicated  by filled
squares and empty triangles correspond to theoretical data obtained for 
$\Gamma=0$. Results shown in
half-open triangles were obtained by solving the Kubo formula for 
$\sigma_{AH}$ with non-zero $\Gamma$ due to scattering off Mn$_{\rm Ga}$ and As-antisites,
 or Mn$_{\rm Ga}$ and Mn$_{\rm I}$ impurities. The calculations explain much of the 
 measured low-temperature AHE in metallic
 (Ga,Mn)As DMSs, especially so for the Mn$_{\rm I}$ compensation scenario. 
 The largest quantitative discrepancy between theory and experiment is  for
 the $x=8$\% material which can be partly explained by a non mean-field-like 
 magnetic behavior of this specific, more disordered sample.    

In DMSs the AHE  has played a key role  in establishing ferromagnetism and in providing
evidence for the hole-mediated coupling
between Mn local moments \cite{Ohno:1992_a,Ohno:1998_a}.
The dominance of the AHE in weak field measurements (see Fig.~\ref{ahe_nott})
allows the Hall resistance to serve also as a convenient proxy for the magnetization. On the other hand,
the same property  can obscure Hall measurements of the hole density.  
If the magnetization is not fully saturated at low
fields, for example, then $\rho_{AH}=\sigma_{AH}\rho_{xx}^2$ will increase with increasing 
external field $B$ through the dependence of $\sigma_{AH}$ on $M(B)$, and  hole densities
derived from the slope of $\rho_{xy}(B)$ will be too low. Accurate determination of hole densities in DMSs is
essential, however, and the Hall  effect is arguably the most common and accurate non-destructive
tool for measuring the level of  doping in semiconductors. Hall experiments performed in high 
magnetic fields to guarantee magnetization saturation seem a practical way for separating AHE
contributions, especially in samples showing weak longitudinal magnetoresistance 
\cite{Ohno:1999_a,Edmonds:2002_a}.  Hole density measurements performed using this technique assume
that the Hall factor,  $r_H=(\rho_{xy}(B)-\rho_{AH})/(B/ep)$ with $\rho_{AH}=\rho_{xy}(B=0)$,
is close to 1 despite the multi-band
spin-orbit coupled nature of hole dispersion in (Ga,Mn)As ferromagnets.
In the following paragraphs we briefly review a theoretical analysis of this assumption  \cite{Jungwirth:2005_b}. 

Microscopic  calculations  in   non-magnetic  p-type  GaAs  with  hole
densities $p\sim 10^{17}-10^{20}$~cm$^{-3}$ have shown that $r_H$ can vary
between 0.87  and 1.75, depending on doping, scattering mechanisms, and
on details of the model used for the GaAs valence band \cite{Kim:1995_a}. 
An estimate of  the influence on
$r_H$ from the  spin-splitting of the valence band  and from the anomalous
Hall term is based on the  KL kinetic-exchange model description of the 
hole band structure. The Hall conductivity has been obtained by evaluating the Kubo formula at   
finite    magnetic   fields that includes  both intra-band and inter-band transitions. The approach
captures the anomalous  and ordinary Hall terms on equal footing \cite{Jungwirth:2005_a}.

\begin{figure}
\ifXHTML\Picture{review/figures/Fig37.png}\else\hspace*{-.5cm}\includegraphics[angle=-0,width=3.8in]{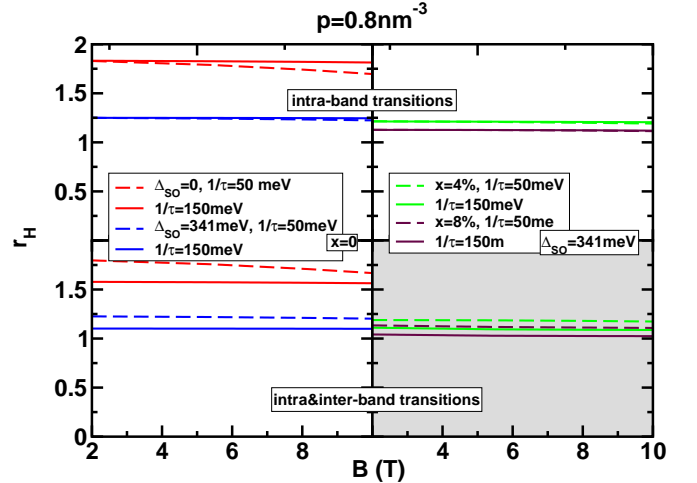}\fi
\caption{Theoretical Hall factors for $p=0.8$~nm$^{-3}$; $\hbar/\tau=50$~meV (dashed lines), 
$\hbar/\tau=150$~meV (solid lines). Top panels: only intra-band transitions
are taken into account. Bottom panels: intra- and inter-band transitions are taken into account.
Left panels: GaAs ($x=0$); zero spin-orbit coupling (red lines), $\Delta_{SO}=341$~meV (blue lines).
Right panels: (Ga,Mn)As with Mn$_{\rm Ga}$ concentration 4\% (green lines), 8\% (brown lines).
$\rho_{xy}(B=0)=0$ in all panels except for the bottom left panel where $\rho_{xy}(B=0)\neq 0$ due
to the anomalous Hall effect. From  \cite{Jungwirth:2005_b}.
} \label{rh8}
\end{figure}

Many  of  the qualitative  aspects  of  the  numerical calculations, shown  in
Fig.~\ref{rh8}, can be explained using a simple model of
a conductor with two parabolic  uncoupled bands. Note that the typical
scattering rate in (Ga,Mn)As epilayers is $\hbar/\tau\sim 100$~meV and that 
the cyclotron energy at  $B=5$~T is $\hbar\omega\sim 1$~meV, {\em i.e.}, the
system is in  the strong scattering limit, $\omega\tau\ll  1$. In this
limit, the two band model gives resistivities:
\begin{eqnarray}
\rho_{xx}&\approx&\frac{1}{\sigma_{xx,1}+\sigma_{xx,2}}
\approx\frac{1}{\sigma_{0,1}+\sigma_{0,2}}\nonumber \\
\rho_{xy}&\approx& -\frac{\sigma_{xy,1}+\sigma_{xy,2}}
{(\sigma_{xx,1}+\sigma_{xx,2})^2}\nonumber\\
&=&\frac{B}{ep_1}\;\frac{1+\frac{p_2}{p_1}(\frac{m^{\ast}_1}{m^{\ast}_2})^2}
{(1+\frac{p_2}{p_1}\frac{m^{\ast}_1}{m^{\ast}_2})^2}\ge\frac{B}{ep}\;, \label{rho_2band}
\label{hall_two_band}
\end{eqnarray}
where  the  indices  1 and  2  correspond  to  the  1st and  2nd  band
respectively,  the  total  density  $p=p_1+p_2$,  and  the  zero-field
conductivity   $\sigma_0=e^2\tau   p/m^{\ast}$.   Eq.~(\ref{rho_2band})
suggests that in the strong  scattering limit the multi-band nature of
the  hole  states   in  (Ga,Mn)As  should  not  result   in  a  strong
longitudinal  magnetoresistance. This  observation is  consistent with
the measured weak dependence of $\rho_{xx}$ on $B$ for magnetic fields
at which magnetization in the (Ga,Mn)As ferromagnet is saturated \cite{Edmonds:2002_a}.

The simple two-band  model also suggests that the  Hall factor, $r_H$,
is larger than one in multi-band systems with different dispersions of
individual  bands.  Indeed, for  uncoupled  valence  bands, {\em i.e.}  when
accounting  for  intra-band   transitions  only,  the  numerical  Hall
factors in  the top panels of Fig.~\ref{rh8}  are
larger  than 1 and independent of $\tau$ as also suggested by 
Eq.~(\ref{hall_two_band}).  The  suppression of  $r_H$  when spin-orbit coupling  is
turned on, shown in the  same graphs, results partly from depopulation
of  the   angular  momentum  $j=1/2$  split-off   bands.  In addition to  this
``two-band  model'' like  effect, the  inter-Landau-level matrix
elements are reduced  due to spin-orbit coupling since the  spinor part of the
eigenfunctions   now  varies   with   the  Landau   level  index.   In
ferromagnetic Ga$_{1-x}$Mn$_x$As  the bands are  spin-split and higher
bands depopulated  as $x$  increases. In terms  of $r_H$,  this effect
competes with  the increase of the  inter-Landau-level matrix elements
since the  spinors are now more  closely aligned within a  band due to
the exchange  field produced by  the polarized Mn  moments. Increasing
$x$ can therefore lead to either an increase or a decrease in $r_H$ depending
on  other  parameters,  such  as the  hole  density
(compare top right panels of Fig.~\ref{rh8}).

Inter-band transitions result in a more single-band like character
of  the  system,  {\em i.e.}  $r_H$   is  reduced,  and  the  slope  of  the
$\rho_{xy}(B)$ curve now depends more strongly on $\tau$. Although the
AHE and OHE contributions to $\rho_{xy}$ cannot
be  simply decoupled,  the comparison  of numerical  data in  the four
panels confirms  the usual assumption
that the  AHE produces a  field-independent off-set
proportional to  magnetization and  $\rho_{xx}^2$. The comparison  also suggests
that  after  subtracting  $\rho_{xy}(B=0)$,   $r_H$  can  be  used  to
determine the hole  density in (Ga,Mn)As with accuracy  that is better
than in non-magnetic GaAs  with comparable hole densities. For typical
hole and Mn densities  in experimental (Ga,Mn)As epilayers 
the error of the Hall measurement of $p$ is estimated to be less than $\pm 20\%$ 
\cite{Jungwirth:2005_a}.

\subsection{Conductivity near and above $T_c$}
\label{bulk-magtransp-finite_t}

\begin{figure}
\ifXHTML\Picture{review/figures/Fig38.png}\else\includegraphics[angle=-90,width=3.4in]{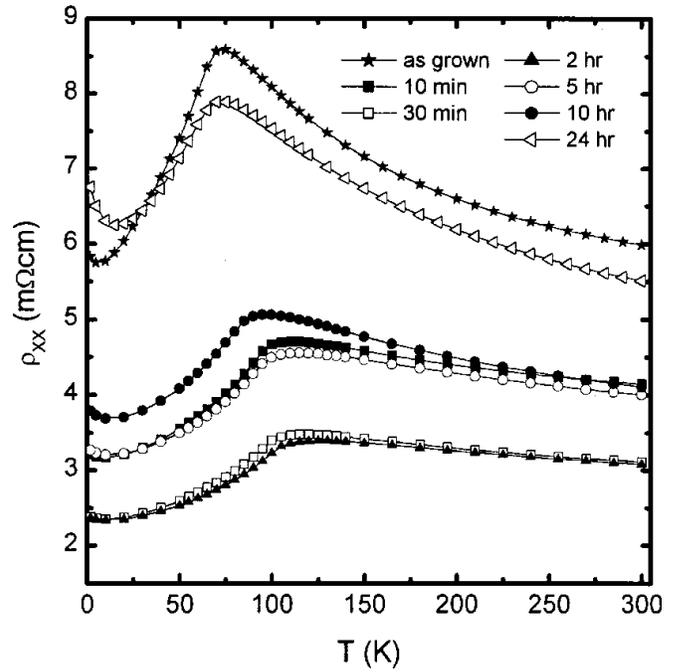}\fi
\caption{Experimental resistivity of Ga$_{1-x}$Mn$_x$As for $x=8\%$ vs. temperature for various annealing times.
From
\cite{Potashnik:2001_a}.}
\label{rho}
\end{figure}

Typical Fermi temperatures, $T_F=E_F/k_B$, in ferromagnetic  (Ga,Mn)As  are much larger than the Curie temperature relegating direct Fermi distribution effects of finite temperature to a minor role in transport. 
The carrier-mediated nature of ferromagnetism implies, however, strong
indirect effects through the temperature dependence of the magnetization. 
A prime example is the AHE which  from the early studies of (III,Mn)V DMSs has served as a practical tool to accurately measure Curie
temperatures  \cite{Ohno:1992_a}. A rough estimate of $T_c$ can be inferred also from the
temperature dependent longitudinal resistivity  which exhibits a shoulder in the more metallic (optimally
annealed) samples and a peak in the less metallic (as-grown) materials near the
ferromagnetic transition
\cite{Edmonds:2002_b,Potashnik:2001_a,VanEsch:1997_a,Matsukura:1998_a,Hayashi:1997_a}.
An example of this behavior is shown in Fig.~\ref{rho} 
for a (Ga,Mn)As material with 8\% nominal Mn-doping \cite{Potashnik:2001_a}.

\begin{figure}
\ifXHTML\Picture{review/figures/Fig39.png}\else\includegraphics[width=3.3in]{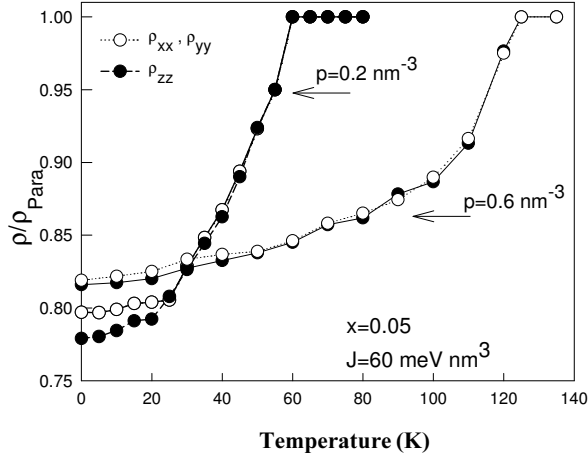}\fi
\caption{KL kinetic-exchange model resistivity, normalized with respect to the value above $T_c$, vs. T for
a Mn doping of 5\% and hole densities $p=0.2 {\rm nm}^{-3}$ and $p=0.2 {\rm nm}^{-3}$
From \cite{Lopez-Sancho:2003_b}.}
\label{rhoT}
\end{figure}

The shoulder in $\rho_{xx}(T)$ has been qualitatively modeled using the mean-field, KL kinetic-exchange 
Hamiltonian given by Eq.~(\ref{kl-pd-ham}). Solutions to the
Boltzmann equation  \cite{Lopez-Sancho:2003_b,Hwang:2005_a}
are shown in Fig.~\ref{rhoT}. The temperature dependence of
the longitudinal conductivity follows in this theory from variations in the parameters derived from the spin-polarized
hole band structure ({\em e.g.} Fermi wavevector) and from variations in screening of impurity
Coulomb potentials.  

The peak in resistance near $T_c$ has been discussed in terms of scattering effects beyond the lowest order
Born approximation and by using a network resistor model \cite{Timm:2004_a}.
It has also  been suggested that this transport anomaly in more highly resistive DMSs is a
consequence of the change in localization length caused by the ferromagnetic transition \cite{Zarand:2004_b}.

Above the Curie temperature, measurements of $\rho_{xx}$ have been used to estimate the value of $J_{pd}$.
Assuming  
scattering off the $p-d$ exchange potential at randomly distributed paramagnetic Mn$_{\rm Ga}$
impurities and parabolic hole bands,   
the corresponding contribution to the resistivity is approximated by,
\begin{equation}
\rho_{xx}=2\pi^2\frac{k_F}{p e^2}\frac{(m^{\ast})^2 J_{pd}^2}{h^3}N_{Mn}
[2\chi_{\perp}(T,B)+\chi_{\parallel}(T,B)],
\label{high_t_rho_xx}
\end{equation}     
where    $\chi_{\perp}$ and  $\chi_{\parallel}$   are  the   transverse   and  longitudinal   magnetic
susceptibilities  \cite{Dietl:1994_a,Omiya:2000_a,Matsukura:2002_a}.   This theory
overestimates critical scattering, particularly near the Curie temperature where the susceptibility diverges. 
Far from the transition on the paramagnetic side, however, fitting Eq.~\ref{high_t_rho_xx} to experimental magnetoresistance  data gives  
an estimate of the  $J_{pd}$  \cite{Omiya:2000_a} which is consistent with values inferred from spectroscopical
measurements  \cite{Okabayashi:1998_a}.

\begin{figure}
\ifXHTML\Picture{review/figures/Fig40.png}\else\includegraphics[width=3.3in,angle=-90]{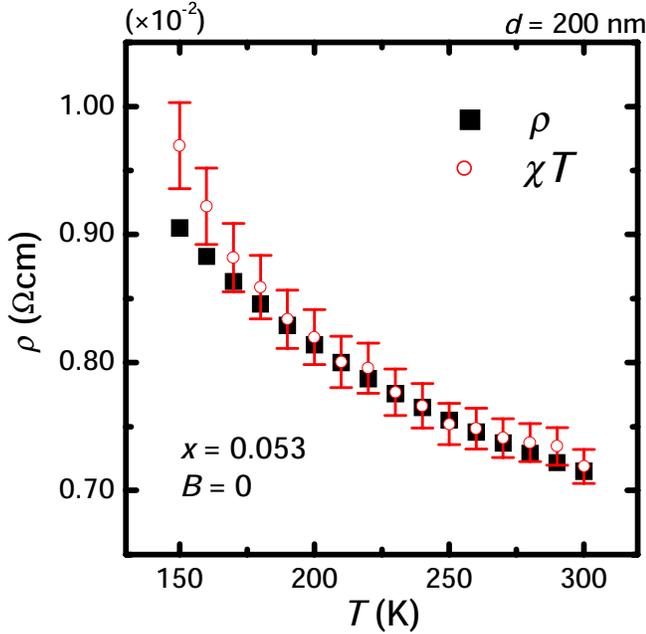}\fi
\caption{Temperature dependence of the resistivity for a
200-nm thick film of Ga$_{1-x}$Mn$_{x}$As with $x = 0.053$ in the
high-temperature paramagnetic region. Solid squares and open
circles show experimental data and the fit using Eq.~(\ref{high_t_rho_xx}),
respectively.
From \cite{Omiya:2000_a}.}
\label{high_t_rho_dietl}
\end{figure}
%%%%%%%%%%%%%%%%%%%%%%%%%%%%%%%%%%%%%%%%%%%%%%%%%%%%%%%%%%%%

\section{Magneto-optics}
\label{bulk-ac_magtransp}
The prospects for new technologies based for example on materials in which the ferromagnetic 
transition can be controlled by light or on (III,Mn)V Faraday isolators monolithically 
integrated with existing semiconductor lasers, have motivated research in magneto-optical 
properties of DMSs
\cite{Munekata:1997_a,Koshihara:1997_a,Sugano:2000_a,Matsukura:2002_a}.
Apart from these applied physics interests, ac probes have been used to study 
DMS materials by a 
wide range of experimental techniques.  In Sections~\ref{micro-subMn_pd} and \ref{bulk-mag-magnet-manyMn}
we mentioned x-ray spectroscopies (core-level photoemission and XMCD) used to characterize    
Mn 3$d$ states and detect the sign and magnitude of the $p-d$ exchange coupling. Dispersions
of hole bands in DMSs have been studied by angle-resolved photoemission  with ultraviolet excitations
\cite{Okabayashi:2002_a,Okabayashi:2001_a,Asklund:2002_a}
and infrared-to-ultraviolet spectroscopic ellipsometry
\cite{Burch:2004_a}. Raman scattering induced by excitations in the visible range was
used as an alternative means of estimating hole densities
\cite{Seong:2002_a,Limmer:2002_a,Sapega:2001_a}.
Spectroscopic studies of isolated Mn($d^5$+hole) impurities in the infrared region
 provided key information
on the valence of Mn in (Ga,Mn)As, as discussed in 
Section~\ref{micro-subMn}, and cyclotron resonance measurements were used to study 
highly Mn-doped DMS materials in
this frequency range \cite{Khodaparast:2003_a,Sanders:2003_a,Mitsumori:2004_a}. The microwave
EPR and FMR experiments, mentioned in Section~\ref{micro-subMn}
and \ref{bulk-mag-micromag}, 
have been invaluable  for understanding the nature  of Mn
in III-V hosts at low and high dopings, and for characterizing magnetocrystalline anisotropies and magnetization
dynamics in ferromagnetic materials.
In this section we review studies of some of the magneto-optical responses 
\cite{Ando:1998_a,Kimel:2005_a,Matsukura:2002_a,Beschoten:1999_a,Szczytko:1999_a,Hrabovsky:2002_a,Lang:2005_a}, 
particularly 
magnetic circular dichroism (MCD), in the visible range and  infrared absorption 
\cite{Singley:2002_a,Singley:2003_a,Hirakawa:2002_a,Hirakawa:2001_a,Nagai:2001_a,Szczytko:1999_a,Burch:2005_a}.

\subsection{Visible magnetic circular dichroism}
\label{bulk-magopt-vismcd}

Optical absorption due to electron excitations across the band gap is a standard characterization technique in
semiconductors. In (Ga,Mn)As, the absorption occurs in the visible range 
and the position of its edge on the frequency axis depends on the circular polarization of the incident light.
Analysis of this magneto-optical effect provides information on the $p-d$
exchange induced band splitting and on doping in the DMS material \cite{Dietl:2002_b,Matsukura:2002_a}. 

\begin{figure}[h]
\ifXHTML\Picture{review/figures/Fig41.png}\else\includegraphics[width=3.0in,angle=-0]{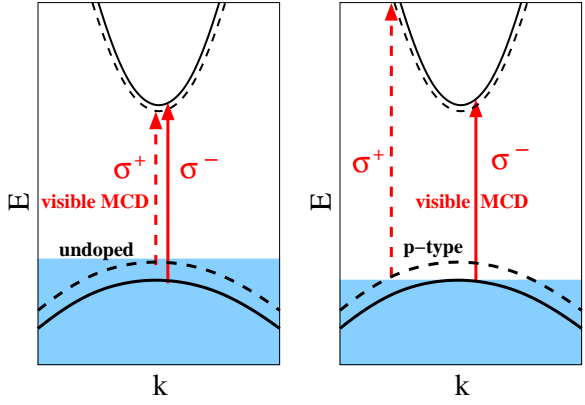}\fi
\caption{Schematic diagrams of electron excitations across the band-gap induced by
circularly polarized light absorption in undoped (left) and p-type (right) DMSs with spin-split
bands. Only the heavy-hole band is shown for illustration. For a given circular polarization of the absorbed light,
optical selection rules allow transitions between 
the heavy-hole valence band and the conduction band 
with one spin orientation only and therefore make it possible to spectrally resolve the   band splitting. 
The corresponding
MCD signal can change sign in the doped system due to the Moss-Burnstein effect, as illustrated in this diagram. 
}
 \label{vis_mcd_schem}
\end{figure}
\begin{figure}[h]
\ifXHTML\Picture{review/figures/Fig42.png}\else%
\hspace*{-0.5cm}\includegraphics[width=3.2in]{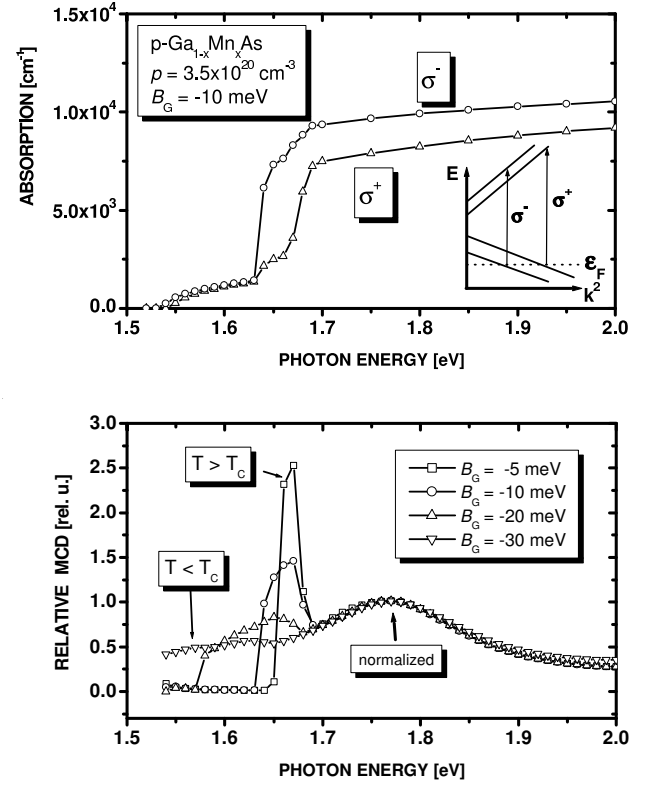}\fi
\caption{Top panel: Theoretical absorption edge for two circular polarizations in
p-(Ga,Mn)As computed for spin-splitting parameter $B_G=-10$ meV (corresponding
to $T=0$ mean-field from local Mn moments of density $x=1.7$\%) and
hole concentration $3.5\times 10^{20}$ cm$^{-3}$. Inset shows how
the Fermi sea of the holes reverses the relative positions of the edges
corresponding to $\sigma^+$ and $\sigma^-$ polarizations in agreement 
with experimental findings. 
Bottom panel: Spectral dependence of magnetic circular dichroism in the optical range in
(Ga,Mn)As computed for hole concentration $3.5\times 10^{20}$
cm$^{-3}$ and various spin-splitting parameters $B_G$. The
magnitudes of MCD at given $B_G$ are normalized by its value at
1.78 eV.
From \cite{Dietl:2001_b}.}
 \label{dietl01}
\end{figure}

The schematic diagram in Fig.~\ref{vis_mcd_schem}
shows that for a given sign of the exchange coupling, the order of  
absorption edges corresponding to the two circular
photon polarizations can reverse  in p-type materials, compared to systems with a completely filled
valence band. Calculations for (Ga,Mn)As that include this Moss-Burnstein effect were 
carried out using the mean-field KL kinetic-exchange model
\cite{Dietl:2001_b}. The resulting absorptions
 $\alpha^{\pm}$ of the $\sigma^{\pm}$ circularly polarized light,
 and  MCD, defined as \cite{Sugano:2000_a}
\begin{equation}
{\rm MCD}\equiv\frac{\alpha^--\alpha^+}{\alpha^++\alpha^-}=-\frac{{\rm
Im}[\sigma_{xy}(\omega)]} {{\rm Re}[\sigma_{xx}(\omega)]},
\end{equation} 
are shown in Fig.~\ref{dietl01}.
As suggested in the cartoon, the sign of the MCD signal in (Ga,Mn)As is opposite to the one obtained in bulk 
(II,Mn)VI DMSs where the
sense of band splittings is the same as in (Ga,Mn)As but Mn substituting for
the group-II element is an isovalent neutral impurity \cite{Dietl:1994_a}. 
(The Moss-Burnstein sign change in MCD
was also observed in co-doped p-type (II,Mn)VI quantum well \cite{Haury:1997_a}.) 

\begin{figure}[h]
\ifXHTML\Picture{review/figures/Fig43.png}\else\includegraphics[width=3.4in,angle=0]{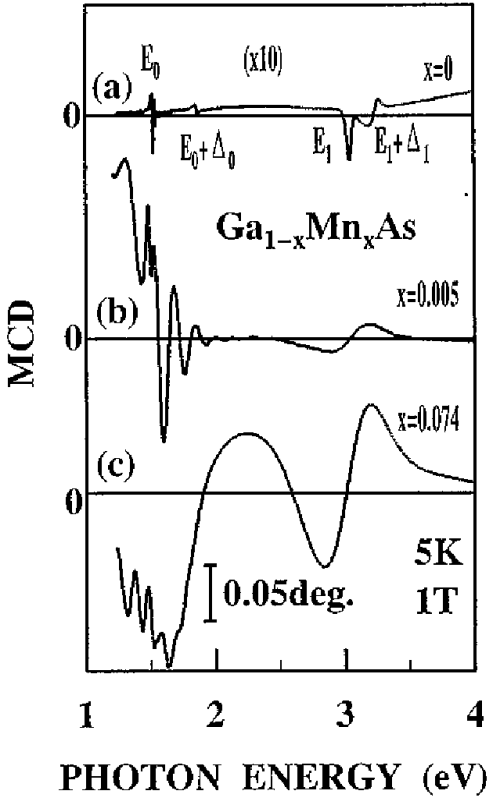}\fi
\caption{Experimental MCD spectra of (a)
undoped semi-insulating GaAs substrate and (b), (c) of
epitaxial Ga$_{1-x}$Mn$_x$As films at T = 5 K and B = 1 T. The
spectrum of GaAs is magnified ten times because the signal
is weaker than that of Ga$_{1-x}$Mn$_x$As.
From \cite{Ando:1998_a}.}
 \label{matsukura_mcd}
\end{figure}

Experimentally, the incorporation of several per cent of Mn in GaAs 
strongly enhances MCD, as
shown in Fig.~\ref{matsukura_mcd} \cite{Ando:1998_a,Matsukura:2002_a},
and the  sign of the signal near band-gap frequencies is consistent with  
the above theory which assumes antiferromagnetic
$p-d$ exchange coupling
\cite{Ando:1998_a,Beschoten:1999_a,Szczytko:1999_a,Dietl:2001_b}.
 
\subsection{Infrared absorption}
\label{bulk-magopt-irabs}
\begin{figure}[h]
\ifXHTML\Picture{review/figures/Fig44.png}\else\includegraphics[width=3.0in,angle=-0]{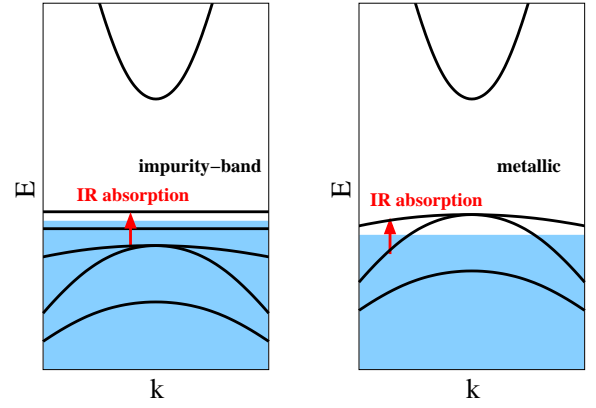}\fi
\caption{Schematic diagram illustrating absorption due to electron excitations to the impurity band (left panel) and due
to intra-valence-band excitations (right panel).
}
 \label{infra_schem}
\end{figure}

In (III,Mn)V DMSs, light absorption can occur  at sub-band-gap frequencies due to valence-band to Mn impurity
band excitations 
\cite{Hwang:2002_a,Alvarez:2003_a,Craco:2003_a}
in the more insulating materials and due to intra-valence-band excitations 
\cite{Sinova:2002_a,Yang:2003_b,Sinova:2003_a}
in the more
metallic systems, as illustrated schematically in Fig.~\ref{infra_schem}. 
The infrared absorption associated
with substitutional Mn impurities
 is spectrally resolved from higher energy excitations to donor levels of 
 the most common unintentional
defects, such as 
 the Mn$_{\rm I}$ interstitials
 and As$_{\rm Ga}$ antisites in (Ga,Mn)As, and therefore represents another valuable probe into intrinsic
 properties of these systems.
Since infrared  wavelengths are much larger than typical (sub-micron) 
DMS epilayer widths, the absorption is 
related to the real  part of the conductivity by \cite{Sugano:2000_a}
\begin{equation}
\alpha(\omega)=2\frac{{\rm Re[}\sigma(\omega)]}{Y+Y_0}\,\,,
\end{equation}
where  $Y$ and $Y_0$  are the  admittances of  the substrate  and free
space,  respectively. 

\begin{figure}[h]
\ifXHTML\Picture{review/figures/Fig45.png}\else\includegraphics[width=3.2in]{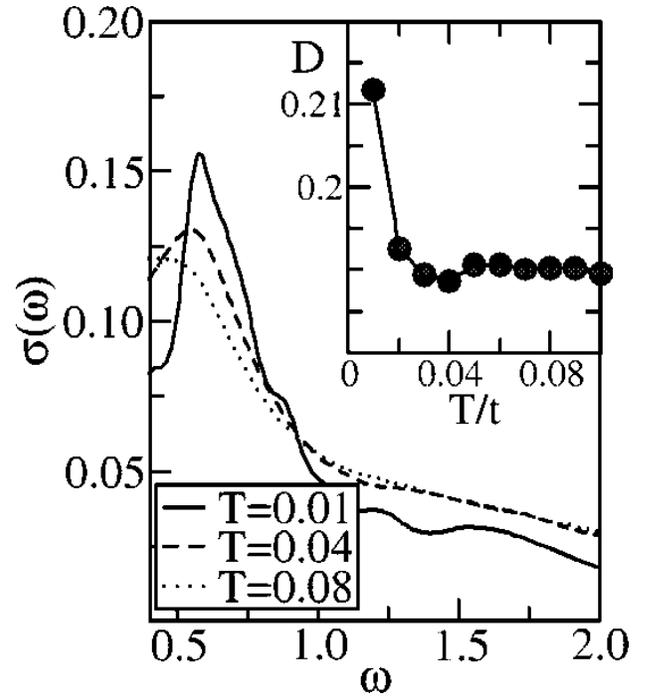}\fi
\caption{Impurity-band model calculations of $\sigma(\omega)$ vs. $\omega$ for a 
10$\times$10 periodic system
with 26 spins ($x\sim0.25$), $J/t=2.5$, $p=0.3$, and  
for different temperatures, as indicated. 
Inset: Drude weight, $D$, vs. temperature $T$ \cite{Alvarez:2003_a}.
\label{alvarez}}
\end{figure}

Model Hamiltonians (\ref{spinlatham}) (see Section~\ref{theory-lattice})
combined with dynamical  mean-field-theory or Monte Carlo simulations were used to 
study the role of the impurity band in infrared absorption 
\cite{Hwang:2002_a,Alvarez:2003_a}.  The impurity band forms in this theory
when the strength of the model effective
exchange interaction $J$ is comparable to the width of the main band, characterized by the hopping parameter $t$.
A non Drude peak is  observed  in the
frequency-dependent  conductivity, associated with transitions from the main band to the 
impurity band.
The behavior is illustrated in Fig.~\ref{alvarez} together with the predicted temperature
dependence of the absorption spectra obtained by the Monte Carlo technique.
As discussed in Section~\ref{theory-lattice},
these model calculations are expected to apply to systems with strong $p-d$ exchange coupling,
like (Ga,Mn)P and possibly also to (III,Mn)V DMSs which are strongly compensated due to the presence
of unintentional donor defects.  (Impurity-band mediated ferromagnetism does not occur in
uncompensated samples.)  

\begin{figure}[h]
\ifXHTML\Picture{review/figures/Fig46.png}\else\includegraphics[width=3.2in]{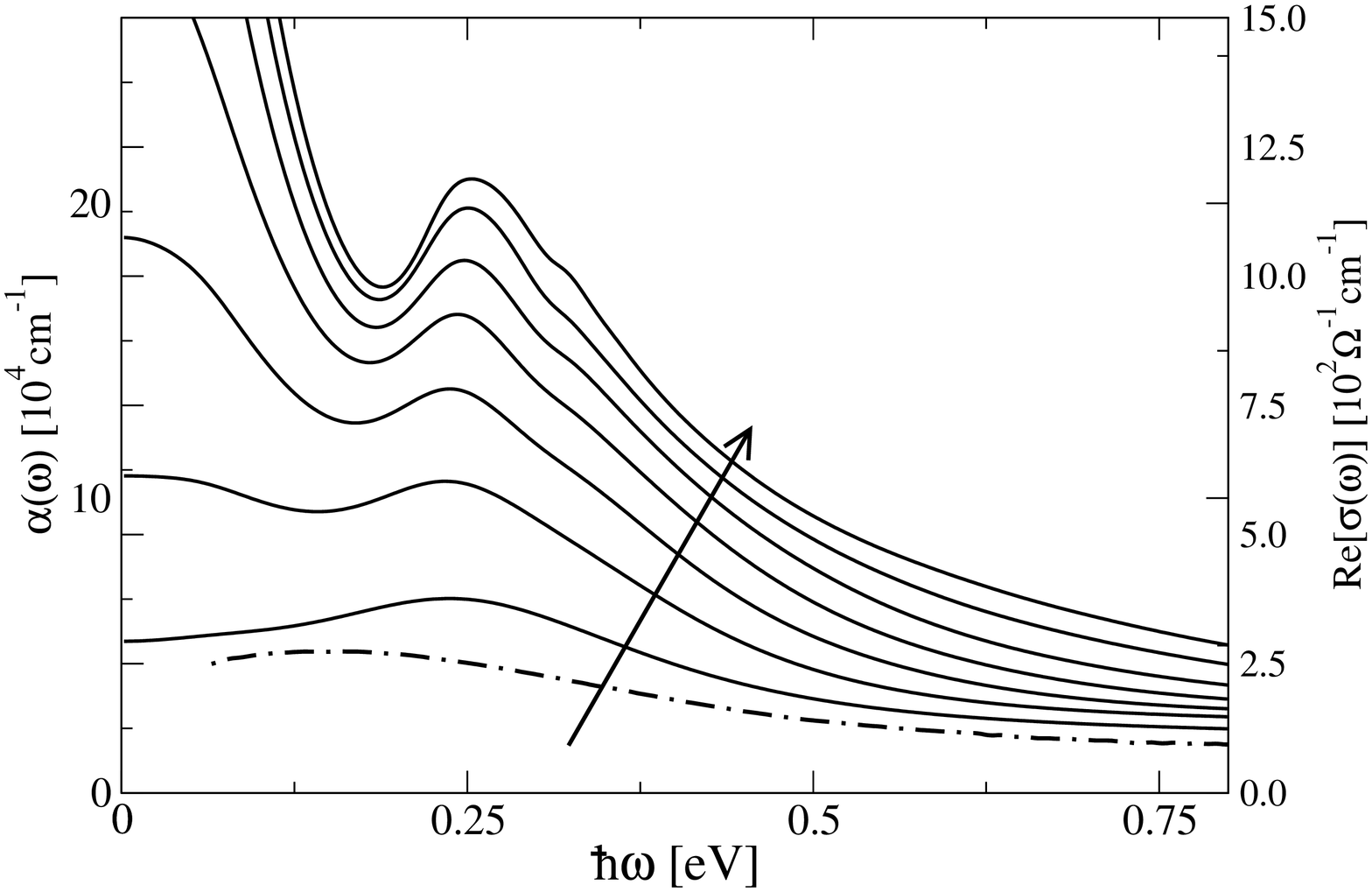}\fi
\caption{KL kinetic-exchange model calculations of 
infrared    conductivity   ${\rm    Re}[\sigma(\omega)]$   and
absorption coefficient  $\alpha(\omega)$ for carrier densities
from $p=0.2$ to $0.8 {\rm  nm}^{-3}$ in the direction indicated by the
arrow,  for  Ga$_{0.95}$Mn$_{0.05}$As \cite{Sinova:2002_a}.  Disorder is treated within
the first order Born approximation. The  dot-dashed  line  is  the
experimental absorption curve for a  sample with 4\% Mn doping 
from \cite{Hirakawa:2002_a}.
} 
\label{infra_sinova}
\end{figure}

\begin{figure}[h]
\ifXHTML\Picture{review/figures/Fig47.png}\else\includegraphics[width=3.4in]{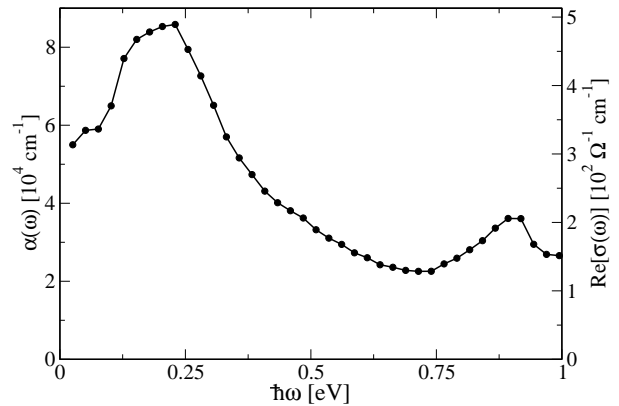}\fi
\caption{Infrared absorption  and conductivity  of a  metallic  (Ga,Mn)As computed
using the KL kinetic-exchange model and exact diagonalization
technique in a finite size disordered system. Here  
$p=0.33$~nm$^{-3}$, $n_{Mn}=1$~nm$^{-3}$ ($x\approx 4.5$\%). 
From \cite{Yang:2003_b}.}
\label{yang2}
\end{figure}

Theoretical infrared absorption spectra calculated using the ${\bf k}\cdot{\bf p}$ model for 
(Ga,Mn)As DMSs with delocalized holes
in the semiconductor valence band, plotted in Fig.~\ref{infra_sinova},
show a similar non-Drude characteristics with a peak near the excitation energy of 220~meV. The underlying
physics is qualitatively different, however, as the peak in these KL kinetic-exchange model calculations
originates from heavy-hole to light-hole intra-valence-band transitions \cite{Sinova:2002_a}.  
These results were obtained by evaluating the Kubo formula for ac conductivity 
assuming non-interacting holes and modeling disorder within the first order Born
approximation (see  Eq.~(\ref{gamma}) in Section~\ref{bulk-magtransp-dc_cond}).  
In Fig.~\ref{yang2} we show theoretical predictions of exact
diagonalization  studies based on the KL kinetic-exchange Hamiltonian
but treating  disorder  effects
exactly in a finite size system. The results correct for the overestimated dc 
conductivity in the former model,
which is a quantitative deficiency of the Born approximation
as already mentioned in Section~\ref{bulk-magtransp-dc_cond}. At finite frequencies, 
the theoretical absorption in these metallic
(Ga,Mn)As DMSs is
almost insensitive to the way disorder is treated in the simulations, as see from Figs.~\ref{infra_sinova}
and \ref{yang2}.

\begin{figure}[h]
\ifXHTML\Picture{review/figures/Fig48.png}\else\includegraphics[width=3.2in]{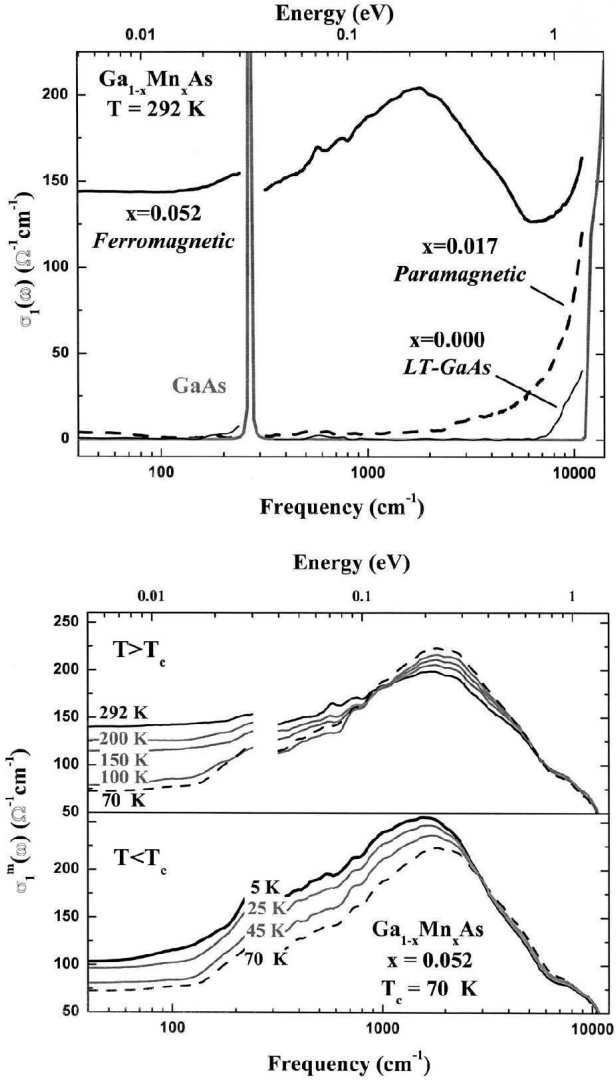}\fi
\caption{Top panel: Real part of the conductivity derived from measured
transmission spectra for a paramagnetic $x=1.7$\% sample (dashed line), a ferromagnetic
$x=5.2$\% sample (thick solid line), and the LT GaAs
film (thin black line). 
Bottom panel:
Temperature dependence of conductivity for  $x=5.2$\% sample
on a log scale for $T>T_c$ and  $T<T_c$ .
From \cite{Singley:2003_a}.
}
\label{singlet}
\end{figure}

Experimental infrared absorption studies in ferromagnetic (Ga,Mn)As epilayers 
exhibit several 
common  features summarized in Fig.~\ref{singlet} 
\cite{Singley:2002_a,Singley:2003_a,Hirakawa:2002_a,Hirakawa:2001_a,Nagai:2001_a,Szczytko:1999_a,Burch:2005_a}.  Ferromagnetic materials ($x=5.2$\% curves in   Fig.~\ref{singlet}) show a 
non-Drude  behavior in  which the conductivity
increases with  increasing frequency in  the interval between  $0$ 
and $220$ ~meV, a broad absorption  peak near $220-260$~meV that
becomes  stronger  as  the  sample   is cooled,  and  a  
featureless absorption up to approximately 1~eV. As  seen in Fig.~\ref{singlet}, the peak is absent  in
the reference, LT-MBE-grown GaAs sample confirming that the infrared 
absorption in ferromagnetic (Ga,Mn)As is related to changes in the band structure near the Fermi energy
induced by Mn impurities.

The presence of a finite-frequency peak in both impurity band and KL
kinetic-exchange
models for the infrared conductivity leads to an ambiguity in the
interpretation of existing data
which has for the most part been taken in as grown, presumably heavily
compensated material.
The metallic behavior of the $x=5.2$\% material below $T_c$,  seen in the
lower panel of Fig.~\ref{singlet},
favors the inter-valence-band absorption scenario. On the other hand, the
large compensation likely present in as-grown low-$T_c$ (Ga,Mn)As suggests
that many holes may be strongly localized and that both
absorption mechanisms may contribute to the measured absorption peak.
Experiments in a series of samples interpolating between as grown and
optimally annealed
limits, analogous to the resistance-monitored annealing studies
\cite{Edmonds:2002_b,Edmonds:2004_a},
should enable a clear interpretation of infrared absorption
spectra in (Ga,Mn)As DMSs. These studies will hint towards necessary
refinements of the simplified theories
used so far, {\em e.g.} inclusion of the energy dependence of $J_{pd}$ and a more
quantitative theory of the impurity
band model. The support for either scenario by these experiments has to be
considered in conjunction with other available
data in a self-consistent picture, {\em e.g.} in an impurity band picture $T_c$ is
predicted to approach zero as the system
reaches zero compensation whereas the KL kinetic-exchange model has an
opposite trend.

%%%%%%%%%%%%%%%%%%%%%%%%%%%%%%%%%%%%%%%%%%%%%%%%%%%%%%%%%%%%

\section{Discussion}
\subsection{Magnetic interactions in systems with coupled local and itinerant moments }
\label{discussion-general}

Systems with local moments coupled to itinerant electrons are  common in 
condensed matter physics and exhibit a wide variety of behaviors.  Ferromagnets
are far from the most common low-temperature states.  
For that reason it is useful to ask how 
(III,Mn)V materials, and (Ga,Mn)As with its robust ferromagnetic order in particular, 
fit in this larger context. This general qualitative analysis
can   help
to identify some of the key factors that might limit the strength of
ferromagnetic interactions  in the highly doped and strongly
$p-d$ coupled DMS ferromagnets for which mean-field theory predicts the highest Curie
temperatures.

An important class of 
materials that has been very extensively studied  is heavy fermions, in which 
f-electron local moments are exchange coupled to band electrons \cite{Stewart:1984_a}.  
Kondo lattice models, which are believed to qualitatively describe heavy fermion systems, assume that 
local moments exist at each lattice site. 
Models of DMS systems which make a virtual crystal approximation
(see Section~\ref{micro-subMn_pd}and \ref{bulk-mag-tc-mf})
place moments on all lattice sites and are therefore Kondo lattice models,
often with specific details that attempt to capture some of the
peculiarities of specific DMS materials.
Theories of Kondo lattice model often start from
the comparison of the RKKY  (see Section~\ref{micro-general})
and Kondo temperature scales \cite{Doniach:1977_a,Degiorgi:1999_a,Tsunetsugu:1997_a}.
The characteristic
RKKY temperature refers to the strength of interactions between local moments mediated by a weakly disturbed 
carrier system and is proportional
to the mean-field $T_c$ given by Eq.~(\ref{tc}). The Kondo scale refers to the temperature below which strong 
correlations are established between an isolated local moment and the carrier 
system with which it interacts.   Standard
scale estimate formulas \cite{Doniach:1977_a} applied to the case of DMS ferromagnets
imply that the Kondo scale is larger than the RKKY scale when the mean-field exchange coupling,
$SN_{Mn}J_{pd}$, is larger 
than the Fermi energy $E_F$ of the hole system, 
in other words in the strong coupling regime.  The Kondo scale falls rapidly
to small values at weaker coupling.  (In heavy-fermion materials the Kondo temperature scale 
is larger than the RKKY temperature scale.) 

\begin{figure}

\ifXHTML\Picture{review/figures/Fig49.png}\else%
\vspace*{-1.0cm}\includegraphics[width=2.5in]{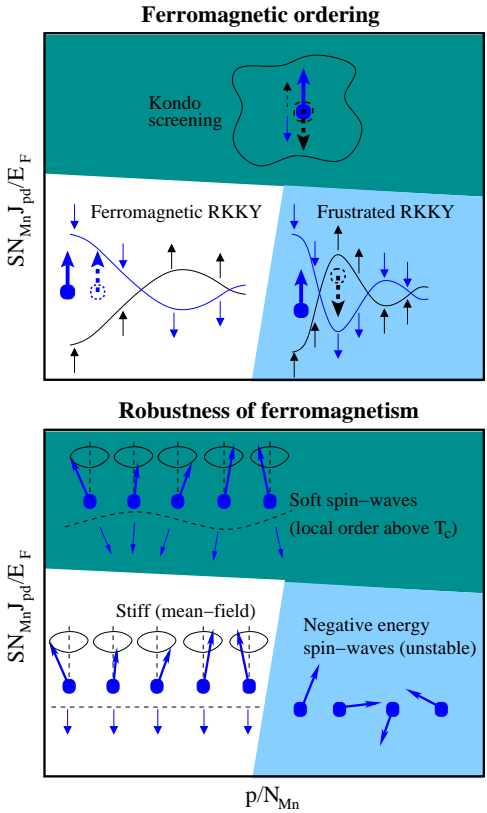}\fi
\caption{Top panel: A schematic qualitative diagram illustrating the requirements for
ferromagnetic ordering in
systems with coupled local and itinerant moments. The y-axis represents the strength of the exchange
coupling relative to the itinerant system Fermi energy and the x-axis the ratio between carrier and local moment
densities.  In the weak coupling regime the interaction between local moments is described by  RKKY theory
in which polarization of band electrons (holes) due to the interaction with the local moment at one site
is propagated to neighboring sites. The mechanism corresponds to out-of-phase Friedel oscillations of carriers
scattering off an attractive impurity potential for one carrier spin orientation (maximum density at the impurity site),
and a repulsive potential for
the opposite carrier spin polarization (minimum density at the impurity site). With increasing carrier density,
the ferromagnetic RKKY state gradually approaches a regime of frustrated RKKY interactions.
The RKKY picture does not apply to strong couplings. Here the tendency to order ferromagnetically
is weakened by correlated (flip-flop) quantum fluctuations of the interacting local and itinerant moments making
the moment disappear eventually. This is the so-called Kondo singlet regime.
Bottom panel: The robustness of the ferromagnetic state is viewed from the ferromagnetic low-temperature side.
With increasing hole density, the ferromagnetically stiff mean-field state will start to suffer from the
occurrence of lower energy spin-wave excitations until finally the energy of 
spin-waves will become negative  signaling the instability of the ferromagnetic
state. In the strong coupling regime, hole spins will tend to align
locally and instantaneously in opposition to the fluctuating
Mn moment orientations. Since the hole 
Fermi energy is relatively small in this regime, 
the kinetic energy cost of these fluctuations is small, resulting in soft spin-wave excitations
for the magnetic system. The long-range ferromagnetic order disappears eventually at
temperatures much smaller than the mean-field $T_c$, while  short-range order 
may still exists above the Curie temperature.
} 
\label{discussion}
\end{figure}

Optimally annealed metallic (Ga,Mn)As materials are on the weak-coupling side of this 
boundary, but (Ga,Mn)N and possibly (Ga,Mn)P may be starting to reach toward the 
strong coupling limit (if the simple S=5/2 local moment model still applies in these materials).
As the strong coupling limit is approached, quantum fluctuations (see discussion
below Eq.~(\ref{two-spin-Ham})) 
will play a
greater role, reducing the saturation moment per Mn and eventually
driving down the ferromagnetic transition temperature.  When the Kondo scale is much larger than the 
RKKY scale, the local moments are  screened out by strongly correlated 
band electron spin-fluctuations and effectively disappear 
before they 
have the opportunity to couple.  

On the weak coupling side, 
RKKY interactions in Kondo lattice models tend to lead to ferromagnetism only
when the number of itinerant electrons per moment is small, {\em i.e.}, only 
when at least the near-neighbor RKKY interaction is ferromagnetic \cite{Tsunetsugu:1997_a}.  One of the 
surprising features of (Ga,Mn)As  is the property that ferromagnetism 
still occurs when the number of itinerant electrons per moment is $\sim 1$.
As we have mentioned in Section~\ref{bulk-mag-tc-sw}, this property follows from the 
specific 
multi-band electronic structure and spin-orbit coupling at the top of the valence band.  Nevertheless, 
frustrating antiferromagnetic RKKY interactions, and 
exchange interactions that promote non-collinear magnetic states 
 (see Section~\ref{bulk-mag-magnet}), 
will eventually become important for sufficiently large carrier densities.
As this regime is approached from the ferromagnetic side, the transition temperature will
be suppressed.  These tendencies are summarized schematically in Fig.~\ref{discussion}.

Similar considerations apply in assessing 
the robustness of ferromagnetism in the 
ordered state as characterized by the spin-stiffness micromagnetic 
parameter \cite{Schliemann:2001_a,Konig:2001_c,Konig:2003_a}.
We have used this approach in Section~\ref{bulk-mag-tc-mf} when analyzing the  limitations of the
mean-field theory in (Ga,Mn)As.
Starting from the ferromagnetic state, 
long-wavelength spin-orientation modulation
will tend to lower the energy of some Mn-Mn interactions for sufficiently high carrier densities.
The spin-stiffness will weaken as the 
frustrated magnetism regime is approached, until finally the energy of ferromagnetic
spin-waves will become negative  signaling the instability of this 
state \cite{Schliemann:2002_a}. 
Similarly for sufficiently strong coupling, the band system will be (nearly when spin-orbit 
coupling is included) fully spin-polarized and the cost of spin-orientation spatial variation 
will be borne mainly by the kinetic energy of the hole system and no longer increase with exchange coupling.
As shown schematically in the lower panel of  Fig.~\ref{discussion}, hole spins in this regime
are locally antiferromagnetically locked to the fluctuating Mn moment orientations. For relatively small hole 
Fermi energies the kinetic energy cost of these fluctuations is small, resulting in soft spin-wave excitations
of the magnetic system \cite{Konig:2001_c}. In this regime long-range ferromagnetic order disappears at
temperatures smaller than the mean-field $T_c$, {\em i.e.}, short-range order still exists above the Curie temperature.

The white bottom-left area in the panels of Fig.~\ref{discussion} qualitatively depicts  
the parameter range in which the ferromagnetic RKKY
mean-field state applies.  Here  $T_c$ increases with the carrier density, local moment density  and  (quadratically) with
the strength of the $p-d$ exchange coupling. For a fixed ratio of $p/N_{Mn}$, increasing $N_{Mn}$ corresponds to
moving only slowly (as $N_{Mn}^{1/3}$) upwards in the diagrams. This may explain why (Ga,Mn)As
materials with larger Mn doping and with similar hole compensations as in the ferromagnetic low
local moment density systems do not show any marked weakening of the ferromagnetic state. With
$p/N_{Mn}$ still fixed, attempts to
increase $T_c$ in (III,Mn)V DMSs by increasing the $J_{pd}$ constant, in {\em e.g.}
ternary host alloys of Ga(As,P), might at some point reach the boundary of the soft spin-wave (Kondo screened) state.
Similarly, the Kondo lattice model allows only a limited space for enhancing the robustness of the ferromagnetic state
by tweaking the carrier and local moment densities independently. In this case moving horizontally from the boundary of the
frustrated RKKY (ferromagnetically unstable) state is accompanied by approaching vertically the  soft spin-wave 
(Kondo screened) regime, and {vice versa}. Viewed from the opposite perspective, however, it is astonishing that
a window in this parameter space has been found by the material research of (III,Mn)V compounds
for robust DMS ferromagnets with Curie temperatures close to 200~K.
The diagrams do not imply any general physical mechanism  that limits $T_c$ in these materials below room temperature. 
  
Our remarks on the cartoons in  Fig.~\ref{discussion} refer to the properties of Kondo lattice 
models which have been adjusted to reflect peculiarities of the 
zinc-blende semiconductor valence band.  We have so far neglected the 
importance of disorder and of Coulomb scattering in DMSs, 
and these can modify some parts of the simple qualitative
picture depicted in Fig~\ref{discussion}.  This is particularly true in the very low-density 
isolated Mn limit, {\em i.e.}, very strong exchange coupling limit. Because of Coulomb attraction between
valence band holes and the charged Mn ion which carries the local moment, 
a total angular momentum  $F=1$ isolated bound state is created, as we have explained
in Section~\ref{micro-subMn_pd}, instead of the strongly correlated Kondo singlet.  
The importance of Coulomb interactions and disorder are lessened 
by screening and Pauli exclusion principle effects when both the Mn density and the carrier density 
are high. 

\subsection{Ferromagnetism in the diluted magnetic semiconductor family}
\label{discussion-ralated}
In this section we will narrow down the discussion of magnetism in local moment systems 
to semiconducting compounds, focusing on the phenomenology of ferromagnetic DMSs
other than the (III,Mn)V materials.
Almost any semiconducting or insulating compound that contains
elements with partially filled $d$ or $f$ shells (local moments) will
order magnetically at a sufficiently low temperature.  
Semiconductors and insulators with high density of magnetic moments
usually order
antiferromagnetically however, although ferromagnetism does occur
in some cases.
A famous example of a ferromagnetic
system that can be regarded as a doped semiconductor is
provided by the manganite family ({\em e.g.} La$_{1-x}$Sr$_{x}$MnO$_3$)
whose ferromagnetism is favored by the double-exchange mechanism and occurs over a wide
range of transition temperatures from below 100~K to nearly 400~K.
The onset of magnetic order in these systems is accompanied by a
very large increase in conductivity. For a review, see for example \cite{Coey:1999_a}. 
Other well known dense moment
(of order one moment per atom) ferromagnetic semiconductors with strong exchange
interaction between itinerant and local spins
include Eu- and Cr-chalcogenides \cite{Kasuya:1968_a,Mauger:1986_a,Baltzer:1966_a,Stapele:1982_a}, 
such as rock-salt EuO  and spinel CdCr$_2$Se$_4$ with Curie temperatures 70~K and 130~K, respectively.

DMS systems  in which
magnetic atoms are introduced as impurities have moments on
only a small fraction of all atomic sites.  The mechanisms that
control magnetic order are therefore necessarily associated with
the properties of these impurities.  The coupling between moments will
generally depend on the locations of the dilute moments in the host
lattice, on the doping properties of the magnetic impurities, and on other dopants and defects present
in the material.  It seems plausible therefore that when DMS systems
are ferromagnetic, their magnetic and magnetotransport properties will be
more sensitive to engineerable material properties.  

This review has concentrated on (Ga,Mn)As and related materials in which, as
we have explained, substitutional Mn acts both as an acceptor and
as a source of local moments.  Ferromagnetism is carrier mediated and it has been
demonstrated that it persists to surprisingly high temperatures.  More may
be achieved in the future by tweaking these materials.  On the other
hand there is a vast array of alternate DMS materials that could be contemplated.
Research to date has only scratched the surface of the volume of possibilities - we are
truly still at the beginning of the road in studying diluted moment magnetism in
semiconductors.  Each system brings its new challenges.  The interpretation of
simple magnetic and transport characterization measurements is often not immediately
obvious, in particular because of the possibility that the moments will segregate into
crystallites of one of a variety of available dense moment minority phases which are
often thermodynamically more stable.  In addition, magnetic properties will
certainly depend in general not only on the dilute moment density, which normally is well
controlled and variable, but also on the partitioning of local moments among
many available  sites in the host crystal which is not always known and is
usually much harder to control.  The search for promising DMS materials would
be simplified if {\em ab initio} DFT methods had reliable predictive
power. Unfortunately this luxury appears to be absent in many cases because
of extreme sensitivity of magnetic properties to details of the electronic
structure and because of strong correlation effects that are often present in these systems.
We mention briefly in the following paragraphs
some of the other classes of diluted magnetic semiconductors which have been studied.

The class of DMS ferromagnets that is closest to (III,Mn)V
materials is (II,Mn)VI compounds co-doped with group V element
acceptors.  Examples include p-(Zn,Mn)Te:N \cite{Ferrand:2001_a}
and p-(Be,Mn)Te:N \cite{Hansen:2001_a}.  These materials differ
from (Ga,Mn)As mainly because the local moments and holes
are provided by different types of impurities  and can be controlled independently.
Although the physics behind ferromagnetism seems to be very
similar in the two classes of materials, the highest ferromagnetic
transition temperatures that have been achieved are much smaller
in the case of co-doped (II,Mn)VI materials, $\sim 2{\rm K}$
rather than $\sim 200{\rm K}$.  The difference is explained partly
by difficulty in achieving the same extremely high hole doping
($\sim 10^{21}{\rm cm}^{-3}$) in (II,Mn)VI that has been achieved
in (Ga,Mn)As and partly by a favorable interplay between
electrostatic and magnetic effects in the (III,Mn)V materials.  In
(III,Mn)V materials, unlike (II,Mn)VI materials, the Mn moment is
charged and attracts holes.  The tendency of holes to have a
higher density near Mn sites tends to increase the effective
strength of the $p-d$ exchange interaction. This
effect is magnified when two Mn moments are on neighboring cation
sites. In (Ga,Mn)As the interaction between Mn moments on neighboring cation positions 
is ferromagnetic, compared to the strongly antiferromagnetic interactions seen in (II,Mn)VI materials.    
(For a discussion
of the interplay between electrostatic and magnetic interactions
see \cite{Sliwa:2005_a}.) In p-doped (II,Mn)VI materials, 
competition between antiferromagnetic near
neighbor interactions and the longer range carrier-mediated
ferromagnetic interactions suppresses the magnetic ordering
temperature.  This competition apparently does not occur in
(III,Mn)V ferromagnets with large hole densities.

(Zn,Mn)O is an interesting II-VI counterpart of the nitride III-V DMS \cite{Liu:2005_e}.
With advances in oxide growth techniques (Zn,Mn)O, can be
considered to be much like other (II,Mn)VI DMS materials and its
investigation was originally motivated by theoretical work
\cite{Dietl:2000_a} that extrapolated from experience with
(III,Mn)V DMS ferromagnets and predicted large $T_c$'s.  
Studies of this material have provided
clear evidence of strong $p-d$ exchange but, so far,  have led to inconsistent
conclusions about the occurrences of long range magnetic order \cite{Sharma:2003_a,Fukumura:2005_a,Petit:2004_a,Lawes:2005_a}. 

Tetrahedral DMS materials doped with  transition metal atoms other than Mn
have shown promising results. For instance
(Zn,Cr)Te is apparently homogeneous, has the required coupling
between local moments and carriers \cite{Mac:1996_a,Saito:2003_a} and
Curie temperature as high as 300~K has been reported for this material.  
It may be, though, that the ferromagnetism is due to superexchange interactions
rather than being carrier mediated since it occurs at very small
ratios of the carrier density to the moment density \cite{Saito:2002_a}.  
Another interesting material with Cr moments is
(Ga,Cr)N which exhibits ferromagnetism at $\sim 900{\rm K}$ \cite{Liu:2004_c}. The
question still at issue in this material is the possible role of
dense moment precipitates.  

Traditional groups of ferromagnetic DMSs also include (IV,Mn)VI
solid solutions with the rock-salt structure \cite{Story:1986_a,Story:1992_a,Eggenkamp:1993_a}.
Although the band structures of IV-VI and III-V
semiconductors are quite different, these DMS ferromagnets ({\em e.g.}  (Pb,Sn,Mn)Te) appear to have a 
carrier-mediated mechanism quite similar to that of (Ga,Mn)As. Holes with densities up to
10$^{21}$~cm$^{-3}$ are  supplied in these materials by cation
vacancies, rather than by Mn substitution for the divalent
cations. The reported Curie temperatures in (Pb,Sn,Mn)Te are below
40~K \cite{Lazarczyk:1997_a}.

DMS ferromagnetism with Si or Ge as the host semiconductor is
obviously attractive because of their greater compatibility with
existing silicon based technology. In Si, Mn impurities favor interstitial position
which significantly complicates the synthesis of a uniform DMS system. Mn in Ge, on the other hand,
is a substitutional impurity and ferromagnetism has been reported
in MBE grown ${\rm Ge}_{x}{\rm
Mn}_{1-x}$ thin film DMSs \cite{Park:2002_a,Li:2005_a}. Careful studies of ${\rm Ge}_{x}{\rm
Mn}_{1-x}$ \cite{Li:2005_a} have demonstrated that slow
low-temperature growth is required to avoid the formation of
thermodynamically stable dense-moment ferromagnetic precipitates;
it is likely that the high-temperature ferromagnetism sometimes
found in these materials is due to precipitates.  The latest
studies \cite{Li:2005_a} appear to indicate that true long-range
order in ${\rm Ge}_{x}{\rm Mn}_{1-x}$ emerges only at 
low temperatures $\sim 12{\rm K}$ and that 
weak coupling between remote moments is mediated by holes which are
tightly bound to Mn acceptors. Further work is necessary to
determine whether this picture of magnetism, reminiscent of the polaronic physics 
discussed in the context of (Ga,Mn)P or low-carrier-density (III,Mn)V systems 
\cite{Scarpulla:2005_a,Kaminski:2003_a}, 
applies to ${\rm Ge}_{x}{\rm
Mn}_{1-x}$ DMSs. 

The possible presence of dense-moment thermodynamically stable
precipitates has also confused studies of oxide semiconductor DMS
systems.  More consistent evidence of above room temperature
ferromagnetism has been reported in Co-doped TiO$_2$  although the origin of ferromagnetism
in this material is still under debate (see, {\em e.g.}, \cite{Matsumoto:2001_a} and
recent reviews \cite{Prellier:2003_a,Fukumura:2005_a}). 
Mn-doped indium-tin-oxide (ITO) is
another promising candidate for a transparent ferromagnetic semiconductor which could be
easily integrated into magneto-optical devices. Particularly encouraging is the observation of a large
anomalous Hall effect showing that charge transport and magnetism are intimately connected 
in  this oxide DMS \cite{Philip:2004_a}. 

Other interesting related materials are the Mn-doped II-IV-V$_2$ chalcopyrites
surveyed theoretically in a first
principles calculation study \cite{Erwin:2004_a}. Three of these compounds,
CdGeP$_2$, ZnGeP$_2$ and ZnSnAs$_2$, have
shown ferromagnetism experimentally. The origin of this ferromagnetic
behavior has not been explored extensively yet.

Finally we mention recent observations of ferromagnetic order up to $\approx 20$~K 
in a layered semiconductor Sb$_2$Te$_3$ doped with V  or Cr 
\cite{Dyck:2002_a,Dyck:2005_a}. 
These highly anisotropic materials combine DMS behavior with strong thermoelectric effects. 
The character of the ferromagnetic coupling in these compounds is unclear at present.

%%%%%%%%%%%%%%%%%%%%%%%%%%%%%%%%%%%%%%%%%%%%%%%%%%%%%%%

\section{Summary}
\label{summary}

This article is a review of theoretical progress that has been achieved in 
understanding ferromagnetism and related electronic properties in (III,Mn)V DMSs. 
The materials we have focused on have randomly located  Mn($d^5$) local moments 
which interact via approximately 
isotropic exchange interactions with itinerant carriers in the semiconductor 
valence band. (Ga,Mn)As is by far the most 
thoroughly studied and the best understood system in this class. Some (III,Mn)V materials 
may exhibit fluctuations in the 
Mn valence between Mn($d^5$) and Mn($d^4$) configurations or have 
dominant Mn($d^4$) character, possibly (Ga,Mn)N for example.  
Magnetic and other properties of materials in the latter class will differ qualitatively from those of 
(Ga,Mn)As and this review makes no attempt to discuss the theory that would describe them.
When we refer to (III,Mn)V ferromagnetism below, it should be understood that any materials that 
prove to be in the latter class are excluded.  

Interest in DMS ferromagnetism is motivated by the vision that 
it should be possible to engineer systems that combine many of the technologically useful 
features of ferromagnetic and semiconducting materials.  This goal has been achieved to 
an impressive degree in (III,Mn)V DMSs, and further progress can be anticipated in the future.
The goal of high temperature semiconductor ferromagnetism flies in the face of fundamental physical
limits, and the fact that so much progress has nevertheless been achieved, is due to a serendipitous
combination of attributes of (III,Mn)V materials.  We have reserved the term ferromagnetic semiconductor 
for materials in which the coupling between local moments is mediated by carriers in the 
host semiconductor valence or conduction bands.  Then magnetic properties can be adjusted over 
a broad range simply by modifying the carrier system by doping, photo-doping, gating, heterojunction 
band-structure engineering, or any technique that can be used to alter other semiconductor electronic 
properties.  Most of these tuning knobs have already been established in (III,Mn)V ferromagnets. 

The progress that has been made in achieving (Ga,Mn)As ferromagnetism and in understanding 
its phenomenology has a few lessons.
The analysis of any DMS should start with understanding 
the properties of isolated defects associated with the magnetic element.  In the case of (Ga,Mn)As materials 
the desirable magnetic defect is substitutional Mn$_{\rm Ga}$, because Mn then both introduces a local moment and acts 
as an acceptor. The holes doped in the system by Mn$_{\rm Ga}$ impurities
provide the glue that couples the moments together.  Understanding the role of other defects 
that are present in real material is also crucial.  Substitutional Mn$_{\rm I}$ is particularly
important in (Ga,Mn)As
because it reduces the number of free moments and reduces the density of the hole gas that mediates 
ferromagnetism.  Learning how to remove defects that are detrimental to strong magnetic order is key to creating
useful materials.  These two steps have been largely achieved in (Ga,Mn)As.  There is every reason to believe that if the 
same progress can be made in other DMS materials, some will be even more magnetically
robust.       

We have reviewed in this article a 
number of theoretical approaches that shed light on what controls key properties of
ferromagnetic (III,Mn)V semiconductors.  First principles 
electronic structure calculations give a good overview of fundamental material trends across the series and explain
many of the structural characteristics of these alloys.  
Semi-phenomenological microscopic tight-binding models provide a convenient way to 
use experimental information to improve the quantitative accuracy of the description.  Another
phenomenological description that successfully models  magnetic, magneto-transport,
and magneto-optical properties is a single parameter theory that
adopts a ${\bf k}\cdot{\bf p}$ description for the host semiconductor valence bands, and assumes that the exchange
interaction between local moments and band electrons is short-ranged and isotropic.  The single exchange 
parameter that appears in this theory can be determined by fitting to known properties of an 
isolated Mn local moment, leading to parameter-free predictions for ferromagnetism.  
Qualitative models which  focus on what kind of physics can occur generically for randomly located 
local moments that are exchange coupled to either localized or itinerant band electrons, also 
provide useful insights for interpreting experiments.

The most important properties of (III,Mn)V materials are their Curie temperature
and ferromagnetic moment which reflect both the strength of the coupling between Mn local spins and its range.  
The highest ferromagnetic 
transition temperatures  in (Ga,Mn)As epilayers
have so far been achieved with substitutional Mn$_{\rm Ga}$ fractions in the 
neighborhood of 5\%  by post-growth annealing which eliminates most of the interstitial 
Mn ions. Achieving 
$T_c$ values close to room temperature in (Ga,Mn)As, which is expected to occur for  10\%
Mn$_{\rm Ga}$ doping, 
appears to be essentially a 
material growth issue, albeit a very challenging one.
In optimally annealed samples, experiment and theoretical considerations indicate that the Mn-Mn 
exchange interactions are sufficiently long range to produce a magnetic state that is nearly collinear and  
insensitive to micro-realization of the Mn$_{\rm Ga}$ spatial distribution. Magnetization and Curie 
temperatures in
these systems are well described by  mean-field theory.

The magnetic and transport properties of 
high-quality (Ga,Mn)As materials 
are those of a low-moment-density, low-carrier-density metallic ferromagnet, with a few 
special twists.  
Because of the strong spin-orbit interactions in the valence band, metallic (Ga,Mn)As shows
a large anomalous contribution to the Hall effect and the source of 
magnetic and transport anisotropies  is the more itinerant electrons, unlike
the transition metal  case in which anisotropies originate primarily in $d$-electron 
spin-orbit interactions.  
The small moment densities lead to a large magnetic hardness reflected in a single-domain-like behavior
of many (Ga,Mn)As thin films.
They also explain a large part of the orders of magnitude reduction in the current
densities required for transport manipulation of the magnetic state through spin-momentum-transfer effects.
The low-carrier density of the itinerant holes responsible for magnetic 
coupling means that they are concentrated around a particular portion of the Brillouin-zone in 
the valence band which has a large oscillator strength for optical transitions to the conduction band.
This property opens up opportunities for optical manipulation of the magnetic state that do not exist in 
transition metal ferromagnets and have not yet been fully explored.  The 
research which we have reviewed here 
that is aimed at an understanding of the optical properties of (Ga,Mn)As
ferromagnets is still incomplete, particularly for the most interesting ideal annealed materials,
and will be important in setting the ground work for the exploration of new effects.   

These conclusions do not 
necessarily apply to all (III,Mn)V ferromagnets.  For example, material trends suggest that wider band-gap
hosts would have stronger exchange scattering that would lower the 
conductivity, shorten the range of Mn-Mn exchange interactions, and increase the importance of quantum fluctuations in 
Mn and band hole spin orientations. This could eventually lead
to Curie temperatures significantly below the mean-field estimates.   
In the opposite limit,
when the exchange interaction is weak enough to be treated perturbatively, sign variations in the RKKY Mn-Mn interaction 
are expected to lead to frustration and weaken ferromagnetism at large carrier densities.  One of the important 
miracles of (Ga,Mn)As ferromagnetism is that this effect is much weaker than would naively be expected because of 
the complex valence band structure.  In (Ga,Mn)As, incipient frustration that limits magnetic stiffness
may be responsible for the weak dependence of ferromagnetic transition temperature on carrier density.
This property of (Ga,Mn)As suggests that little progress on the $T_c$ front is likely to be gained by 
non-magnetic acceptor co-doping. 

Disorder is an inevitable part of the physics of all DMS ferromagnets
because of the random substitution of elements possessing moments for host semiconductor elements.
Even in metallic, ideal annealed samples that 
have only substitutional  Mn$_{\rm Ga}$ impurities, randomness in the Mn microstructure leads to both Coulomb and 
spin-dependent exchange potential scattering.  For (Ga,Mn)As, Coulomb scattering 
dominates over exchange potential scattering, limiting the conductivity to 
$\sim 100-1000\,\Omega^{-1}{\rm cm}^{-1}$.
Frustration and disorder are certainly very important near the onset of ferromagnetism at low Mn 
density, where the network of exchange interactions that lead to long range order is still tenuous.  Studies of 
well characterized materials with low Mn fraction near the metal-insulator transition
are now possible because of progress in understanding the role 
of defects and are likely to exhibit complex interplay between disorder, and Coulomb and 
exchange interactions.  

\section*{Acknowledgments}
We thank all colleagues who have given us the
permission to show their results in this review. Among the many
stimulating discussions we would like to especially acknowledge our
interactions with experimentalists at the University of Nottingham. 
Work on the article was supported by by the Grant Agency
of the Czech Republic through Grant No.  202/05/0575, by the 
Academy of Sciences of the  Czech  Republic  through 
Institutional Support No. AV0Z10100521, by the Ministry of Education of the Czech Republic Center
for Fundamental Research LC510, by the UK EPSRC through Grant GR/S81407/01,
by the Welch Foundation, by the US Department of Energy under Grant DE-FG03-02ER45958,
and by the US Office of Naval Research under Grant ONR-N000140610122.
%\bibliography{MSWEBpublications}

\end{document}